\documentclass[a4paper, 12pt,reqno]{amsart}
\usepackage[top=1in, bottom=1in, left=1in, right=1in]{geometry}
\usepackage{floatrow}
\newfloatcommand{capbtabbox}{table}[][\FBwidth]

\reversemarginpar
\usepackage{todonotes}
\usepackage{xcolor}\definecolor{tum_blue}{RGB}{0, 115, 207}\colorlet{col_section }{tum_blue}

\usepackage{tikz}
\tikzset{
  treenode/.style = {shape=rectangle, rounded corners,
                     draw, align=center,
                     top color=white, bottom color=blue!20},
  root/.style     = {treenode, font=\normalsize, bottom color=blue!30},
  decision/.style      = {treenode, font=\normalsize, bottom color=red!30},
  env/.style      = {treenode, font=\normalsize},
  dummy/.style    = {circle,draw}
}
\usetikzlibrary{shapes,arrows}
\usepackage{
	booktabs
	,lmodern
	,multirow
	,subfigure
	,graphicx
	,amssymb
	,amsfonts
	,amsmath
	,amsthm
	,amssymb
	,stmaryrd
	,color
	,array
	,enumerate
    ,microtype
    ,geometry
    ,graphics
	,threeparttable
	,longtable
	,rotating
	,lscape
	,tabularx
	,epsfig
    ,epstopdf
	,setspace
}
\usepackage{floatrow}

\usepackage{chngcntr}
\counterwithin{table}{section}
\counterwithin{figure}{section}
\usepackage[tableposition=top]{caption}
\newcolumntype{Y}{>{\centering\arraybackslash}X}

\usepackage[multiple]{footmisc}

\usepackage{hyperref}
\usepackage{breakurl}
\hypersetup{
	pdfstartview = FitH,
	pdfauthor = {...},
	pdftitle = {...},
	pdfkeywords = {...; ...; ...; ...},
	linktocpage=true,
	colorlinks=true,      
    linkcolor=blue,       
    citecolor=blue,       
    filecolor=blue,       
    urlcolor=blue         
}

\usepackage{xurl}
\usepackage{chapterbib}
\usepackage{natbib}
\newcommand\blfootnote[1]{%
  \begingroup
  \renewcommand\thefootnote{}\footnote{#1}%
  \addtocounter{footnote}{-1}%
  \endgroup
}
 \usepackage{amsaddr}
\newcommand\independent{\protect\mathpalette{\protect\independenT}{\perp}}
\def\independenT#1#2{\mathrel{\rlap{$#1#2$}\mkern2mu{#1#2}}}
\newtheorem{theorem}{Theorem}
\newtheorem{definition}{Definition}
\newtheorem{remark}{Remark}

\newtheorem{assumption}{Assumption}
\newtheorem{lemma}{Lemma}

\begin{document}

\title[Ex ante Returns]{Just Ask Them Twice:\\ Choice Probabilities and Identification of Ex ante returns and Willingness-To-Pay}
\author{Romuald M\'{e}ango\textsuperscript{1}, Esther Mirjam Girsberger\textsuperscript{2}}
\address{\textsuperscript{1} University of Oxford and CESifo, \textsuperscript{2} University of Technology Sydney and IZA}
\noindent \blfootnote{\scriptsize{Correspondence address: Romuald M\'eango, Department of Economics, University of Oxford, Manor Road Building, Manor Road, Oxford, OX1 3UQ, United Kingdom; email: romuald.meango@economics.ox.ac.uk. A previous version of this paper has been circulated under the title: `Identification of Ex ante Returns using Elicited Choice Probabilities: an Application to Preferences for public sector jobs'. Romuald acknowledges the support from the Higher Studies Fund - Travel Fund for the Economics of Developing Countries of the University of Oxford, the Centre for Research on Structural Transformation and Growth (CROSTAG) in the Department of Economics at the University of Oxford, which in turn receives funding from the Centre for Study of African Economies (CSAE) at Oxford and the research programme on Structural Transformation and Economic Growth (STEG). Romuald acknowledges support from the AfOx Catalyst Grant, grant number AfOx-233. We thank Richard Moussa for supervising the data collection,  Adeline Delavande, St\'ephane Bonhomme, Marc Henry, Isma\"el Mourifi\'e, Fran\c{c}ois Poinas, Sami Stouli, and Martin Weidner and participants to various seminars for useful comments. The Institutional Review Board at the Department of Economics of the University of Oxford approved this study. All remaining errors are ours.}}
\date{This version: \today} 

\maketitle

\begin{abstract}
One of the exciting developments in the stated preference literature is the use of probabilistic stated preference experiments to estimate semi-parametric \textit{population distributions} of ex ante returns and willingness-to-pay (WTP) for a choice attribute. This relies on eliciting several choices per individual, and estimating separate demand functions, at the cost of possibly long survey instruments. This paper shows that the distributions of interest can be recovered from at most two stated choices, with significantly weaker assumptions. Hence, it allows for significantly shorter survey instruments. The paper also shows that eliciting probabilistic stated choices allows identifying much richer objects than we have done so far, and therefore, provides better tools for ex ante policy evaluation. Finally, it showcases the feasibility and relevance of the results by studying the preference of high ability students in C\^ote d'Ivoire for public sector jobs by exploiting a unique survey on this population.  Our analysis supports the claim that public sector jobs might significantly increase the cost of hiring elite students for the private sector.

\end{abstract}
\onehalfspacing

\clearpage

\section{Introduction}

Stated preference analyses aim to learn about human behaviour by asking people directly what they would do or choose in a hypothetical situation. Several types of stated choice experiments exist, which, in their standard format, present agents/consumers with two or more choice options/products with different attributes, and ask them to pick one.\footnote{\textit{Choice-based conjoint} experiments are a leading method for market research. They present consumers with two or more products with different attributes, and ask them to pick one. The experiment is usually repeated with varying choice attributes \citep[see a historical review in][]{mcfadden2017}. The same principle has been applied to several human capital investments or choices by social scientists \cite[see discussions in][]{hainmueller2014,maestas2023, kocsar2023, giustinelli2023}. \textit{Strategic Survey Questions} present hypothetical scenarios that tend to be very elaborate, with the aim of retrieving precise model-based parameters \citep[see an example in][]{ameriks2020a}. Some studies, termed \textit{vignette analyses} supplement choice alternatives with detailed, often visual story descriptions of alternatives. Environmental economists have developed independently the \textit{contingent valuation method} that applies to valuing natural resources, where transactional markets typically do not exist and the non-use value of the choice alternatives is of interest \citep[see, for example,][]{cummings1986}.}  
The interest in the stated preference approach has accelerated recently, and it is increasingly used to (1) describe individual preferences by estimating the willingness-to-pay (WTP) for choice attributes and (2) to conduct counterfactual policy analyses, that is, evaluating the impact of a policy that changes the distribution of choice attributes. The main advantages of the stated preference approach over the traditional revealed preference approach are that the analyst can design experiments with rich variations in choice attributes, observe individuals' stated choices in counterfactual scenarios and explore preferences over policies never implemented before. Thus, it becomes easy to estimate causal effects of choice attributes because there is no concern of self-selection on unobserved characteristics.\footnote{Recent applications are wide-ranging and include education choices \citep{arcidiacono2020, delavande2019}, mobility decisions \citep{gong2022, kocsar2022, batista2025}, long-term care and health investments \citep{kesternich2013, ameriks2020b,boyer2020}, parental investments \citep{attanasio2019,almaas2024}, marriage preference \citep{adams2019, low2024}, occupational choices \citep{maestas2023, wiswall2015, wiswall2018}, retirement decisions \citep{ameriks2020a, giustinelli2024}, voters' preference toward asylum seekers \citep{bansak2016}, and irregular migration \citep{meango2022}. \cite{almaas2024} predict that `stated preference and answers to hypothetical questions' will become part of the normal toolkit in economics.} 

An application that demonstrates the advantages of the stated preference approach is the occupational choice of jobseekers. Take, for instance, the choice of working in the public sector or in the private sector. 
A growing literature cautions against the misallocation effects that a generous public sector generates: attracted by the stability and sometimes wage premium offered by the public sector, talented individuals sort out of productive activities \citep[see, for example,][]{algan2002, albrecht2019, burdett2012, cavalcanti2020, DufloEtal2021WP, girsberger2022, mangal2024}. However, it is difficult to assess the displacement effect of the public sector, in part because it is difficult to estimate how large a pay cut a jobseeker is willing to accept to work in the public sector rather than in the private sector \citep[see some evidence in][]{gindling2020}. The challenge is that workers choose to work in the public or private sector based on the offered wage, but also on their unobserved (latent) ability and preference for one sector over the other. However, the ability and preferences are related to the wage they receive. Thus, it is difficult to isolate the response of workers to a \textit{ceteris paribus} pay cut (everything else being equal), just by observing workers, their choice of sector and the corresponding wage.

To solve the unobserved heterogeneity issue, researchers have relied on Randomised Controlled Trials (RCT) or exogenous variations that mimic a random assignment (quasi-experiments) to learn about the causal effect of an intervention/treatment on a set of individual outcomes.\footnote{For example, \cite{mangal2024} used a partial public sector hiring freeze in the state of Tamil Nadu, India, to understand how young jobseekers redirected their search effort.} However, in many settings including the occupational choice example, such exogenous variations are not readily available and randomisation is not feasible. Furthermore, with revealed preference data, the analyst only observes realised \textit{ex post} outcomes, and typically infers ex post returns of the choice. However, what drives the decision are the \textit{ex ante} returns, the perception that jobseekers have of their returns when they choose their sector.

The stated preference approach allows one to learn the preferences of jobseekers ex ante, by \textit{asking directly} people what they would choose if presented with hypothetical job offers that vary, for example, the wage, the number of hours worked, or the chance of losing the job in each sector \citep[see, for example,][]{maestas2023, wiswall2018}. 

The current state-of-the-art approach to a stated choice experiment is to ask each respondent their probability of choosing a given option, for example, the probability of choosing a public sector job, in several hypothetical scenarios. This creates a panel of choice probabilities.

Asking for a choice probability, that is, on a scale from 0 to 100 rather than a binary answer, allows respondents to express \textit{uncertainty about their choice}. \cite{juster1966} and, more recently, \cite{manski1999} note that scenario descriptions are often incomplete: for example, the job offer scenario might miss a specific amenity related to the public sector, which will be known by the jobseeker when making the decision in real life. Thus, respondents are uncertain and form beliefs about this amenity. \cite{blass2010} have coined this uncertainty a `resolvable uncertainty' and show that asking for choice probabilities places less restrictions on the underlying utility. They propose a least absolute deviation (LAD) estimator that estimates the average (or median) WTP in the population from the probabilistic stated choices, under a parametric assumption on the resolvable uncertainty.

Eliciting several choices allows one to estimate the \textit{population distribution} of WTP for a choice attribute, not just the population average.  Based on a random coefficient model, \cite{wiswall2018} estimate separate WTP parameters, one for each individual, without imposing parametric assumptions on the WTP parameters. The population distribution follows from individual-specific estimates. This approach has now been replicated several times \citep{aucejo2023, folke2022, kocsar2022, ugalde2022}. There are two crucial requirements: first, the procedure requires that the number of elicitation points exceeds the dimension of attributes considered. Second, it maintains the parametric assumption on the resolvable uncertainty \citep[see also a discussion in][]{kettlewell2024}.

This paper improves on the existing literature by showing that for very general random utility models of stated choice, the population distribution of WTP parameters can be identified and estimated from probabilistic stated choices with just \textit{two elicitation scenarios per individual} and without a parametric assumption on the resolvable uncertainty. As we explain in more detail below, it also provides new econometric tools for conducting ex ante policy evaluation. Finally, it demonstrates the feasibility and relevance of the results by studying the preference of high-ability students in C\^{o}te d'Ivoire for public sector jobs.

This paper makes three main contributions. First, building on a very general random utility model of probabilistic stated choice, it presents identification results and an estimation procedure for the population distribution of WTP parameters for a stated choice experiment with just two scenarios. The main requirements are that the choice attributes are randomly assigned across respondents during the choice experiment (Assumption \ref{ass:independence}) and that the experiment is \textit{ceteris paribus} (Assumption \ref{ass:ceteris_paribus}), as often assumed in the literature. Unlike the existing literature, the model considered here allows for nonlinearity and non-separability of the preferences with respect to choice attributes. Furthermore, the dimension of the unobserved heterogeneity is unrestricted.

The result is of practical importance. This means that one can learn about the population distribution of the WTP with significantly shorter survey instruments. \cite{wiswall2018} elicit up to 16, \cite{kocsar2022} up to 22, \cite{ugalde2022} up to 30, and \cite{aucejo2023} up to 42 per individual. With long and cognitively demanding questionnaires, there is concern that attention drops and response quality deteriorates \citep{krosnick1991, lenzner2010}. The result also opens the possibility for researchers to include short stated choice experiments within well-established panels, which sometimes impose a limit on the number of questions a researcher can propose.

Second, the paper provides new tools for ex ante policy evaluation. More specifically, the original motivation for asking the choice probabilities was that, because the scenarios are incomplete, an agent $i$ does not know exactly the returns on the hypothetical choice, say $S_i$. Instead, the perception of the agent can be described as a private distribution $F_{S,i}$. Heuristically, the assumption in the existing literature is that $F_{S,i}$ follows an extreme value type I distribution with mean $\mu_i$ and variance $\sigma_i$. Under this assumption, the procedure proposed by \cite{wiswall2018} allows estimating $\mu_i$ separately for each individual and reconstructing a population distribution. Thus, the distributions of the WTP parameters that are routinely estimated are for the \textit{individual-specific average WTP}, say $F_\mu$. This paper shows that we can provide a more complete description of $F_{S,i}$. That is, within our random utility models with probabilistic stated preferences, we can identify and estimate the population distribution of quantiles of $F_{S,i}$, with a single elicitation round and without the parametric assumption of the resolvable uncertainty.

This result matters for ex ante policy evaluation, where it is key to understand what drives agents' choices. When the uncertainty is resolved, agent $i$ will act on the \textit{realised} $S_i$, not on $\mu_i$. If there is a great deal of uncertainty, $\mu_i$ could be completely irrelevant to the policy maker, even under the `best-case hypothesis' of rational expectation \citep{manski1999}.\footnote{To see this, think of the following example: Assume that the distribution of returns that will be realised tomorrow is a uniform distribution on the unit interval. The policy maker is interested in estimating this distribution ex ante, for example, because they are concerned about redistribution in the population. Assume that all agents will receive a draw from the same distribution and have rational expectations. The distribution of ex ante mean returns is degenerate at 0.5, since all agents have a correct perception. This is not informative for the policy maker. It is even misleading because it suggests that there will be no need for redistribution.} Instead of using $F_{\mu}$ for predictions, the paper proposes to predict a distribution of realised returns, say $\bar{F}_S$, by aggregating the distributions $F_{S,i}$. If agents have rational expectations about the resolvable uncertainty, aggregating those distributions with uniform weights provides a valid predictor of the realised returns. If agents have biased beliefs about the resolvable uncertainty, the weights may be different from uniformity. In this case, the policy maker may (i) use auxiliary data to determine the weights or (ii) derive bounds on the policy parameters of interest. The use of auxiliary data is an interesting research avenue that we leave for future research. In this paper, we discuss the second option and derive a lower bound on the parameter of interest. It is important to note that the realised returns may differ from the stated returns for reasons other than a failure of the respondents to correctly predict the resolvable uncertainty. Respondents may not act as they say.\footnote{To quote \cite{juster1964}: `Survey questions [asking prospective borrowers to indicate their preferences among a set of hypothetical financing arrangements] are dangerous, mainly because it is doubtful that the respondent would act exactly as he says. From such evidence, forecasts of the future behavior of consumers cannot be made with any degree of precision, although the pitfalls can be minimized.' (p. 20). Challenging this view, \cite{mcfadden2017} provides evidence that carefully designed stated choice experiments have predictive power for consumer choices.} This discrepancy is known as the \textit{hypothetical bias}.  In this paper, we focus on the task of predicting a distribution of realised returns under the assumption that the hypothetical bias is negligible but the agent may be wrong about the distribution of resolvable uncertainty.\footnote{\cite{bernheim2022} and \cite{meango2023b} study the question of what can be learnt from combining stated choice experiments and revealed preference data even in the presence of a hypothetical bias and a wrong belief about the resolvable uncertainty.}

As a third contribution, the paper uses a stated choice experiment to analyse the preference for public sector jobs of a sample of high-ability students from two highly selective universities in C\^{o}te d'Ivoire. In the Ivorian labour market, the public sector provides most formal jobs and high labour costs due to a high skill premium have been identified as one possible constraint for firms to grow \citep{christiaensen2017}. Within the choice experiment, respondents are presented with two job offers, one from the public sector and one from the private sector, with attributes that vary exogenously. 
The methodology proposed in this paper is particularly suited for this analysis for three reasons: (i) the perception of public and private sector jobs is very different in the population; therefore, we should not except the resolvable uncertainty to be symmetric. In fact, the test procedure developed in \cite{meango2025} rejects the assumptions of a random coefficient model with a type I extreme value or symmetric resolvable uncertainty.
(ii) The interest of the policy maker is in the competition induced by the public sector on the private sector. But the decision to choose the public or private sector is only in the future for students. Therefore, the policy maker who elicits students' preferences must account for their uncertainty and the possibility that students have a wrong perception of the resolvable uncertainty. The proposed framework for ex ante policy evaluation allows one to derive a lower bound on the predicted cost elasticity of hiring top-skilled workers in the private sector. The lower bound does not require students to have rational expectations. 
(iii) Finally, the survey collects preferences over ten attributes but has only five scenarios. This makes the estimation of separate demand functions unfeasible.

\vskip12pt
\paragraph{\textbf{Plan of the paper}} Section \ref{sec:framework} formalises the choice model and our econometric framework. We achieve all three contributions by focussing on the identification of a \textit{reservation value} and \textit{ex ante returns}. The reservation value we consider is the value of the numeraire that makes an agent indifferent between two sectors. For example, we consider the wage in the public sector that makes a jobseeker indifferent between the public and private sectors. The ex ante returns on an offer is then the difference between the wage in a given offer and the reservation wage. In other words, this is the maximum price the agent is willing to pay to secure the offer. 
At the \textit{time of decision}, an agent $i$'s choice is characterised by their ex ante returns, $S_i$, which depend on the choice attributes. The WTP for any specific choice attribute is also characterised by the ex ante returns: it is measured by the difference in returns resulting from a change of the choice attribute. During the stated preference experiment, the agent does not know $S_i$ because the experimental scenario is incomplete. The agent only has a private distribution of ex ante returns $F_{S,i}$. The spread of the distribution $F_{S,i}$ characterises the resolvable uncertainty. To describe agents' preferences, our results describe the quantiles and mean of $F_{S,i}$. It also describes the WTP by measuring the effect that changing choice attributes produces on the quantiles and mean of $F_{S,i}$. 

Section \ref{sec:identification} contains the identification result for $F_{S,i}$. Within it, Theorem \ref{th:characterisation} shows that information about elicited choice probabilities $P_{i}$ can be translated into information about (i) the distribution of quantiles of $F_{S,i}$. 
A predictor of the realised returns can then be derived by aggregating the distributions of the quantiles of $F_{S,i}$ (Section \ref{subsec:counterfactual}). Theorem \ref{th:characterisation_2} presents additional identification results for the distributions of (ii) mean returns in the population, $F_\mu$, (iii) quantile effects, and (iv) mean effects of choice attributes. The latter two quantify the WTP for specific choice attributes. For identifying distributions (i) - (ii), a single stated choice is sufficient. For distributions (iii) and (iv), a pair of stated choices is necessary. Our results are related to \cite{lewbel2011} who study a referendum contingent valuation model in resource economics, where one is interested in features of the distribution of WTP placed by consumers on a public good such as endangered species. They show that by eliciting a binary decision on the choice of preserving the resource at exogenously set costs, one can interpret the cost as a special regressor and retrieve nonparametrically the distribution of WTP in the population. This paper considers in addition uncertainty in the decision, which is expressed by stating choice probabilities rather than discrete choice. This allows a richer analysis of ex ante returns. 

 
Section \ref{sec:estimation} considers the estimation of the objects of interest. It proposes a novel strategy for calculating the distribution of ex ante returns and WTP in the population. The identification result is constructive, and the nonparametric estimation is conceptually feasible. However, in practice, the researcher often faces the curse of dimensionality, because the choice attributes can be many, their variation in choice experiments is often restricted to salient values, and the typical sample size is limited. Thus, the paper suggests a semiparametric alternative: In the first step, it estimates the quantiles of the stated choice conditional on scenario attributes by several quantile regressions as in \cite{chernozhukov2013}. In the second step, this quantity is `plugged in' the characterisation of Theorem \ref{th:characterisation} (or Theorem \ref{th:characterisation_2}, depending on the object of interest). 

Section \ref{sec:results} analyses the preference for public sector jobs of a sample of high-ability students from two highly selective universities in C\^{o}te d'Ivoire. The survey reveals that beliefs about the characteristics of the public and private sectors are consistently different in the population. On the one hand, the private sector is perceived as a dynamic sector with more abundant and better paid jobs and more opportunities for progression. However, these benefits are tied to stressful work conditions. On the other hand, the public sector appears as a secure and stable alternative, even if less well paid. 
Given these beliefs, it is interesting to note that preferences for specific sectors vary significantly between students. The estimated ex ante returns are very heterogeneous and sometimes large in absolute value. 
There is also considerable uncertainty about these returns.

Given these preferences, Section \ref{sec:policy} estimates the cost elasticity associated with a private sector expansion for the market of high-skilled jobseekers using a predicted distribution of returns. Because we do not have auxiliary information to point identify the distribution, we derive a lower bound for the cost elasticity, thanks to our characterisation of the distribution of quantiles. 
Under the assumption that the private sector cannot discriminate among the workers, our calculation suggests a cost elasticity of at least 1.34 (for a 1 percent increase of workers, a 1.34 percent increase of the total wage bill), an economically large cost. 

\section{Econometric Framework}\label{sec:framework}
This section presents the econometric framework. Section \ref{subsec:example} introduces a job-choice model as a motivating example. Section \ref{subsec:framework} describes the choice model in the case of a binary option. Section \ref{subsec:assumptions} details the key notations and assumptions about the preferences elicited. Section \ref{subsec:multinomial} makes the connection with the general case of polychotomous choice. Describing the binary choice model first has two advantages: (i) the notations are lighter, and (ii) one can move seamlessly to the multinomial model by considering them as a collection of binary choices. 

\subsection{Motivating example: job-choice model between public and private}\label{subsec:example} Consider a job choice model in which individuals have preferences over jobs characterised by a bundle $(y,a)$ of income $y$ and sector-specific amenities $a$. Public and private sector jobs offer different wages and amenities, such as job security. Let option 0 be a job in the private sector and option 1 a job in the public sector. Suppose that the individual utility is described by a utility function: $U_i(y,a)$, which is strictly increasing with income $y$. For example, a CES utility gives: $$U_i(y,a) = \bigl({\alpha_{i} y^{\beta_i} + (1-\alpha_{i}) a^{\beta_i} }\bigr)^{\frac{1}{\beta_i}}.$$
In the classical revealed preference framework, when deciding between a job offer in the private sector, $(y_0,a_0)$, and the public sector, $(y_1,a_1)$, the individual chooses the one that maximises their utility. During a stated choice experiment, jobseeker $i$ is presented with a pair of wages associated with a job in the private sector ($y_{0i}$) and a job in the public sector ($y_{1i}$). They are asked to state their probability of choosing option 1 over option 0. Importantly, the stated choice experiment does \textit{not} specify the associated amenities. Thus, the respondent does not know the vector of amenities $(a_0,a_1)$, but holds beliefs about its distribution $F_{a_0,a_1}(.;\rho_i)$, which depends on a parameter $\rho_i$ (e.g., agents' beliefs about the correlation of amenities offered in each sector). The jobseeker knows $\eta_i:=(\alpha_i,\beta_i,\rho_i)$, but these parameters are not observable for the analyst. Thus, $\nu = (a_{i0},a_{i1})$, represents the resolvable uncertainty, the information that agent $i$ does not have from the scenario but that will be revealed at the time of decision. The vector $\eta^* = (\eta,\nu)$ represents the unobserved heterogeneity at the time of decision. 
This framework allows describing the choice probability of agents: When presented with scenario $(Y_{1},Y_{0})$, the respondent reports their probability of choosing option 1, $P_i = \Pr(U_i(Y_1,a_1) \ge U_i(Y_0,a_0)\vert \eta_i).$

For the identification results, we will pay particular attention to the \textit{reservation wage} $R_i$ and to the \textit{ex ante returns}, $S_i$. In our context, the reservation wage corresponds to the wage in the public sector, which, given $(y_0,a_1,a_0)$, makes the individual indifferent between the two job offers: $U_i(R_i,a_1) = U_i(y_0,a_0)$.
The ex ante returns correspond to a reservation price, the maximum price that $i$ is willing to pay to secure the public sector offer. It can be negative or positive and makes $i$ indifferent between options 0 and 1 at the time of decision, that is, $U_i (y_0, a_0) = U_i (y_1 - S_i, a_1)$. By the monotonicity of $U_i$ with respect to income, it follows that: $$\begin{array}{ll}
    R_i =&  U_i^{-1}(U_i (y_0, a_0),a_1),   \\
     S_i =&  y_1 - R_i,
\end{array}$$ where $U^{-1}$ is the inverse with respect to the first argument. For example, with a CES utility, this corresponds to: $$R_i = \bigl[{y_0^{\beta_i} + \dfrac{(1-\alpha_{i})}{\alpha_i} (a_{0i}^{\beta_i} - a_{1i}^{\beta_i})}\bigr]^{\frac{1}{\beta_i}}.$$
There are three features that make $R_i$ and $S_i$ interesting: First, they characterise the agent choice. Agent $i$ chooses option 1 if and only if $y_1$ exceeds $R_i$ or equivalently $S_i \ge 0$ and reports $P_i = \Pr(S_i\ge 0\vert \eta_i)$ during the choice experiment.
Second, unlike the utility function, $R_i$ and $S_i$ have the scale of the \textit{numeraire} (here income) and the intuitive interpretation of reservation wages and prices. $S_i$ answers directly the question: `How large a pay cut is a jobseeker willing to take to work in the public sector rather than in the private sector?' Furthermore, when considering additional choice attributes below, we can show that the WTP for these attributes is related to $S_i$. Third, $S_i$ is linear in $y_1$.  This is crucial for our identification result.

The next section presents a more general framework. It includes the possibility of (i) other choice attributes manipulated by the researcher, (ii) unobserved heterogeneity $\eta$ and resolvable uncertainty $\nu$ of unrestricted dimension, and (iii) some \textit{unresolvable} uncertainty. The extension to (iv) polychotomous choices is discussed in Section \ref{subsec:multinomial}. To simplify the language and in line with our empirical application, we will still use the term `income' to refer to the numeraire and to `reservation wage' to refer to $R_i$.

\subsection{The binary choice model}\label{subsec:framework}

Consider an economic agent $i$, a binary choice alternative $0$ or $1$, and two consecutive periods: a time of preference elicitation (today) and a time of decision (tomorrow). The utility of agent $i$ in option $d \in \{0,1\}$ is given by: $U_d(y_d,z_d,\eta_i^*)$. As is customary in stated preference analyses that translate preferences into pecuniary values, the utility depends on a \textit{numeraire} that we will interpret here as expected income $(y_0,y_1)$ in option 0 and option 1 respectively.  We assume that the utility is continuous and strictly increasing with respect to the \textit{numeraire}. The vector $z = (z_0,z_1)$ represents further choice characteristics that can be manipulated within a hypothetical choice experiment. We will use the shorthand $x:= (y_0,y_1,z)$ and denote the support of $x$ by $\mathcal{X}$. Without loss of generality, we will assume that the support of $(y_0,y_1)$ is $(0,y_{\max})^2$.

In the empirical application, the choice is between a private sector job and a public sector job. $X$ represent the job attributes: for example, the monthly wage, the number of hours, the probability of being laid off in each sector. The wage serves as the numeraire.

The random vector $\eta^*$ represents the unobserved individual heterogeneity at the time of decision. It represents possibly unobserved characteristics of individual $i$ and unobserved choice attributes that matter in the decision, but are neither specified nor altered by the choice experiment. They are never observed by the analyst, but are known by the agent at the time of decision (tomorrow). This could include a private taste for one of the two options or some private information about the returns to choosing one option over the other.

At the time of decision, $i$ chooses between option $0$ and option $1$ based on a rule:
\begin{equation}\label{eq:choice_equation_utility}
    D_i(x) = 1\{U_1(y_1,z_1,\eta^*) \ge U_0(y_0,z_0,\eta^*) \} 
\end{equation}
$1\{A\}$ takes value 1 if condition $A$ holds and 0 otherwise. $D_i(x)$ is the choice of individual $i$. The notation borrows from the potential outcome framework, as $D_i(x)$ represents $i$'s choice, when the first set of choice attributes are exogenously set to $x$. Noting that the utility is continuous and strictly increasing in income, this is equivalent to the threshold-crossing rule:
\begin{eqnarray}
    D_i(x) &=& 1 \bigl\{{S\left({x,\eta_i^*}\right) \ge 0}\Bigr\}, \text{ where } \label{eq:choice_equation}\\
    S(x,\eta_i^*) &=& y_1 - R(y_0,z,\eta_i^*) \label{eq:definition_S} \text{, and } \\
    R(x,\eta_i^*) &=& U_1^{-1}\bigl({U_0(y_0,z_0,\eta^*),z_1,\eta^*  }\bigr)  \label{eq:definition_R}
\end{eqnarray}
$R(x,\eta_i^*)$ is the reservation wage. The agent chooses option 1 if and only if the offer wage $y_1$ exceeds the reservation wage (given the choice attributes). It satisfies: $U_1(R(x,\eta^*_i),z_1,\eta^*_i) = U_0(y_0,z_0,\eta^*_i)$. $S \left({x,\eta_i^*}\right)$ is $i$'s returns to choosing option 1 over option 0 at the time of decision, when the choice attributes are described by $(x,\eta_i^*)$. Thus, ex ante returns are interpreted as the agent's perceived returns before they take the decision. They are the fundamental drivers of individual decisions \citep{heckman2006}.\footnote{Most contributions that attempt to understand (the distribution of) ex ante returns use a revealed preference approach \citep{carneiro2003,cunha2005, heckman2006,cunha2007,heckman2007,trachter2015,lee2015,eisenhauer2015, bhuller2022}. The definition of returns adopted in this paper differs from the use of the term in the literature surveyed by \cite{cunha2007} that refers to returns as only to ex ante \textit{earnings} returns. Appendix \ref{app:comparison} discusses in appropriate length the differences between the approach taken in this paper and the revealed preference approach of the seminal contributions in this literature.}

Note that although we use the notation $R(x,\eta_i^*)$ for concision, $R$ depends on $(y_0,z)$ and not on $y_1$. Note also that $S$ satisfies the following property, $S(y_0  ,y_1-S(y_0,y_1,z,\eta^*),z,\eta^*) = S(y_0 , R(y_0,z,\eta^*),z,\eta^*) = 0$.

\begin{remark}
It is possible to circumvent the definition of a random utility model and start directly from a threshold crossing model defined as:\begin{eqnarray}
    D_i(x) =1 \bigl\{{S\left({x,\eta_i^*}\right) \ge 0}\Bigr\}, \text{ where } \label{eq:choice_equation_threshold}\\
    S(y_0 ,y_1 - S(y_0,y_1,z,\eta_i^*),z,\eta_i^*) = 0 \label{eq:char_S}
\end{eqnarray}
Under the assumption that the mapping $y_1 \mapsto S(.,y_1,.)$ is continuous and strictly increasing, equations (\ref{eq:choice_equation_threshold}) and (\ref{eq:char_S}) uniquely characterise $S$. See the proof in Appendix \ref{app:threshold-crossing}. 
\end{remark}

\subsection{Elicited preferences in the binary choice} \label{subsec:assumptions} Prior to their decision and at the time of elicitation (today), $i$ is asked to state their preference over the binary choice alternatives, $0$ or $1$, in a hypothetical scenario characterised by a vector of observable, manipulable characteristics $X_{i} := (Y_{0,i},Y_{1,i},Z_{i})$, where $Y_{d,i}$ is the hypothetical income for option $d$, and $Z_{i}$ collects the remaining hypothetical characteristics.

The stated choice model considers an environment of sequential resolution of uncertainty in which the value of $\eta_i^*$ is only revealed tomorrow. Due to the time gap between the time of elicitation and the time of decision, and since $\eta_i^*$ is not specified by the scenario, $i$ does not know the realisation of $\eta_i^*$, although they might know some subcomponent. In the motivating example, the individual has knowledge of their own preference parameters $(\alpha_i,\beta_i)$ and of the correlation of unspecified amenities ($\rho_i$). The model assumes that $i$ entertains a probabilistic distribution over their reservation wage and returns based on their available information, characterised by a vector $\eta_i$. Define for any $r,s$ in $\mathbb{R}$: \begin{equation}\label{eq:definition_FS}
\begin{array}{lll}
    F_{R,i}(r;x) &:=& \Pr(R \left({x,\eta_i^*}\right) \le r \vert \eta_i).\\
    F_{S,i}(s;x) &:=& \Pr(S \left({x,\eta_i^*}\right) \le s \vert \eta_i).
    \end{array}
\end{equation}
The perceived distributions $F_{R,i}$ and $F_{S,i}$ will be the main object of analysis. In traditional stated preference analyses, $S$ (and thus $R$) is assumed to be known, and the distribution is degenerate. In this case, $S$ is recovered as a WTP parameter from the stated choices. However, when the uncertainty is resolved sequentially, agent $i$ does not know $S$ or $R$. In other words, because hte scenario is incomplete, the respondent is unsure of their reservation wage. Instead, they entertain a distribution $F_{R,i}$, equivalently $F_{S,i}$. As the agent reveals their probabilistic choice, they also reveal information about $F_{R,i}(s;x)$.

More formally, during a survey experiment, $i$ is presented with the scenario $X_{i}$. The next assumption makes explicit that the scenarios are determined exogenously.
\begin{assumption}[Independence]\label{ass:independence}
For any individual $i$ in the population, $X_i \independent \eta_i$ and has support $\mathcal{X}$.
\end{assumption}
Assumption \ref{ass:independence} is achieved by design in most stated preference experiments. Choice attributes are chosen by the analyst independently of the individuals characteristics. The support of the experiment determines the support on which identification is achieved.

For the scenario presented, respondents are asked to state their chance of choosing option 1 over option 0, say $P_{i}$. Assumption \ref{ass:ceteris_paribus} makes explicit a common assumption in the stated preference literature: the stated choice experiment is construed as a \textit{ceteris paribus} experiment.

\begin{assumption}[Ceteris paribus]\label{ass:ceteris_paribus}
    Presented with scenario $X_{i}$, respondent $i$ states:
\begin{equation}
    \label{eq:definition_m}
P_i = m(X_i,\eta_i) \text { where } m(x,\eta_i):= \Pr(S \left({x,\eta_i^*}\right) \ge 0 \vert \eta_i) = 1- F_{S,i}(0;x).
\end{equation}
\end{assumption}
The mapping $x \mapsto m(x,\eta)$ defines the \textit{stated demand function} for an individual with characteristic $\eta$. Assumption \ref{ass:ceteris_paribus} requires that the respondents are able to report their stated choice as if $X_{i}$ was determined exogenously from the unspecified attributes. They do not infer new unspecified attributes as the specified attributes change.  In practice, to ensure that it holds, the survey design includes explicit or implicit instructions for the respondents.

To see the importance of this assumption, it is instructive to compare equation (\ref{eq:definition_m}) with a more general definition of elicited preferences:
    \begin{equation}\label{eq:definition_non_cpa}
        P_{i} = \Pr \left({S \left({X_{i},\eta_i^*}\right) \ge 0 \Bigr| X_{i},\eta_i }\right).
    \end{equation}
In equation (\ref{eq:definition_non_cpa}), the perceived distribution of $\eta^*$ changes with $X_{it}$ and the respondent uses it to infer a distribution of resolvable uncertainty. This is sometimes called the \textit{fill-in problem} \citep[see, for example,][]{hudomiet2018}. It makes it impossible to distinguish between the effect of the choice attributes on the preferences and the beliefs. Assumption \ref{ass:ceteris_paribus} is key because it ensures that the analyst can take advantage of the variation in the choice attributes to understand the preferences. In this paper, we do not relax this assumption, which is maintained in the stated preference literature. See a further discussion in Section \ref{subsec:comparison}.
\begin{remark}[Unresolvable uncertainty]  Some uncertainty may remain unresolved at the time of decision (tomorrow). For example, the jobseeker may learn about some job amenities only after they have taken the decision (the day after tomorrow): the work atmosphere, the financial situation of the firm, etc. In this case, the ex post utility of the agent can be characterised by the utility function $\tilde{U}_d(y_d,z_d,\eta^*,\xi_d)$, where $\eta^*$ is known at the time of decision, but $\xi_d$ only materialises after the decision is taken. The framework accommodates unresolvable uncertainty very naturally. The analyst can assume that the agent maximises the expected utility given by:
\[U_d(y_d,z_d,\eta^*) = \mathbb{E}\left({\tilde{U}_d(y_d,z_d,\eta^*,\xi_d) \vert y_d,z_d,\eta^*}\right)\]   
\end{remark}
\subsection{Polychotomous choice} \label{subsec:multinomial} This section extends the framework to a general discrete choice. Thanks to the flexibility of the framework, the main changes are mainly notational.

Consider an economic agent $i$ with choice alternatives $\mathcal{J} = \{0,1, \ldots, J\}$. As before, the utility of agent $i$ in option $j \in \mathcal{J}$ is given by: $U_j(y_j,z_j,\eta_i^*)$, continuous and strictly increasing in $y_j$. At the time of decision, $i$ chooses between the $J+1$ options based on a rule:
\begin{equation}\label{eq:choice_equation_utility_multinomial}
    D_i(x) = j \textrm{ iff } 1\Bigl\{{U_j(y_j,z_j,\eta^*) \ge \max_{k \in \mathcal{J}\backslash{j}} U_k(y_k,z_k,\eta^*) }\Bigr\} 
\end{equation}
Under the monotonicity assumption, this is equivalent to the model:
\begin{eqnarray}
    D_{i}(x) &=& \sum_{j = 0}^J j \;1 \bigl\{{S_j\left({x,\eta_i^*}\right) \ge 0}\Bigr\} \text{, where } \label{eq:choice_equation_j}\\
    S_j(x,\eta_i^*) &=& y_j - R_j\left({y_{-j},z,\eta_i^*}\right) \label{eq:definition_S_j} \text{ with } y_{-j} = (y_0, \ldots,y_{j-1},y_{j+1}, \ldots, y_J)\text{, and } \\
    R_j(x,\eta_i^*) &=& U_j^{-1}\left({\max_{k \in \mathcal{J}\backslash{j}}U_k(y_k,z_k,\eta^*),z_j,\eta^*  }\right)  \label{eq:definition_Rj}
\end{eqnarray}
The polychotomous framework can be construed as a collection of binary choices in which the individual chooses between the option $j$ and the alternative that maximises the utility of the remaining options. This maximisation is equivalent to checking whether the wage offered in option $j$ exceeds the reservation wage $R_j(x,\eta_i^*)$, where the reservation wage is defined considering all other alternatives. As before, $S_j(x,\eta_i^*)$ represents the maximum price that $i$ is willing to pay to secure option $j$.

The interest of the analyst is in the quantity define for $j \in \mathcal{J}$ and $r,s$ in $\mathbb{R}$: \begin{equation}\label{eq:definition_FS_j}
\begin{array}{lll}
    F_{R,j,i}(r;x) &:=& \Pr(R_j \left({x,\eta_i^*}\right) \le r \vert \eta_i).\\
    F_{S,j,i}(s;x) &:=& \Pr(S_j \left({x,\eta_i^*}\right) \le s \vert \eta_i).
    \end{array}
\end{equation}
Presented with scenario $X_{i} = (X_{0i}, \ldots,X_{Ji})$, respondent $i$ states:
\begin{eqnarray}
    \label{eq:definition_m_multinomial}
P_{ji} &=& m_j(X_i,\eta_i), j \in \mathcal{J},
\text { where }\\ m_j(x,\eta_i) &:=& \Pr(S_j \left({x,\eta_i^*}\right) \ge 0 \vert \eta_i) = 1- F_{S,j,i}(0;x).
\end{eqnarray}
The above equations show that the flexibility of the framework allows us to accommodate the general discrete choice model with minimal change in notation from the binary case: (i) indexing the reservation wage and ex ante returns, the choice probabilities and the stated demand function with the subscript $j$, and (ii) replacing $y_0$ with $y_{-j}$. Therefore, the remainder of the text presents the results for the binary case. The results for the general case follow \textit{mutatis mutandis}.

\section{Identification of the Population Distribution of Ex Ante Returns}\label{sec:identification}
This section provides novel nonparametric identification results for the population distribution of ex ante returns. Section \ref{subsec:main_results} provides the main identification result for the distribution of reservation wages and ex ante returns. Section \ref{subsec:additional_results} presents additional results for other parameters of interest, including the WTP. Then, Section \ref{subsec:counterfactual} shows how the results can be used to perform an ex ante prediction. Finally, Section \ref{subsec:comparison} compares the framework and identification results with the existing literature.

\subsection{From the stated preferences to the distribution of returns}\label{subsec:main_results} 
Theorem \ref{th:characterisation} builds on the following Lemma.
\begin{lemma}\label{lem:characterisation}
In the model described by equations (\ref{eq:choice_equation})-(\ref{eq:definition_R}) and (\ref{eq:definition_FS}) (or equations (\ref{eq:choice_equation_threshold}) - (\ref{eq:definition_FS})), and under Assumptions  \ref{ass:independence} and \ref{ass:ceteris_paribus}, the following holds:
\begin{eqnarray}
F_{R,i}(r;x) &=&  m(y_0,r,z,\eta_i), \text{ for any } r \in (0,y_{\max}), \label{eq:char_fri}\\ 
F_{S,i}(s;x) &=&  1 - m(y_0,y_1-s,z,\eta_i), \text{ for any } s \text{ such that }\in y_1-s \in (0,y_{\max}). \label{eq:char_fsi}
\end{eqnarray}
\end{lemma}
All proofs are in Appendix \ref{app:proofs}. Lemma \ref{lem:characterisation} articulates the link between the private distribution of reservation wage and returns (LHS), our objects of interest, and the stated demand function (RHS). It states that knowledge of the stated demand function $m$ is enough to recover $F_{R,i}$ and $F_{S,i}$. One way of understanding this result is to reinterpret the stated choice experiment. When the analyst asks: `what is the chance that you choose option 1?', this is equivalent to asking: `what is the chance that the offered wage in option 1 exceeds your reservation wage?' or equivalently, `what is the chance that your returns are positive?' For example, the chance that an individual accepts the private job offer is exactly the chance that the offer from the private sector exceeds their reservation wage. If we could ask this question for the same individual and at different wage offers, we could map for any given individual the distributions of reservation wage and ex ante returns. The main point of Theorem \ref{th:characterisation} below is that the analyst does not need to ask this question several times to the same person if their objective is to understand characteristics of the distributions of $R_i$ and $S_i$ at the aggregate level. For example, the literature has been interested in the distribution of medians and means of $S_i$. According to Theorem \ref{th:characterisation}, it suffices to randomise the wage offers (and other choice attributes) to understand the quantiles of the stated demand function $m$.

More specifically, in the general case where $\eta$ is of unrestricted dimension, the stated demand function is not identified from the data. Hence, the individual-specific distributions $F_{R,i}$ and $F_{S,i}$ are not available using equations (\ref{eq:char_fri}) and (\ref{eq:char_fsi}). However, by averaging $\eta_i$ out, one can learn about the distribution of quantiles of $F_{S,i}$ in the population. Averaging is possible because the quantile treatment response of $m(x,\eta)$ is identified under Assumption~\ref{ass:independence}.


Theorem \ref{th:characterisation} is one of the main results. It shows how information about the stated choices $P$ can be harnessed to learn about the distribution of quantiles of $F_{R,i}$ and $F_{S,i}$. 

\begin{definition}
     Let $F_{\tilde{X}}$ be the cumulative distribution function of the variable $\tilde{X}$, which is of interest to the analyst. For example, the analyst may be interested in setting all observed attributes to be the same across choice options, except for one. This allows understanding the WTP for this attribute.

    Denote by $Q_{P|X}$ the conditional (on X) quantile of the variable $P$.
    Define: $Q_{R,i}(\tau;\tilde{X}):=\inf \{r: F_{R,i}(r;\tilde{X}) \ge \tau\}$ and
    $Q_{S,i}(\tau;\tilde{X}):=\inf \{s: F_{S,i}(s;\tilde{X}) \ge \tau\}$, the quantile functions associated to $F_{R,i}$ and $F_{S,i}$.
    
    Define also $F_{Q,R}(r;\tau, F_{\tilde{X}}):= \Pr\left({Q_{R,i}(\tau;\tilde{X}) \le r }\right)$ and $F_{Q,S}(s;\tau, F_{\tilde{X}}):= \Pr\left({Q_{S,i}(\tau;\tilde{X}) \le s }\right)$ the distribution of $\tau$th-quantile of ex ante returns in the population.
    
    Finally, denote $t(s,x) = (y_0,y_1-s,z)$.
\end{definition}

\begin{theorem}\label{th:characterisation}
In the model described by equations (\ref{eq:choice_equation})-(\ref{eq:definition_R}) and (\ref{eq:definition_FS}), suppose that Assumptions \ref{ass:independence} and \ref{ass:ceteris_paribus} hold, and that $F_{\eta^*|\eta}$ is a continuous distribution.
For any real value $r$ and $\tau \in [0,1]$,
\begin{eqnarray}\label{eq:ident_FQ}
\nonumber F_{Q,R}(r;\tau, F_{\tilde{X}}) = \Pr\left[{F_{R,i}(r;\tilde{X}) \ge \tau}\right] = \int_{\mathcal{X}} \int_0^1 1 \bigl\{ {Q_{P \vert X}(a \vert y_0,r,z) \ge \tau }\bigr\} da \;dF_{\tilde{X}}(y_0,z).
\end{eqnarray}
For any real value $s$ such that $(y_0+s,y_1,z) \in \mathcal{X}$ and $\tau \in [0,1]$,
\begin{eqnarray}\label{eq:ident_FQ}
\nonumber F_{Q,S}(s;\tau, F_{\tilde{X}}) = \Pr\left[{F_{S,i}(s;\tilde{X}) \ge \tau}\right] = \int_{\mathcal{X}} 1 \bigl\{{ Q_{P \vert X}(a \vert y_0,y_1-s,z) \le 1- \tau }\bigr\} da\;dF_{\tilde{X}}(y_0,y_1,z).
\end{eqnarray}
\end{theorem}
Theorem \ref{th:characterisation} recovers the distribution of the cdf $F_{R,i}(s;\tilde{X})$ and $F_{S,i}(s;\tilde{X})$ in the population from the \textit{observed} conditional quantile $Q_{P\vert X}(a|x)$.  Hence, it describes individuals' perceived reservation wage and returns, accounting for their perceived uncertainty. Alternatively, it can be represented as a set of distributions of quantiles $F_{Q,R}(s;\tau, F_{\tilde{X}}), F_{Q,S}(s;\tau, F_{\tilde{X}}), \tau \in (0,1)$. This has an intuitive interpretation: for example, $F_{Q,S}(s;\tau, F_{\tilde{X}})$ is the proportion of the population for whom the $\tau$-quantile of returns is lower than some given value $s$.

\subsection{Additional Results}\label{subsec:additional_results}
The identification of ex ante returns and their distribution in the population is of interest in and of itself. The ex ante returns are also fundamental to characterise further parameters that describe individual preferences.\footnote{The discussion being similar when using the reservation wage, this section concentrates on the ex ante returns for concision.} This section presents identification results for additional parameters of interest: mean ex ante returns, quantile effects, and mean effects. The latter two describe the WTP for choice attributes.

Complementing the definition of quantiles, it is straightforward to define parameters routinely encountered in the literature:
\begin{enumerate}
\item Mean returns: $ \mu_S(x,\eta) := \mathbb{E}\left[{S(x,\eta^*) \vert \eta }\right] $;
\item Quantile effects: $qWTP(h,\tau;x,\eta):= Q_{S}(\tau;x+h,\eta) - Q_{S}(\tau;x,\eta)$, where $h \in \mathbb{X}$.
\item Mean effects: $mWTP(h;x,\eta):= \mu_S(x+h,\eta) - \mu_S(x,\eta)$.
\end{enumerate}
Quantile effects ($qWTP$) and mean effects ($mWTP$) measure the contribution of specific choice characteristics to agents' utility; in other words, the willingness to pay for these attributes. For example, take $h = c\boldsymbol{e}_j$, where $c$ is a constant and $\boldsymbol{e}_j$ is a vector of zeros except for the $j$-th component, which equals one. The quantile effect measures the quantile change that is attributable to a change $c$ in the $j$-th component of the vector of choice characteristics. In our empirical application, it would measure, for example, how the $\tau$-quantile of returns changes with an increase in the chance of being laid off. Note that, unlike the usual definition of quantile effects in a population, this effect is individual-specific: It measures the difference in the quantiles of perceived returns for the same individual. The mean effect has a similar interpretation.


We start our discussion with the following remark:

\begin{remark}\label{rem:average}
Given knowledge of the distribution of $F_{S,i}(s;x)$ (Lemma \ref{lem:characterisation}), one can derive the first moment, say $\mu_S(x,\eta)$, and conditional quantiles, say $Q_{S}(\tau;x, \eta)$, as in \cite{chernozhukov2020} \citep[cf. also][pp. 113-114]{karr1993}. Let $\mathcal{S}$ denote the support of $S$, and assume that $\mu_S$ is bounded. Denote by $t(s,x) := (y_0,y_1-s,z)$. Then: 
\begin{eqnarray}
\nonumber \mu_S(x,\eta) &=& 
\int_{\mathcal{S}} \left[{m(t(s,x),\eta) - 1\{s\le 0\} }\right] ds,\\
Q_{S}(\tau;x,\eta) &=& \int_{\mathcal{S}} \left\{{\left[{ 1- m(t(s,x),\eta)}\right] \le \tau}\right\} - 1\{s\le 0\} ds.
\label{eq:average_return}
\end{eqnarray}
\end{remark}

Building on Remark \ref{rem:average}, we can provide a characterisation for the distributions of interest as functionals of the stated demand function.
Recall that $\mathcal{S}$ denotes the support of $S$. Under the conditions of Theorem \ref{th:characterisation} and for $\mu_S$ bounded, the population distributions of interest are obtained by integration. For example:
\begin{eqnarray}
\Pr\left({\mu_S(X,\eta) \le y }\right)
&=&  \int_{\mathcal{X}} \int_{\mathcal{H}}  1 \left\{{\mu_{S}(x,n) \le y }\right\}  dF_{\eta}(n) dF_{\tilde{X}}(x), \label{eq:def_mu}
\end{eqnarray}
where $\mu_S$ is characterised by equation (\ref{eq:average_return}).
Equation (\ref{eq:def_mu}) 
implies that the identification of the function $m(x,\eta)$ and the distribution of $\eta$ is sufficient for the identification of the distributions of interest. 
As before, the quantile treatment response, $Q_{m(x,\eta)}(a)$ is identified from the single cross-section by $Q_{P|X}(a|x)$. This turns out to be sufficient for the identification of the population distribution of mean returns ($\mu_S$).

Things are more challenging with mean effects ($mWTP$) and quantile effects ($qWTP$). This is because, in general, the difference of two quantile functions is not equal to the quantile of the difference. Hence: $Q_{m(x+h,\eta) - m(x,\eta)} \neq Q_{m(x+h,\eta)} - Q_{m(x,\eta)}$. An exception is when there is rank invariance. This would imply that $\eta$ is unidimensional, a hypothesis we do not wish to impose in the general case. Instead, Theorem \ref{th:characterisation_2} uses a pair of scenarios, $(P_1,X_1,P_2,X_2)$, to learn about the joint distribution of quantiles.

\begin{theorem}\label{th:characterisation_2}
Let:
\[
\begin{array}{lll}
q(x,a) &=& Q_{P \vert X}(a \vert x) \\
A^{\mu}(x,a) &=&  \int_{\mathcal{S}} \left[{ q(t(s,x),a) - 1\{s\le 0\}}\right]ds \\
A^\tau(x,a) &=& \int_{\mathcal{S}}  \left[{1 \left\{{\left[{ 1- q(t(s,x),a)}\right] \le \tau}\right\} - 1\{s\le 0\}}\right]  ds,\\
\end{array}
\]
Under the conditions of Theorem \ref{th:characterisation} and for $\mu_S$ bounded, the population distributions of mean returns and IQR are identified by:
    \begin{eqnarray}
&&\Pr\left({\mu_S(X,\eta) \le y }\right)
=  \int_{\mathcal{X}} \int_{0}^1  1 \left\{{A^{\mu}(x,a) \le y }\right\}  da\; dF_{\tilde{X}}(x), \label{eq:ident_mu}
\end{eqnarray}
\vskip4pt
Furthermore, assume that the analyst can elicit stated choices in two scenarios, such that $(P_1,X_1,P_2,X_2)$ are random variables, and for each $k$, $(P_k,X_k,\eta)$ satisfies equations (\ref{eq:choice_equation}) - (\ref{eq:definition_R}) and (\ref{eq:definition_FS}). $(X_1,X_2,\eta)$ are jointly independent. Define $V_k = F_{P_k \vert X_k}(P_k \vert X_k), k = 1,2$. For given $h$, the population distributions of mean effect and quantile effects are identified by:
\begin{eqnarray*}
&&\Pr\left({mWTP(h;X,\eta) \le y }\right) \\
&&=\int_{\mathcal{X}} \int_{(0,1)^2}  1 \left\{{A^{\mu}(x+h,v_2) - A^{\mu}(x,v_1)  \le y }\right\}   dF_{V_1,V_2 \vert  X_1,X_2}(v_1,v_2 \vert x,x+h)\;dF_{\tilde{X}}(x), \label{eq:quantile_ME}
\end{eqnarray*}
\begin{eqnarray*}
&&\Pr\left({qWTP(h,\tau;X,\eta) \le y }\right)\\
&&=\int_{\mathcal{X}} \int_{(0,1)^2}  1 \left\{{A^\tau(x+h,v_2) - A^\tau(x,v_1) \le y }\right\}   dF_{V_1,V_2 \vert  X_1,X_2}(v_1,v_2 \vert x,x+h)\;\;dF_{\tilde{X}}(x). \label{eq:quantile_QTE}
\end{eqnarray*}
\end{theorem}
Theorem \ref{th:characterisation_2} shows how to use a pair of stated preferences to estimate the population distribution of WTP. It uses the joint distribution of $V_1,V_2$, the so-called integral probability transforms of $P_1, P_2$. This overcomes the problem of taking differences between quantiles. Heuristically, the problem arises as individuals `switch ranks' as we change the value of $x$. That is, for any $v_i$, which characterises the rank of individual $i$ in the distribution of $m(x+h,\eta) - m(x,\eta)$, there exists a pair $(v_{1i},
v_{2i})$, such that $v_{1i}$ is the rank of the individual in the distribution of $ m(x,\eta)$, $v_{2i}$ the rank of $i$ in the distribution of $ m(x+h,\eta)$, and the following holds: $Q_{m(x+h,\eta) - m(x,\eta)}(v_i) = Q_{m(x+h,\eta)}(v_{2i}) - Q_{m(x,\eta)}(v_{1i})$. It does not hold in general that $v_{1i} = v_{2i}$. Fortunately, integral probability transforms of $(P_1, P_2)$ `record' the rank of the individuals in each distribution. Observing a pair of choices allows us to understand the rank switching patterns, because we can understand the joint distribution of the ranks $(V_1,V_2)$.


\begin{remark}
    In some cases, it is even possible to obtain the distribution of WTP with one scenario. An example is \cite{wiswall2018}, where the jobs are not labelled, the attributes are the same for each job, and the respondent is instructed to consider that the unspecified attributes are identical. In this case, the distribution of $F_{Q,S}(s;\tau,F_{\tilde{X}})$ delivers exactly the $\tau$-quantile of the WTP, if $F_{\tilde{X}}$ is set such that all attributes are identical except the one attribute of interest. 
\end{remark}

\subsection{From the distribution of returns to a predicted distribution of realised returns} \label{subsec:counterfactual} Under assumptions that we clarify in the following, the distributions of quantiles allow us to form predictions for the counterfactual distributions of realised returns tomorrow, the main object of interest for ex ante policy evaluation.

From an ex ante perspective, the resolution of the uncertainty will generate a mixture of shocks $\eta^* \vert \eta$ in the population. For some individuals, the realised uncertainty $\eta^* \vert \eta$ will correspond to lower quantiles $\tau$, whereas for others, it will correspond to higher quantiles $\tau$. Thus, if preferences are stable, and for a given distribution $F_{\tilde{X}}$, the realised distribution of returns will be a mixture of the ex ante distribution of quantiles $F_{Q,S}(s;\tau, F_{\tilde{X}})$. Let: 
\begin{equation}\label{eq:best_predictor_equal}
F_S(s,F_{\tilde{X}}):= \int_0^1 F_{Q,S}(s;\tau,F_{\tilde{X}})\omega_\tau d\tau.
\end{equation}
where $\omega_\tau$ are weights averaging to 1.

What determines the weights? If shocks are i.i.d. and the policy maker considers that the distribution $F_{S,i}$ is correct in the sense that when uncertainty is resolved, $S(x,\eta_i^*)$ is a draw for the distribution $F_{S,i}$, the best predictor for the realised distribution of returns is a mixture of the distributions of quantiles  $F_S(s,F_{\tilde{X}})$ with equal weights $\omega_\tau=1$.

If agents have biased beliefs about the resolvable uncertainty, the weights may be different from uniformity. In this case, the policy maker may use auxiliary data to determine the weights. One possibility is to exploit a similar choice where data exist for stated and revealed preferences. The weights would be those that match the predicted distribution from the stated preferences to the distributions obtained from the revealed preferences. Unfortunately, in our application, we do not have data on similar revealed preferences. Therefore, we leave this option for future research.

Without auxiliary data, the weights cannot be determined from the stated preferences. A second option for the policy maker is to explore a range of priors about the correctness of respondents' perception, which result in different weighting schemes for $F_{Q,S}(s;\tau,F_{\tilde{X}})$. For example, if the policy maker considers that individuals are too optimistic about the resolvable uncertainty, they can weight the upper quantiles of $\tau$ down. Section \ref{sec:policy} shows how to implement this strategy in the empirical application.

A third, less ad-hoc solution adopted in this paper is to remain agnostic about the weights and to provide bounds for our parameter of interest. Section \ref{sec:policy} studies the cost elasticity of hiring high-skilled workers for the private sector, under the assumption that the private sector cannot discriminate among workers. It shows that an informative lower bound for this parameter can be approximated using the set $\{F_{Q,S}(s;\tau,F_{\tilde{X}}), \tau \in(0,1)\}$.

\subsection{Comparison with the existing literature}\label{subsec:comparison}
The model and restrictions from equations (\ref{eq:choice_equation})-(\ref{eq:definition_R}), (\ref{eq:definition_FS}) and Assumptions \ref{ass:independence} and \ref{ass:ceteris_paribus} on the data generating process, the decision mechanism, and the resolvable uncertainty are significantly milder than the restrictions commonly imposed in the literature. For example, a common representation following \cite{blass2010} is to impose: 
\begin{equation} \label{eq:param_S}
    S(x,\eta_i^*) = (y_1-y_0) + \gamma_i + (z_1 - z_0)'\delta_i+ \nu_i,
\end{equation} where $\nu_i$ is the (additively separable) resolvable uncertainty, and follows a logistic distribution with variance $\sigma_i^2$. If the parametric model would map one-to-one with the true model, one should have $\eta_i = (\gamma_i, \delta_i, \sigma_i)$ and $\eta_i^*=(\eta_i,\nu_i)$. 

The linearity restriction of the above representation facilitates identification and estimation, because the WTP parameters are identified by $\delta_i$. In practice, the analyst uses the ratio between two coefficients of a linear regression \citep[see, for example,][equation (8), p.488]{wiswall2018}.\footnote{The theoretical motivation for the linear form is an additive random utility model (ARUM). One can see a parallel between the ARUM model used for stated choice experiments and the subjective expected utility model (SEU) of structural models of decision using subjective expectations \citep[see, for example, ][]{manski2004}. For instance, some of the choice attributes $x$ pertain to uncertain outcomes as in the case of the probability of being laid off in the future. These uncertain outcomes enter the utility as an additive term, which consists of the product between the probability of the event and the (dis)utility associated to the event. \cite{giustinelli2022} notes two limitations of the linear-SEU model: (1)``An implication of the linear SEU specification is that decision makers are assumed to be risk neutral with respect to continuous outcomes, notably expected earnings'', and (2) ``Although a standard feature of canonical SEU [...], multiplicative separability rules out the possibility that a person's subjective probability of an event depends on his/her (dis)utility for the event, as in models of utility-based or motivated beliefs.'' By remaining silent on the underlying utility model and allowing for unrestricted decision mechanisms, the present framework avoids these limitations.}
The parametric restriction on the resolvable uncertainty is made for technical convenience and has no compelling motivation, as acknowledged by \cite{blass2010}. \cite{meango2025} provides a test for this assumption in the random coefficient model of equation (\ref{eq:param_S}) and rejects it in several empirical applications including the one presented in this paper.

The framework in this paper proposes several important relaxations: it permits nonlinearity and nonseparability of the returns with respect to choice attributes, so that choice attributes can freely interact between them and with other unobservable characteristics. For example, the returns to option 1 depend not only on the net pecuniary returns $(y_1-y_0)$ as for risk neutral individuals, but also on the income level in option 0, $y_0$. The resolvable uncertainty is not restricted to additive separability from the ex ante returns. Hence, beliefs about an event can depend on the (dis)utility of the event (contained in $\eta_i$), as in models of motivated beliefs \citep[see][]{benabou2016}. Furthermore, the resolvable uncertainty is not parameterised, allowing for unrestricted forms of individuals beliefs about the uncertainty that will be resolved at the time of decision. 

Beyond the flexibility of the framework, Theorems \ref{th:characterisation} and \ref{th:characterisation_2} show that the population distribution of ex ante returns and WTP can be retrieved with just two elicitation rounds. With discrete choice data, the mixed logit model of \cite{mcfadden2000} imposes strong parametric assumptions on the distribution of WTP to recover the parameters of these distributions with only one elicitation. For example, $\delta_i$ in equation (\ref{eq:param_S}) would be normally distributed. With probabilistic stated choices, \cite{blass2010} shows that with parametric assumptions on the resolvable uncertainty only, the median WTP can be recovered by a median regression (Least Absolute Deviation, LAD). The innovation of \cite{wiswall2018} is to note that because stated choices can be elicited several times, one can estimate individual-specific LAD on the panel dimension. This allows estimating $\delta_i$ for each individual and thus recovers the entire distribution. However, this requires that the panel dimension is greater than the number of attributes considered. Instead of estimating separate regressions for each individual, we note that the distribution of ex ante returns $S$ can be recovered by quantile regressions performed at different quantiles, not just at the median. Theorems \ref{th:characterisation} and \ref{th:characterisation_2} explain how to recover the WTP. Therefore, for the purpose of identification, it is not necessary to generate very many scenarios for the same individual, as long as the scenarios are randomly assigned. Identification does not rely on parametric assumptions. 

Note that if measurement error is a concern, for example due to inattention, misunderstanding the survey instrument, and/or lack of effort, repeated elicitation can help controlling for the effect of measurement error. In this case, it is possible to apply results like \cite{evdokimov2010} to perform nonparametric quantile regressions/distribution regressions with measurement error. This would require only two rounds of elicitation.

The model retains one common assumption about stated preference experiments: they are \textit{ceteris paribus} experiments. This is a common assumption, often implicit in the stated preference literature \cite[see, for example,][]{benakiva2019, hudomiet2018}. An exception is \cite{wiswall2018}, who entertain the possibility that individuals presented with various manipulated choice attributes infer different unobserved choice attributes. Thus, the inferred distribution of resolvable uncertainty is not stable across experiments. To achieve identification, they need to assume that, within each scenario, the unobserved choice attributes are exactly the same across the choice options. To ensure this, they explicitly instructed the respondents to consider that within a scenario, the unobserved choice attributes are exactly the same across the choice options. Crucially, they also need to assume separable returns/utilities. Under these two assumptions, the effects of these unobserved attributes `cancel out'.

From the theory side, nothing precludes adopting the same strategy: one could augment equations (\ref{eq:choice_equation})-(\ref{eq:definition_FS}) to add separable unobserved choice characteristics that change across experiments, but are identical within pairs of choices. However, while \cite{wiswall2018} consider `anonymous jobs', the empirical application in the paper considers public sector jobs against private sector jobs that respondents perceive to be very different. It is not plausible to consider that the unobserved choice attributes can be perceived as exactly the same across sectors. To ensure that respondents perceived choice attributes as stable, the choice experiment explicitly instructs that across scenarios, the only choice attributes that change are the ones defined in the experiment. The remaining ones are exactly the same and correspond to a `typical job' in each sector. \cite{hudomiet2018, hudomiet2021} provide encouraging evidence that the concern of changing unobserved heterogeneity is minor in some contexts and can be mitigated by carefully designed elicitation procedures.

Our results relate to \cite{lewbel2011} who study a referendum contingent valuation model in resource economics. They consider an experiment in which an analyst elicits binary decisions about the choice of preserving a natural resource at exogenously set costs. They show that one can interpret the cost as a \textit{special regressor}, a continuous, separable variable, independent of the unobserved heterogeneity, with a large support. They retrieve nonparametrically the distribution of WTP to preserve the resource in the population. In their framework, there is no uncertainty about the valuation. This paper considers in addition uncertainty in the decision, which is expressed by stating choice probabilities rather than discrete choice. This allows a richer analysis of ex ante returns. In our empirical application, the wage in the public sector serves as a special regressor. Its variations help tracing out the distribution of reservation wages (Theorem \ref{th:characterisation}). Then, using assumption \ref{ass:ceteris_paribus}, the analyst can learn about the resolvable uncertainty. 

Finally, several papers have been interested in the task of predicting actual choice behaviour from stated preferences. For example, \cite{manski1999} notes that just as in the case of revealed preference: `extrapolation requires the researcher to combine the available empirical evidence on realized choices with prior information restricting the population distribution of choice functions and treatments. Similarly, extrapolation using expected choice data requires prior information.'
His work studies some strong and weak prior restrictions about the resolvable uncertainty and the econometric model such as normality, symmetry, and independence. Relatedly, \cite{arcidiacono2020} proposes estimators for an ex ante treatment effect from stated preferences \citep[see also][]{giustinelli2024}. For the ex ante treatment effect to match the ex post effect, they require `that students have rational expectations about their future earnings and that there are no unanticipated aggregate earnings shocks.'  Our work goes one step further by providing predicted realised returns even when individuals have wrong expectations about the resolvable uncertainty. We do not require that the policy maker has prior knowledge about the biases of agents, but derive bounds on the policy parameter of interest.

\section{Estimation and inference}\label{sec:estimation}
The identification result is constructive and suggests a multistage estimation procedure. Nonparametric estimation is conceptually feasible. However, in practice, the researcher often faces the curse of dimensionality, because the choice attributes can be many, their variation in choice experiments is often restricted to salient values, and the typical sample size is limited. For example, for each block of their stated choice experiment, \cite{wiswall2018} have a support of 16 observation points. For these reasons, the paper suggests a semiparametric alternative. This section provides a description of each step based on a Quantile Regression (QR) estimation for the conditional quantiles $Q_{P \vert X}$. This semiparametric estimator provides a feasible alternative to a Kernel estimator when the dimension of $X$ is large, as in our empirical application.

\subsection{Estimation}\label{subsec:estimation} The multistage procedure is related to the multistage estimation and inference of \cite{chernozhukov2020} (henceforth, CFNSV), which notations are adopted here. Inference is conducted using a weighted bootstrap procedure. The replication package provides a flexible routine for implementing all our estimators.

\subsubsection*{First stage: Estimation of the conditional distribution $Q_{P \vert X}$ }
Estimate by quantile regression the quantile treatment response of $P$ given $X$,  $Q_{P\vert X}(a|.)$, in the original data $\widehat{Q}_{P\vert X}(.|.)$ and in the bootstrap sample $\widehat{Q}^e_{P\vert X}(.|.)$, for the quantiles $a \in (0,1)$. More formally, denoting by $\rho_a(y)$ the check function $(y - 1\{y<0\})a$, we estimate
\begin{eqnarray}
   \widehat{Q}_{\ell(P)|X}^e(a\vert x) &=& R'\beta(a), R = r(x), a \in (0,1), x \in \mathcal{X},\\
   \widehat{Q}_{P|X}^e(a\vert x) &=& \ell^{-1}\bigl({\widehat{Q}_{\ell(P)|X}^e(a\vert x)}\bigr)\\
   \widehat{\beta}^e(a) &\in& \arg\min_{\beta \in \mathbb{R}^{\dim{R}}} \sum_{i =1}^n e_i \rho_a (\ell(P_i) - R_i' \beta)
\end{eqnarray}
In the original sample, $e_i=1$. In the bootstrap sample, the weights $\{e_i\}_{i=1}^n$ satisfy Assumption 3 in CFNSV (p. 518). The monotone transformation $\ell(P)$ is not necessary, but it guarantees that the predicted quantile remains in the unit interval. $r(x)$ is a flexible function of the choice attributes. CFNSV shows that any quantile function can be approximated with a flexible choice of $r(x)$.

In the empirical application,  we take $\ell(P) = \log(P/(1-P))$, and replace $P=0$ and $P=1$ with $P=0.001$ and $P=0.999$ respectively. $\widehat{Q}_{\ell(P)\vert X}(a|x) = r(x)'\hat{\beta}(a)$ is estimated by performing quantile regressions on a grid of the unit interval. To reduce the computation time and mitigate the effect of rounding, the regression is estimated on quantiles 0.01, 0.05, 0.15, \ldots, 0.85, 0.95, 0.99, using the routine `fitrqlinear' of Matlab. For the remaining quantiles, the regression coefficients $\widehat{\beta}_a$ are interpolated, using linear interpolation of coefficients with the Matlab routine `griddedInterpolant'. Note that we reorder the quantiles to avoid quantile crossing \citep{chernozhukov2010}.

In the application, the variation of attributes is limited to salient values. For example, the set of hours worked or the chance of losing the job is limited to three values in each sector. We consider two models: Model (1) with $r(x)$ that linearly includes the vector of choice attributes, Model (2) with $r(x)$ that includes up to cubic functions of the wages in each sector, and interaction terms of the public sector wage with all other choice attributes. This is close to the limit of flexibility that the data allow without causing a rank deficiency of the matrix $r(X)$.

It is interesting to note that the procedure of \cite{wiswall2018} performs a similar quantile regression for $a=0.5$ for each individual and interprets $\beta(0.5)$ as WTP parameters. Our proposed procedure performs quantile regressions for all quantiles and for the pooled sample. These are the building blocks for the second stage.
\vskip4pt

\subsubsection*{Estimation of the distribution of quantiles $F_{Q,.}$ } Let $F_{\tilde{X}}$ be a distribution of interest for the policy maker. For example, the policy maker can be interested in offers where the choice attributes are identical for both sectors. Given the estimator $\hat{Q}_{P \vert X}$ and their bootstrap draws $\hat{Q}^e_{P \vert X}$, we can form estimators of the distribution of quantiles $F_{Q.}$ as functionals of these building blocks. For any $\tau \in (0,1)$:
\begin{eqnarray}\label{eq:ident_FQ}
\nonumber \hat{F}_{Q,R}^e(r;\tau, F_{\tilde{X}}) &=& \int_{\mathcal{X}} \int_0^1 1\bigl\{{ \hat{Q}^e_{P \vert X} ( a \vert y_0,r,z) \ge \tau }\bigr\} da\; dF_{\tilde{X}}(y_0,y_1,z),\\
\nonumber \hat{F}_{Q,S}^e(s;\tau, F_{\tilde{X}}) &=& \int_{\mathcal{X}} \int_0^1 1\bigl\{{ \hat{Q}^e_{P \vert X} ( a \vert y_0,y_1-s,z) \le 1- \tau }\bigr\} da\; dF_{\tilde{X}}(y_0,y_1,z).
\end{eqnarray}
In practice, the integral over $a$ is calculated by averaging the values inside the integral, for example, $1\{\hat{Q}^e_{P \vert X} ( a \vert y_0,y_1-s,z) \ge \tau\}$, calculated on the grid $0.01,0.02,\ldots,0.99$.

\subsubsection*{Estimation of the distribution of means $F_{\mu}$}
For $ y \in [0,y_{\max}]$ estimate the empirical counterpart of the population distribution of mean defined Theorem \ref{th:characterisation_2}.
\begin{eqnarray}
    \widehat{F}_{\mu}^e(y; F_{\tilde{X}}) &=& \int_\mathcal{X} \int_0^1 1 \left\{{ \widehat{A}_n^{\mu,e}(x,a) \le y }\right\} da\; dF_{\tilde{X}}(x).\label{eq:Gtau} \\
    \widehat{A}_n^{\mu,e}(w,a) &=& \sum_s \delta_s \bigl[{\widehat{Q}_{P\vert X}^e(a|t(s,x)) - 1\{s\le 0\}}\bigr] \label{eq:Amu}
\end{eqnarray}
The sum over $s$ in the expression of $\widehat{A}^{\mu}(x,a)$ is on a fine grid of $[0,y_{\max}]$ where $K_S$ is a large integer such that $y_{\max} = K_S\times \delta_s $. In the empirical application, we choose $y_{\max}$ at about twice the average expected wage in the public sector.

\subsubsection*{Estimation of the distribution of WTP}
Here, we consider a pair $((P_1,X_1),(P_2,X_2))$ of elicited preferences. An additional step is to acquire an estimate of $V_{t},t=\{0,1\}$. Given our estimate of $\widehat{Q}^e_{P|X}$ as:
\begin{equation}
    \widehat{V}_{it}^e = \epsilon + \int_{\epsilon}^{1-\epsilon}1\Bigl\{\widehat{Q}^e_{P|X} (a \vert X_{it}) < P_{it}\Bigr\}
\end{equation}
This gives us a pair $(\widehat{V}_{i1}^e,\widehat{V}_{i2}^e)$. If rank invariance holds, the correlation between the two variables should be (close to) 1. From these, we can estimate and predict the quantile functions: $\widehat{Q}_{\widehat{V}_{i1}^e \vert X_1} (.|x)$ and $\widehat{Q}_{\widehat{V}_{i2}^e \vert X_2,\widehat{V}_{i1}^e} (.|x,v_1)$.
Given our estimate of $\widehat{A}_n^{\mu,e}(x,a)$ (equation (\ref{eq:Amu})), we can obtain for any $h$ such that $x+h \in \mathcal{X}$:
\begin{eqnarray}
        \nonumber\widehat{F}_{mWTP}^e(y; h, F_{\tilde{X}}) &=& \dfrac{1}{N} \sum_{i=1}^N \int_\mathcal{X} 1 \left\{{ \widehat{A}_n^{\mu,e}(x+h,\widehat{v}_2(a_2)) - \widehat{A}_n^{\mu,e}(x+h,a_1) \le y }\right\} \;da_1\;da_2\;dF_{\tilde{X}}(x)\\
       \text{ with }\hat{v}_2(a_2) &=& \widehat{Q}_{\widehat{V}_{i2}^e \vert X_1,X_2,\widehat{V}_{i1}^e} (a_2|x,x+h,v_1).
\end{eqnarray}
The estimator tends to have a bias, possibly due to the first step estimation of $V_{it}$. Fortunately, the bias can be quantified by noting that for $h=0$, the distribution should be a step function at $y=0$ (the WTP is equal to 0). The distance between the step function and the estimated function at $h=0$ provides an estimate of the bias and a basis for a bias-corrected estimator.

Let $\widehat{Q}_{mWTP}^e(.; h, F_{\tilde{X}})$ be the quantile function (the inverse) associated to $\widehat{F}_{mWTP}^e(.; h, F_{\tilde{X}})$. The bias corrected estimate is given by:
\begin{equation}
    \widehat{F}_{mWTP}^{e,bc}(y; h, F_{\tilde{X}}) = \int_0^1 1\Bigl\{{\widehat{Q}_{mWTP}^e(a; h, F_{\tilde{X}}) - \widehat{Q}_{mWTP}^e(a; 0, F_{\tilde{X}}) \le y}\Bigr\} \;da
\end{equation}
If there were no bias, $\widehat{Q}_{mWTP}^e(a; 0, F_{\tilde{X}})$ would always be zero. The bias in the first step estimation shifts this quantity away from 0. The correction shifts the quantile back to the correct value.

The steps are identical for the distribution of $qWTP$ after replacing $A^{\tau}$ with its empirical counterpart.

\subsection{Inference}\label{subsec:inference}
This section considers inference over regions of values $s \in \tilde{\mathcal{S}} \subset \mathbb{R}$. In the empirical application, $\tilde{\mathcal{S}} = [-y_{\max},+y_{\max}]$. The weighted bootstrap versions of the distributions of interest are obtained by rerunning the estimation procedure in Section \ref{subsec:estimation} with sampling weights that satisfy Assumption 3 in CFNSV (p.518). They are used to perform uniform inference. For example, a $(1-\alpha)$-confidence region for ${F}_{Q,S}(s;\tau,F_{\tilde{X}})$ over the region $\tilde{\mathcal{S}}$ is given by 
\begin{equation}
\left[{\hat{F}_{Q,S}\left({s;\tau, F_{\tilde{X}}}\right) \pm \hat{k}_{\tau}\left({1-\alpha}\right) \hat{\sigma}_{\tau}, s \in \tilde{\mathcal{S}} }\right],
\end{equation}
where $\hat{\sigma}_{\tau} = IQR[\hat{F}_{Q,S}^e\left({s;\tau,F_{\tilde{X}}}\right)]/1.349$ and $\hat{k}_{\tau}\left({1-\alpha}\right)$ denotes the $(1-\alpha)$-quantile of bootstrap draws for the maximal $t$-stat \[ \lVert{t_{\tau}^e(s)}\rVert_{\tilde{\mathcal{S}}} = \sup_{s \in \tilde{\mathcal{S}}} \left|{\dfrac{\hat{F}_{Q,S}^e\left({s;\tau,F_{\tilde{X}}}\right) - \hat{F}_{Q,S}\left({s;\tau,F_{\tilde{X}}}\right)}{\hat{\sigma}_{\tau}^e}}\right| \]

\subsection{Asymptotic Theory}\label{subsec:asymptotic}
Asymptotic results for the estimators builds on existing results in \cite{chernozhukov2010, chernozhukov2013,melly2015} and CFNSV pertaining to the QR estimator and the operators involved at each step of the estimation. Starting from the functional central limit theorem (FCLT) for the DR estimator \citep{chernozhukov2013}, it suffices to show that each operator map involved at each subsequent stage of estimation is Hadamard differentiable. The proof for each of these operators can be found, for example, in \cite{chernozhukov2010, chernozhukov2013, melly2015} and CFNSV. A FCLT and a bootstrap FCLT for the estimators of the distributions of interest follow by applying the chain rule of Hadamard differentiability and the functional Delta method. The exposition here omits a full derivation of the FCLT, which is tedious and mainly requires careful bookkeeping about the cascading stochastic processes. It provides little additional insight into the main point of the paper, beyond providing a theoretical background for conducting inference by using the exchangeable bootstrap.

\section{Preference for public sector opportunities among young Ivorians} \label{sec:results} This section uses the methodology developed in the previous sections to estimate the value of a job in the public sector for a population of high-ability students from two highly selective universities in C\^{o}te d'Ivoire.
The attraction of a generous public sector that offers secure jobs and sometimes a wage premium is thought to cause misallocation in the economy. 
This concern is even more relevant for developing countries that only have a modest proportion of their workforce with higher education. 
This section uses a stated choice experiment to analyse students' preference for public sector jobs. Within it, respondents are presented with two job offers, one from the public sector and one from the private sector, with attributes that vary exogenously. Section \ref{subsec:data} presents the data. Section \ref{subsec:perceptions} details the experiment set up. Section \ref{subsec:result_ex_ante_returns} and \ref{subsec:results_wtp} present the estimation results for the ex ante returns and the WTP for non-wage amenities. Given these preferences, Section \ref{sec:policy} estimates the cost elasticity associated with a private sector expansion for the market of high-skilled jobseekers using a predicted distribution of returns.


\subsection{Data}\label{subsec:data} The survey was conducted in the first week of February 2024 at two elite higher education institutions in Ivory Coast (\textit{Institut National Polytechnique Houphou\"{e}t-Boigny}, INP-HB and \textit{Ecole Nationale de Statistiques et d'Economie Appliqu\'ee}, ENSEA), that train students in STEM (Science, Technology, Engineering and Mathematics) degrees, and Statistics, Business and Economics, respectively. Entry in each of these institutions is through a selective exam. Together, they bring about 900 to 1,000 new graduates on the labour market each year. A number of well-known alumni of these two universities serve in high-ranked positions both in the Ivorian government and in private corporations.\footnote{Accurate proportions of those who are civil servants and those working in the private sector are not available yet, as these information have only started to be systematically collected recently. Among tertiary educated, the public sector offers about half of formal wage employment in C\^ote d'Ivoire \citep{girsberger2022}.} Hence, the sample represents some of the best students in the country.

The context of the study is one of skill shortage where firms struggle to satisfy their demand for labour. Quotes from two human resources directors illustrate vividly the skill shortage on the one hand: `Industrial engineers are rare, only the INP-HB trains good profiles but not enough of them are being trained to meet the needs of all the companies in Côte d'Ivoire', and the comparative advantage of students from these institutions on the other hand: `There are other schools that train industrial engineers, but we will not entrust our factories to these young people trained in schools other than the INP-HB.' \citep[Both cited from][]{iom2023}.

The target population was students in their last year, after either of a 3-year, 4-year or 5-year degree. The survey was implemented as a Computer Assisted Personal Interview, with five enumerators meeting students in or around their dorm-rooms. The usable sample covers 587 interviews. The vast majority of respondents are between 21 and 25 years old, with a proportion of 30 percent women, which is representative of this population.

The survey is unique in its scope, as it provides unprecedented insight into the perception and preferences of some of the best and brightest young Africans about their prospects in the labour market. It collects novel information on young Ivorians' perception about job characteristics in the public and private sectors. The perceptions garnered consist of beliefs about offer arrival rates, wage distributions, job destruction rates, and the likelihood of moving up in the hierarchy or obtaining a wage rise. The results of this exercise are detailed in Appendix \ref{app:perception}. The general insight is that public and private sector jobs are perceived very differently in the population. The \textit{private} sector appears as a dynamic sector with more abundant, better paid jobs and more opportunities for progression. For example, the perceived (monthly) wage  in the private sector is, on average, about 600,000 CFA Francs compared to 525,000 in the public sector. The perceived chances of obtaining a 20 percent wage rise within two years and of a job promotion are, respectively, 0.55 and 0.41 in the private sector and 0.42 and 0.35 in the public sector. However, these benefits are tied to stressful work conditions. In contrast, the \textit{public} sector appears to be a secure and stable alternative, even if it is less well paid. The perceived chance of being laid off after two years is, on average, 0.20 in the private sector and 0.08 in the public sector. It is important to note that beyond these general trends, there is heterogeneity in individual perceptions. This partly explains the heterogeneity in preferences, which we discuss in the next section.

\subsection{Stated choice experiment: public or private sector jobs}\label{subsec:perceptions}
The survey conducts a stated choice experiment in which respondents are invited to state their preferences about two job offers, one in the public and one in the private sector. Jobs differ by their wage and non-wage characteristics (employer, hours worked, likelihood of losing the job, and likelihood of obtaining a wage rise). Survey participants are presented with a choice experiment with five pairs of hypothetical job offers, each pair consisting of one public sector offer, and one private sector offer. Each pair specifies:
\begin{itemize}
    \item [(i)] the type of employer: public administration or public sector firm for the public sector job, SME or large-size firm for the private,
    \item[(ii)] the typical number of weekly hours associated with the job, to capture the flexibility of the job,
    \item[(iii)] the percentage of employees in this firm/administration who lose their job after two years,
    \item[(iv)] the percentage of employees who are promoted after two years,
    \item[(v)] the monthly wage attached to the job.\footnote{In C\^{o}te d'Ivoire, wages for salaried work are typically expressed in monthly unit.}
\end{itemize}
For each pair/scenario, they are asked to state the probability of choosing either sector. The first scenario always refers to identical offers, that is, the scenario equates all items (ii) - (v) above for an offer from the public administration against an offer from an SME.\footnote{The employers are `public administration' and `SME' respectively, the weekly hours worked 40, the chance of losing job is 5 percent, the chance of job job promotion, 10 percent, and the starting salary 750,000 CFA Francs (25 percent above the average perceived wage).}  For the remaining pairs, the attributes are randomly drawn from a support displayed in Table \ref{tab:support} in Appendix \ref{app:survey}.

Figure \ref{fig:prob_first_scenario} displays the distribution of the probability of accepting the public sector offer for the first scenario, where the offers are identical. The histogram shows evidence of rounding at the nearest 10 percent. Otherwise, there is no conspicuous use of extreme values or the value `0.5' that is sometimes found in stated preference data. The relative quality of these data can be partly attributed to the respondents' experience with the concept of probability, which is higher than in the average population.
\begin{figure}
    \centering
    \includegraphics[width=0.6\linewidth]{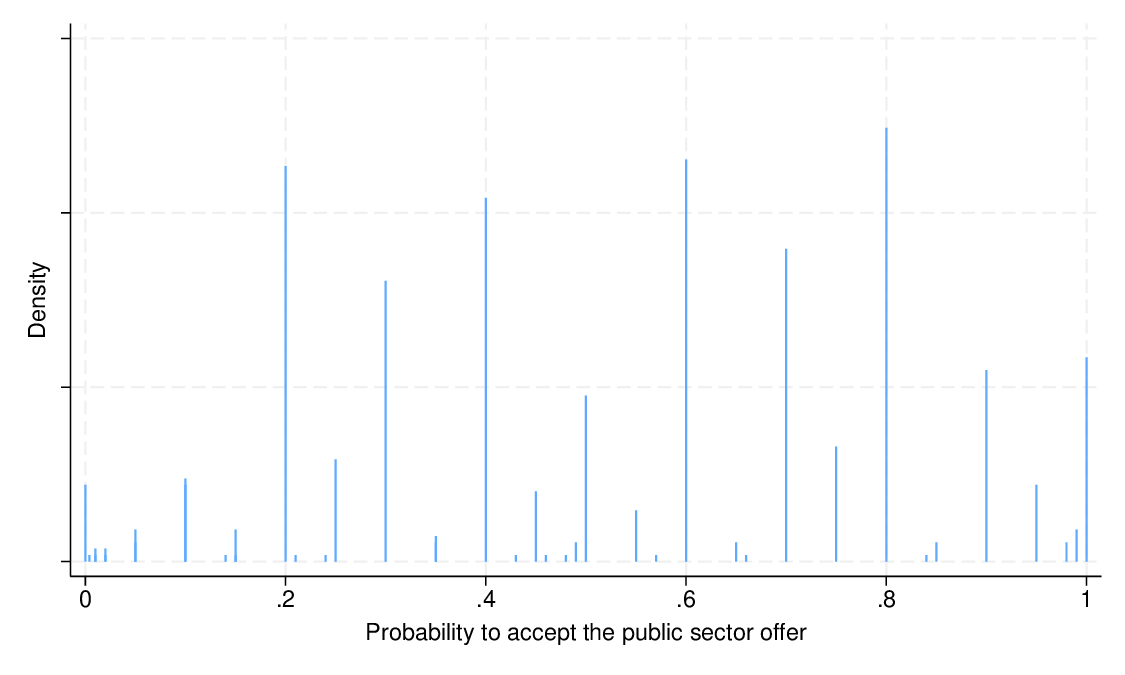}
    \caption{Probability of choosing the public sector for two identical offers}
    \label{fig:prob_first_scenario}
    \floatfoot{Note: The figure represents the reported probability of choosing a public sector offer, when offers are otherwise identical. The employers are `public administration' and `SME' respectively, the weekly hours worked is 40, the chance of losing the job is 5 percent, the chance of a job promotion is 10 percent, and the starting salary is 750,000 CFA Francs (25 percent above the average perceived wage). }
\end{figure}
A striking insight is that the answers are balanced on both sides of the 0.5 mark. The median answer is 0.55. About 35 percent of the sample states a probability between 0.15 and 0.45, whereas about 36 percent gives an answer between 0.55 and 0.85. Thus, it appears that, for identical offers, there is no sweeping preference for one sector over the other. In contrast, preferences are very heterogeneous in the population. Besides, even when respondents express a preference for one sector over the other, the great majority does not exclude the possibility of accepting the competing offer (91 percent gives an interior solution). This suggests a high prevalence of (resolvable) uncertainty. The stated choice probabilities are the main inputs to infer the distribution of perceived ex ante returns for a public sector job.

\subsection{Distribution of perceived ex ante returns for a public sector job}\label{subsec:result_ex_ante_returns}
Following the development of Section \ref{sec:framework}, this section defines ex ante returns to a public sector job $S$ as the maximum price the individual is willing to pay to secure a public sector job. For concision, we report the results from the more parsimonious Model (1). The results of Model (2) are presented in Appendix \ref{app:additional_figures} and are qualitatively similar. 
 

\begin{figure}
    \centering
    \includegraphics[width=0.65\linewidth]{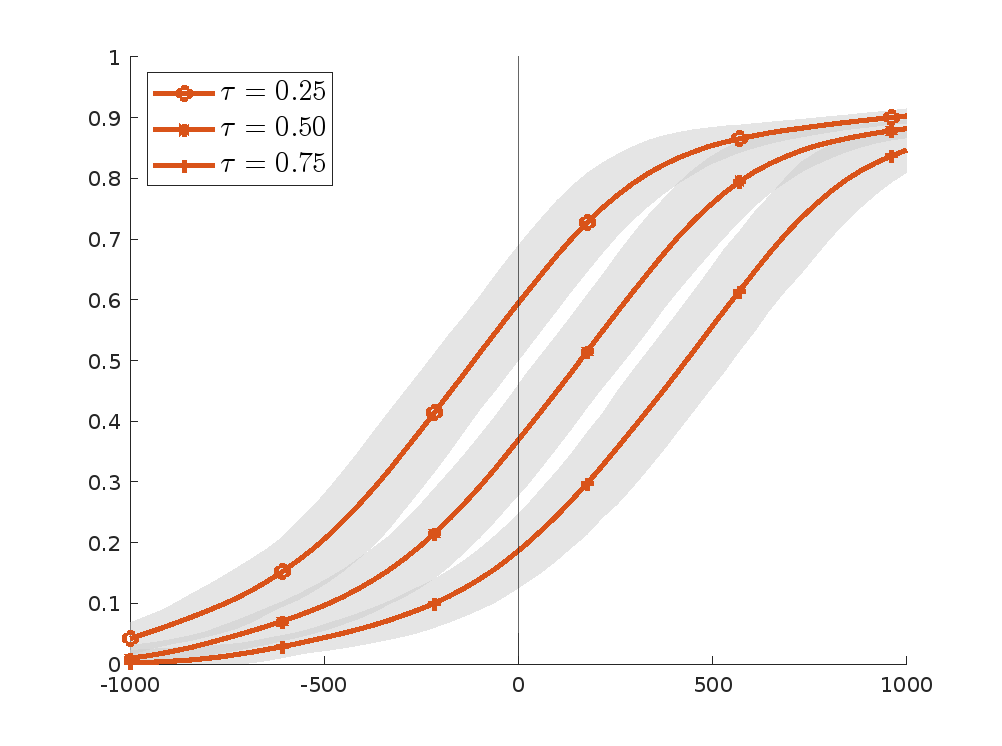}
    \caption{Distribution of quantiles of ex ante returns to a public sector job, $\hat{F}_{Q,S}\left({.;\tau, F_{\tilde{X}}}\right), \tau =0.25, 0.50, 0.75$ for perceived offer distribution.}
    \floatfoot{Note: The x-axis is in 1,000 of CFA Francs. Offers mimic the individual-specific average perception for each sector. Working hours are set to 45.8 hours in the private sector and 41.4 in the public sector. The dark-grey area shows the 90 percent pointwise confidence interval. The standard deviation is obtained with 200 bootstrap replications.}
    \label{fig:distr_quantile_FQ}
\end{figure}
Figure \ref{fig:distr_quantile_FQ} represents the distributions of the quartiles of ex ante returns to a public sector job, that is $\widehat{F}_{Q,S}\left({.;\tau, F_{\tilde{X}}}\right),$ for $ \tau =0.25, 0.50, 0.75$. It shows these distributions for offers that correspond to the individual-specific average belief. The construction of offers uses the respondents' perceptions described in Appendix \ref{app:perception} to compute average values of the public sector and the private sector job attributes. $F_{\tilde{X}}$ is set to take these individual-specific average values. 
The distribution of median returns (middle line) suggests that for their average expected offers, a majority of students (64 percent) would perceive positive returns from the public sector. Thus, the evidence points toward some desirability of the public sector offers. However, returns are very heterogeneous and can be large in absolute values. For the distribution of median, at the top quartile, the returns are as large as 90 percent of the average perceived wage in the public sector, whereas at the bottom quartile they represent a loss of about 26 percent of the average perceived wage in the public sector.


The large variance of the distribution suggests that the students may have very different preferences. Some show a high distaste for a public sector job. In contrast, some others display a high taste for public sector jobs. Is it likely that these individuals can be induced to change their preferred sector? The answer to this question depends in part on the resolvable uncertainty. If there were no resolvable uncertainty, the distribution of the median returns would capture the population distribution of ex ante returns $F_S$. However, if respondents are very uncertain about their preferences, a shock in the resolvable uncertainty might move them above or below the zero-return threshold.


In Figure \ref{fig:distr_quantile_FQ}, the horizontal distance between the first ($\tau=0.25$) and third-quartile ($\tau=0.75$) distribution gives an indication of the individual-specific resolvable uncertainty. The distributions are well separated, which hints at a high prevalence of uncertainty about ex ante returns. To get a sense of the importance of the resolvable uncertainty, one can look at the crossing of the distributions with the zero-line. For example, if everyone was subject to the same shock, and this shock would correspond to the first-quartile of the resolvable uncertainty close to 60 percent of the population would perceive negative returns from a public sector job. This represents a change by about 25pp from the distribution of median returns.
\begin{figure}
    \centering
    \includegraphics[width=0.65\linewidth]{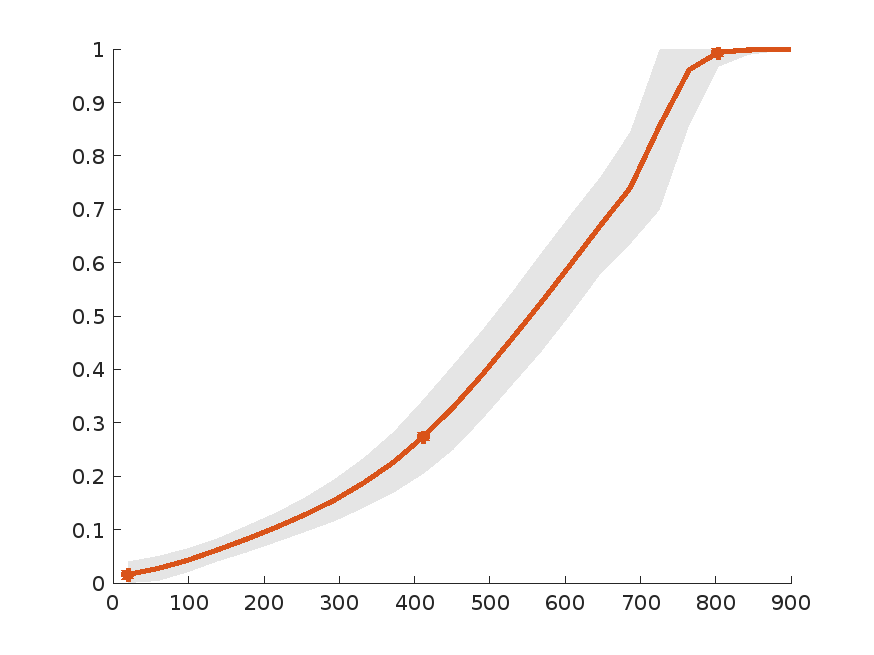}
    \caption{Distribution of individual-specific IQR}
    \label{fig:distr_IQR}
    \floatfoot{Note: Offers mimic the average perception about a public sector job offer. Working hours are set to 45.8 hours. The grey area shows the 90 percent pointwise confidence interval.}
\end{figure}

We can describe resolvable uncertainty more formally with the distribution of individual-specific Inter-Quartile Range (IQR) of the ex ante returns (Proposition 1, M\'eango, 2025). Figure \ref{fig:distr_IQR} shows that for most of the population, the estimated value of the IQR ranges between 0.4 and 1.4 times the average perceived wage in the public sector. These results confirm the high prevalence of uncertainty in the population. This result has two implications: (1) ex ante policy evaluation needs to account for the high level of resolvable uncertainty. Looking just at the mean/median could be misleading. (2) It appears that jobseekers have a high value of waiting and collecting more information on each sector's offer instead of committing ex ante to one sector \citep[on the option-value arising from collecting information, see, for example,][]{gong2020,meango2022}.

\subsection{WTP for non-wage amenities}\label{subsec:results_wtp}
What is the value of non-wage amenities, such as job security or lesser working hours? Figure \ref{fig:distr_quantile_FQ_indiv} in Appendix \ref{app:additional_figures} shows the distributions of returns for perceived amenities: wages are set to be equal in both sectors so that there is no direct pecuniary gain from being in the private sector. Only perceived amenities vary.\footnote{The wage is set at the individual-specific average perceived wage in the public sector. The survey missed a question about the number of weekly working hours expected in the public or the private sector. The number of working hours in the public sector is set to 41.4 and the working hours for the private sectors are set to 45.8 according to \cite{christiaensen2017}. The average working time for salaried workers calculated from the publicly available Ivorian labour force survey (ERI-ESI 2017) is 48 hours. Comparable numbers for other countries of francophone west Africa range between 45 and 51 hours.} 
At the median returns, 55 percent of the students perceive positive amenities from being in the public sector. Close to 15 percent of the population perceives amenities larger than the average wage in the public sector. While in our context, many students perceive positive and potentially large amenities in the public sector, the magnitude is somewhat smaller than reported for other contexts.\footnote{\cite{mangal2024} uses a sample of 147 candidates preparing for competitive exams for government jobs to infer a lower bound on the total value of a government job, including amenities. He finds that the amenity value of a government job in Pune, India, comprises at least two-thirds of total compensation. The above estimates of median returns are lesser in magnitude and provide a more nuanced picture.}

\begin{figure}
    \centering
        \subfigure[Being laid off]{ 
        \includegraphics[width=0.45\linewidth]{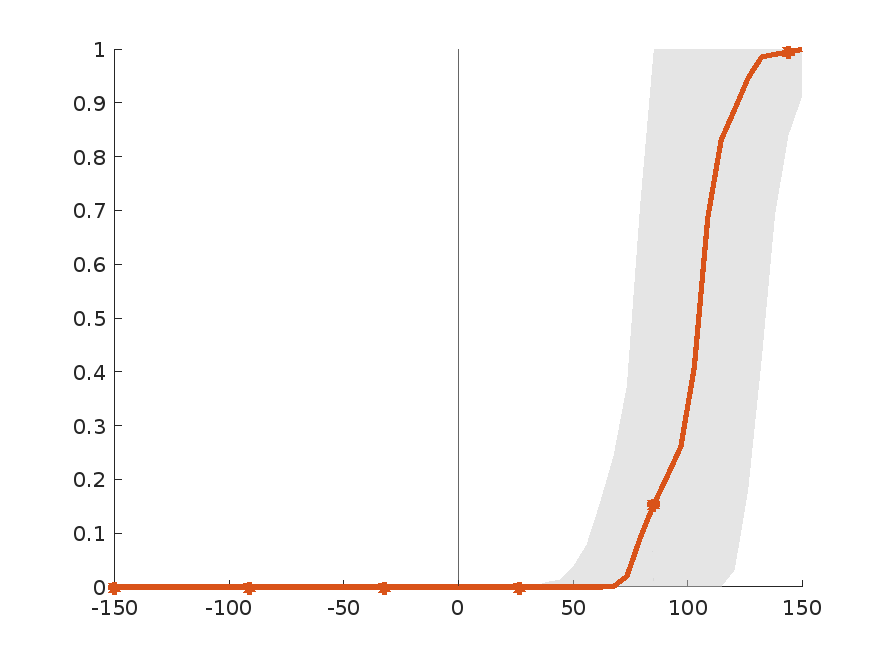}
        }
        \subfigure[Obtaining a job promotion]{ 
        \includegraphics[width=0.45\linewidth]{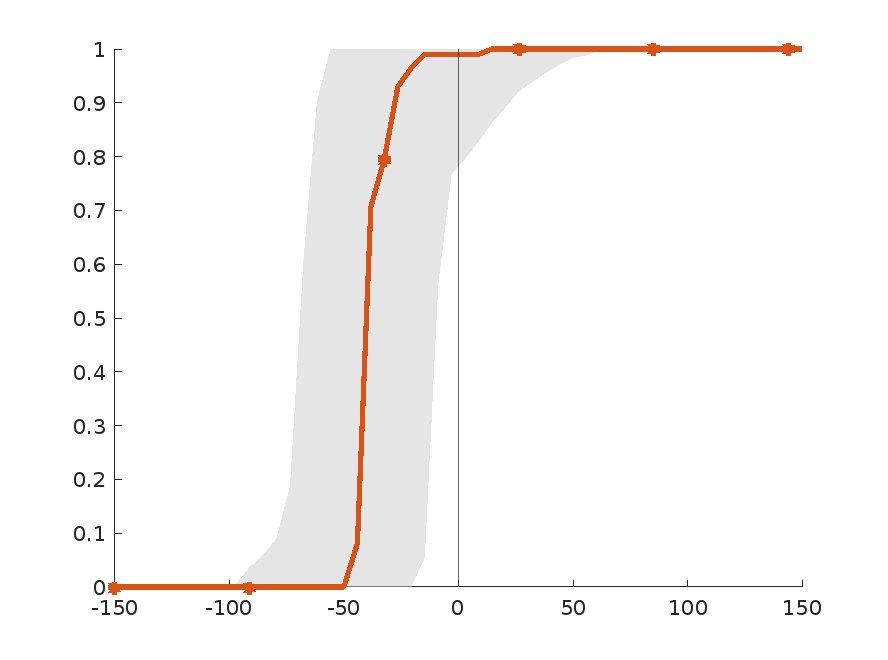}
        }
        \subfigure[Working hours]{ 
        \includegraphics[width=0.45\linewidth]{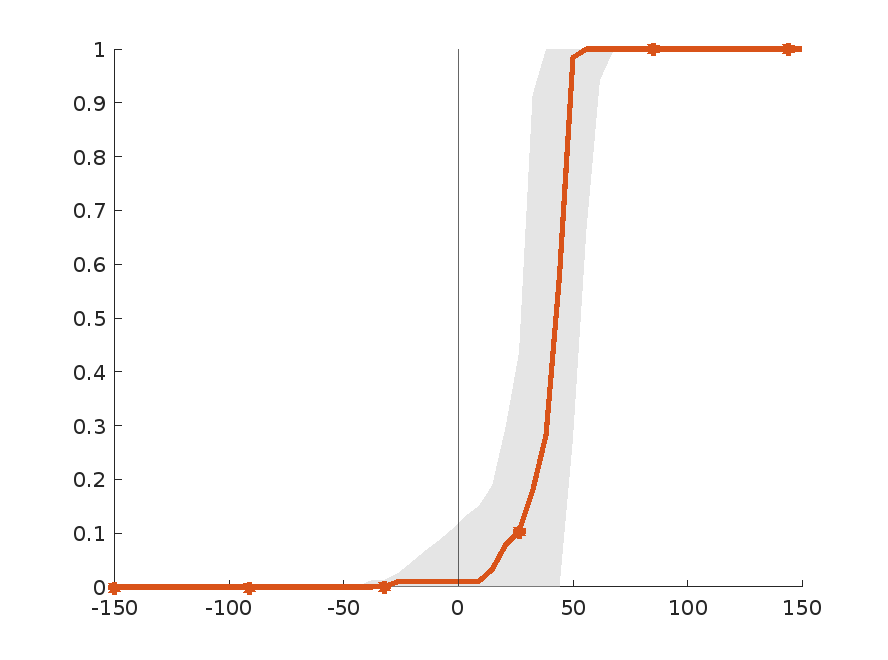}
    }
    \caption{Mean WTP, $mWTP$, for choice attributes : $\Pr\left({\mu_{S}(x+h\boldsymbol{e}_j,\eta) - \mu_{S}(x,\eta) \le y}\right)$}
    \floatfoot{Note: The counterfactual consists in closing the perceived gap between choice attributes one by one. The starting point is job offers that are identical and equal to the typical private sector offers. Then, panel (a) changes the probability to be laid off in the public sector from the perceived average in the private sector (0.20) to the perceived average in the public sector (0.08). Similarly, panel (b) changes the perception about the probability of obtaining a job promotion in the public sector from the average perceived level in the private sector (0.4) to the same average for the public sector (0.30). Finally, panel (c) changes the working hours in the public sector from 45.8 as in the private sector, to 41.4 as in the public sector offer.}
    \label{fig:qtes}
\end{figure}

To further investigate the value of amenities, one can compute the distribution of the mean effect ($mWTP$) of changing the observed choice attributes using Theorem \ref{th:characterisation_2}. 
They answer the question: `How does the variation in a specific choice attribute change the mean returns for any given individual?' This measures the WTP for this attribute, taking into account the resolvable uncertainty. Figure \ref{fig:qtes} displays the distribution of $mWTP$ after equating the individual's perceptions in both sectors. 
The starting point is job offers that are identical and equal to the typical private sector offers. Then, panel (a) changes the probability to be laid off in the public sector from the perceived average in the private sector (0.20) to the perceived average in the public sector (0.08). Similarly, panel (b) changes the perception about the probability of obtaining a job promotion in the public sector from the average perceived level in the private sector (0.4) to the same average for the public sector (0.30). Finally, panel (c) changes the working hours in the public sector from 45.8 as in the private sector, to 41.4 as in the public sector offer.
As the probability of being laid off drops in the public sector, the mean returns to a public sector offer increase by close to one-fifth of average wages in the public sector for the majority of the population. This reflects the importance of job stability in these returns.
The perceived probability of a job promotion is lower in the public sector and the decrease causes a fall in returns by about 5 to 9 percent  of the average wage in the public sector. Finally, in our counterfactual exercise, where the number of hours in the public sector decreases from 45.8 to 41.4, the returns to a public sector offer increase by up to 9.5 percent of average perceived wages in the public sector. Thus, higher job stability is a key non-wage amenity in the public sector.

Overall, the above results are consistent with the verbal description of the students who value the public sector for its stability and flexibility, and identify the private sector first with stressful but pecuniary rewarding working conditions. Beyond the choice attributes controlled for, the perceived returns are very heterogeneous, implying that there exist inherently different preference types in the population. Some people have a high value for the public sector offer, whereas others have an equally high value for the private sector offer. However, there is considerable uncertainty about these returns, which leaves room for substantial changes between the time of elicitation and the time of decision.

\section{Implications for the private sector's cost of hiring}\label{sec:policy}
Given the context of a high-skilled worker shortage, this section investigates the externalities imposed by the public sector on the private sector in the labour market for top-skilled workers. One metric for these externalities is the cost for the private sector to attract an additional fraction of top-skilled workers. One question of interest is: if the private sector wants to attract one percent additional workers, by how much should it increase the offered wage? If, all things being equal, an increase of its workforce by one percent would increase the private sector wage bill by a similar proportion, then the private sector is not affected much by the public sector hiring. Conversely, if this cost elasticity is significantly greater than one, then the competition from the public sector generates high labour costs for the private sector and potentially limits its expansion.

\begin{figure}
    \centering
        \includegraphics[width=0.6\linewidth]{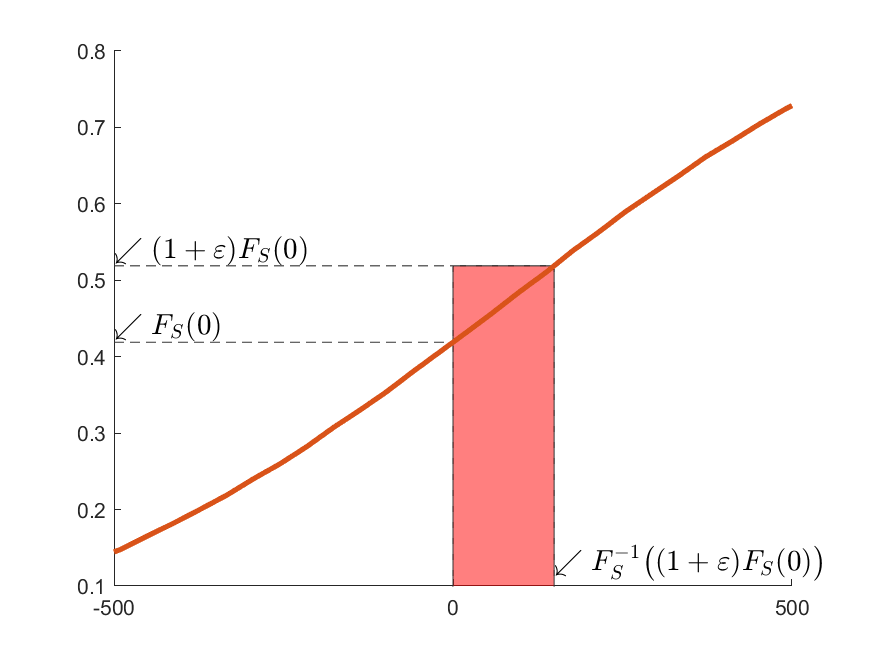}
    \caption{Illustrative example of the transfers to attract additional workers in the private sectors.}
    \label{fig:cf_analysis}
    \floatfoot{Note: $F_S(0)$ represents the proportion of individuals who would have negative ex ante returns. The marginal individual, i.e. who is indifferent between the private and public sector job, is at the $F_S(0) - th$ percentile. To attract a fraction $x$ of additional job-seekers, the private sector must compensate up to the job-seeker at $(1+\varepsilon)F_S(0) - th$ percentile by transferring $F_S^{-1}\bigl[{(1+\varepsilon)F_S(0)}\bigr]$ to this individual. The cost of an expansion of the private sector is the shaded area: $ F_S^{-1}\bigl[{(1+\varepsilon)F_S(0)}\bigr] \times ((1+\varepsilon)F_S(0))$.}
\end{figure}

Answering this question requires a prediction of realised returns $F_S$. Indeed, note that any jobseeker chooses the public sector if they perceive positive ex ante returns. The marginal individual, i.e. who is indifferent between the private and public sector job, is at the $F_S(0) - th$ percentile. To attract a fraction $\varepsilon$ of additional jobseekers, the private sector must compensate up to the jobseeker at $(1+\varepsilon) F_S(0)- th$ percentile by transferring $F_S^{-1}\bigl[{(1+\varepsilon)F_S(0)}\bigr]$ to this individual (see Figure \ref{fig:cf_analysis}).

If shocks are i.i.d. and the policy maker considers that the distribution $F_{S,i}$ is correct in the sense that when uncertainty is resolved, $S(x,\eta_i^*)$ is a draw for the distribution $F_{S,i}$, the best predictor for the realised distribution of returns is given by equation (\ref{eq:best_predictor_equal}) with $\omega_{\tau} =1$ for all $\tau$. However, if individuals have biased expectations about the resolvable uncertainty, a prediction of the realised distribution returns should account for the bias. For example, if students are too optimistic about the resolvable uncertainty, realised shocks would be overwhelmingly concentrated on the lowest quantiles. In this case, students expect the public job offer to be more attractive than it will actually be at the time when they will make a decision. Thus, they expect higher quantiles of returns to realise than it will actually be the case. Thus, predicted returns should assign low weights to the highest quantiles. If, instead, individuals place too much weight on extreme quantiles, the policy maker should discount extreme quantiles. A policy maker who bases their analysis solely on the distribution of median returns assigns a weight one to the median and zero to all other quantiles.\footnote{Previous literature has emphasised the estimation of the mean (or, given the assumptions, the median), which is not the most relevant from the perspective of ex ante policy evaluation. Figure \ref{fig:compare_distributions} in Appendix \ref{app:additional_figures} shows the discrepancy using the empirical application. The counterfactual distribution of returns using equation (\ref{eq:best_predictor_equal}) with equal weights is flatter than the distributions of average and median returns, especially in Model (2). Even if the perceived distribution $F_{S,i}$ was correct, both the mean and median distributions would underestimate the cost elasticity of an expansion. Intuitively, they do not account for the fact that some members of the population will receive extreme values of their perceived returns rather than average or median values.}

\begin{figure}[htbp]
\subfigure[Weights]{ 
        \includegraphics[width=0.45\linewidth]{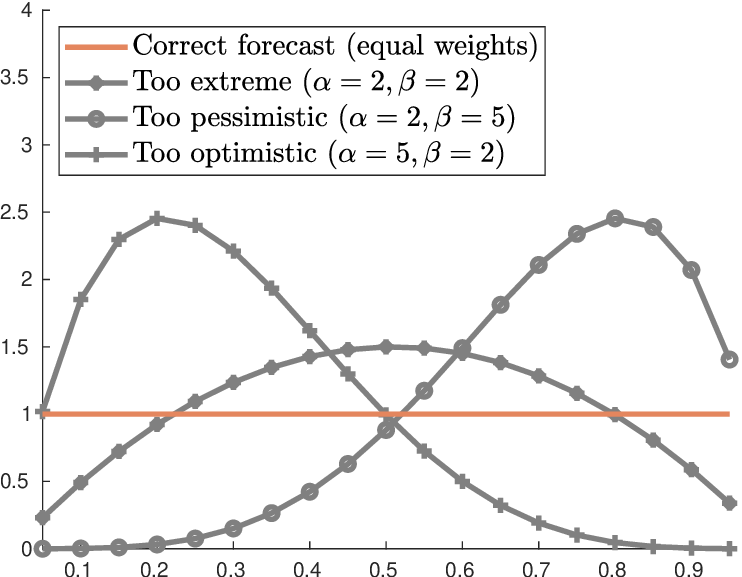}
        }
        \subfigure[Weighted counterfactual distributions $\bar{F}_S$ using different weights]{ 
        \includegraphics[width=0.45\linewidth]{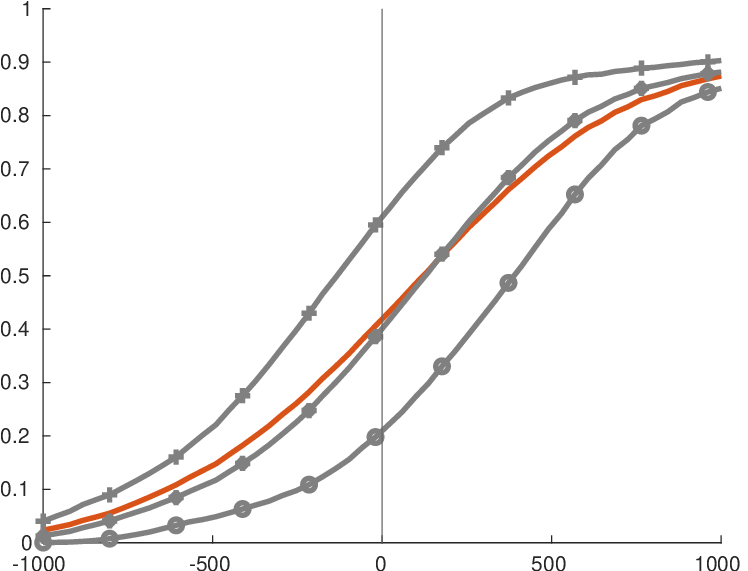}
        }
        \subfigure[Estimated transfers using different weights]{
        \includegraphics[width=0.45\linewidth]{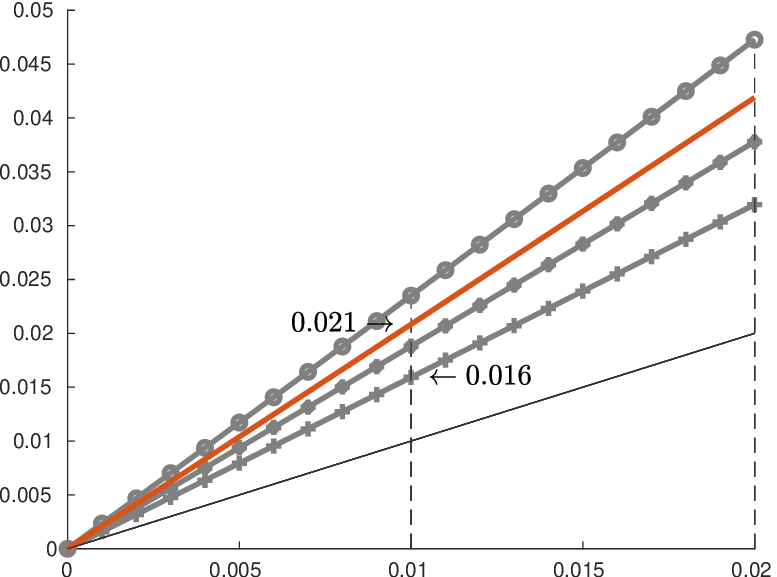}
        }
    \caption{Estimated transfers to attract additional workers in the private sectors.}
    \floatfoot{Note: The predicted distribution of realised returns is calculated by equation (\ref{eq:best_predictor_equal}), using quantiles $\tau = 0.05, 0,10, \ldots, 0.95$. $F_{\tilde{X}}$ is set for average offers in the public and the private sector: In the public sector, the wages are set at $525,000$ CFA Francs, the working hours to 41.4, the probability to lose a job in the public sector is set to 0.08, and the chance to obtain a job promotion to 0.30. In the private sector, the wages are set at $600,000$ CFA Francs, the working hours to 45.8, the probability to lose a job is set to 0.20, and the chance to obtain a job promotion to 0.40. Panel (a) shows the weights applied for four scenarios: (i) the baseline corresponds to a correct forecast of the distribution of resolvable uncertainty, (ii) corresponds to respondents' forecasts that give too much weight to the extreme quantiles, thus the tails need to be weighted down compared to the baseline. (iii) corresponds to respondents's forecasts that are too pessimistic, giving too much weight to the lower quantiles, and thus the lower tail needs to be weighted down. (iv) corresponds to respondents's forecasts that are too optimistic, giving too much weight to the upper quantiles, and thus the upper tail needs to be weighted down. Panel (b) shows the effect of different weights on the predicted distribution of return. For each of these distributions, Panel (c) shows the relationship between the private sector expansion (percentage of additional workers) on the X-axis and the transfers as a proportion of the original costs/wage bill on the Y-axis. In the baseline, a one percent increase of the private sector would entail a 2.1 percent increase of the wage bill. In the case of respondents being too optimistic, the cost elasticity is 1.6. The calculated lower bound is 1.36\%.}
    \label{fig:cf_analysis2}
\end{figure}

Figure \ref{fig:cf_analysis2} illustrates the effect of different weightings. In panel (b), the orange line shows the baseline prediction, where all quantile distributions are averaged with equal weights. In this case, the population is almost equally split between those with positive returns (43.2 percent) and negative returns to the public sector offer. This reflects well the equal split on the actual labour market, where each sector provides about one-half of the formal wage employment \citep[see][Table A.1, p.43]{christiaensen2017}. Instead, if we think that students are too optimistic about the resolvable uncertainty and weight the higher quantiles down, this results in comparatively lower predicted returns of a public sector job. Given the weights we consider, close to 80 percent of the population would have negative returns from a public sector job. This appears to be a rather extreme case, given the observed split of the labour market. 

Without auxiliary data, it is not possible to form a correct prior about the weights. Instead of interpreting just the mean/median, or assuming that individuals have a correct perception (equal weights), we remain agnostic about the weights and provide a lower bound for the cost elasticity for all possible weights. 

To characterise the cost elasticity, we make the following assumption: the private sector is unable to observe workers' private returns and discriminate among them. Therefore, it must offer the same amount even to infra-marginal jobseekers. The wage bill before the extension is $y_0F_S(0)$. When adding a fraction $x$ of new workers, the new wage bill is given by:
$$\text{New wage bill} =y_0 F_S(0) + y_0 \varepsilon F_S(0) + F_S^{-1}\bigl[{(1+\varepsilon)F_S(0)}\bigr] (1+\varepsilon) F_S(0).$$
The first term is the original wage bill. The second term is the increase resulting from the workforce by $\varepsilon$ at the fixed labour price. The third term is the extra cost that comes from the fact that the private sector faces higher prices when they wish to attract more workers. The latter term is illustrated by the shaded area in Figure \ref{fig:cf_analysis}.
If the workforce increases by a fraction $\varepsilon$, the cost increases by:
\begin{equation}\text{Elasticity}(\varepsilon)=\varepsilon\left({1 + \dfrac{1+\varepsilon}{\varepsilon y_0}}F_S^{-1}\bigl[{(1+\varepsilon)F_S(0)}\bigr]\right).\end{equation}
In Appendix \ref{app:proofs}, we show that a lower bound for this quantity can be approximated by:
\[\varepsilon\left({1 + \dfrac{1+\varepsilon}{\varepsilon y_0} \left({\min_{\tau \in (0,1)}\;\dfrac{ F_{Q,S}(s;\tau)}{f_{Q,S}(s;\tau)}}\right) }\right)\]
where $f_{Q,s}(s;\tau)$ is the associated density function.

We compute this lower bound in the data by taking the numerical derivative of $F_{Q,S}$. We restrict the space of $\tau \in (0.05,0.95)$ to avoid estimation of extreme quantiles. For $\varepsilon=1\%$ the lower bound is $1.36\%$ when using Model (1) and $1.34\%$ with Model (2).\footnote{If students have rational expectations, the corresponding figure is $2.1\%$ in Model (1) and $3.0\%$ in Model (2). See Panel (c) in Figure \ref{fig:cf_analysis2} and Panel (d) in Figure \ref{fig:model_2}.} This means that for an increase of $1\%$ of the workforce, the wage bill increases by at least $1.34\%$: Each additional hire costs at least one third more than the previous worker.

This analysis ignores equilibrium effects and the ability of the private sector to discriminate among some workers. Yet, it still gives a hint that public sector jobs are a valuable outside option for elite students, and hiring more of them can come at an economically significant cost for private sector firms. Thus, it supports the claim that the labour market competition from a public sector might constrain the expansion of the private sector \citep{christiaensen2017, girsberger2022}. 

\section{Conclusion}\label{sec:conclusion}
The paper offers innovative tools to the stated preference literature. Stated preference analyses have served to describe individual preferences over choice attributes, presented in the form of WTP parameters. This task is more complex in an environment where agents are uncertain about their returns and uncertainty is sequentially resolved. This paper provides identification results for the population distribution of ex ante returns, the related WTP, and a policy relevant distribution for the purpose of ex ante policy evaluation. These results are obtained while accounting for the uncertainty of agents but without appealing to ad-hoc parametric assumptions, for example, that the resolvable uncertainty is extreme value type I. \cite{meango2025} shows that this assumption is problematic in all four stated choice experiments that he considers. The findings of our paper provide a novel identification and estimation alternative to the current state-of-the-art methodology.

A major innovation is that all of these results rely on at most two experimental scenarios to identify the distribution of interest. There might be good rationales to elicit more scenarios per individual: to increase power, control for measurement error, and/or test for consistency of answers. However, it does not require the number of elicitations to exceed the number of attributes considered. This allows the researcher to use significantly shorter survey instruments and reduce the cost of surveys.

The paper shows how to perform an ex ante policy evaluation while accounting for the possibility that individuals are wrong about the resolvable uncertainty. In our empirical application, we derive a lower bound on the main parameter of interest. 

Finally, the paper showcases the feasibility of the proposed procedure by exploiting a survey of elite students in C\^ote d'Ivoire to understand their perception of job opportunities in the public and the private sector. There is heterogeneity in these perceptions, but the main view is that public sector jobs offer stability and flexibility, while private sector jobs are tied with stressful but pecuniary rewarding working conditions. Despite the relative homogeneity of the \textit{perceptions}, we find that \textit{preferences} are very heterogeneous: even when wage offers are identical, some students have a high value for a public sector job, whereas other have an equally high value for a private sector job. There is also substantial uncertainty that the students expect to be resolved at the time of their decision. Given these preferences, and even when accounting for the uncertainty, there is evidence that the presence of competing public sector offers possibly increases the labour costs for the private sector and limits its expansion.




\newpage

\bibliographystyle{apalike}
\bibliography{ref_expectations}

\begin{thebibliography}{}

\bibitem[Adams-Prassl and Andrew, 2019]{adams2019}
Adams-Prassl, A. and Andrew, A. (2019).
\newblock Preferences and beliefs in the marriage market for young brides.
\newblock CEPR Discussion Paper No. DP13567.

\bibitem[Albrecht et~al., 2019]{albrecht2019}
Albrecht, J., Robayo-Abril, M., and Vroman, S. (2019).
\newblock Public-sector employment in an equilibrium search and matching model.
\newblock {\em The Economic Journal}, 129(617):35--61.

\bibitem[Algan et~al., 2002]{algan2002}
Algan, Y., Cahuc, P., and Zylberberg, A. (2002).
\newblock Public employment and labour market performance.
\newblock {\em Economic Policy}, 17(34):7--66.

\bibitem[Alm{\aa}s et~al., 2024]{almaas2024}
Alm{\aa}s, I., Attanasio, O., and Jervis, P. (2024).
\newblock Presidential address: Economics and measurement: New measures to
  model decision making.
\newblock {\em Econometrica}, 92(4):947--978.

\bibitem[Ameriks et~al., 2020a]{ameriks2020b}
Ameriks, J., Briggs, J., Caplin, A., Lee, M., Shapiro, M.~D., and Tonetti, C.
  (2020a).
\newblock Older americans would work longer if jobs were flexible.
\newblock {\em American Economic Journal: Macroeconomics}, 12(1):174--209.

\bibitem[Ameriks et~al., 2020b]{ameriks2020a}
Ameriks, J., Briggs, J., Caplin, A., Shapiro, M.~D., and Tonetti, C. (2020b).
\newblock Long-term-care utility and late-in-life saving.
\newblock {\em Journal of Political Economy}, 128(6):2375--2451.

\bibitem[Arcidiacono et~al., 2020]{arcidiacono2020}
Arcidiacono, P., Hotz, V.~J., Maurel, A., and Romano, T. (2020).
\newblock Ex ante returns and occupational choice.
\newblock {\em Journal of Political Economy}, 128(12):4475--4522.

\bibitem[Attanasio et~al., 2019]{attanasio2019}
Attanasio, O., Cunha, F., and Jervis, P. (2019).
\newblock Subjective parental beliefs. their measurement and role.
\newblock Technical report, National Bureau of Economic Research.

\bibitem[Aucejo et~al., 2023]{aucejo2023}
Aucejo, E.~M., French, J., and Zafar, B. (2023).
\newblock Estimating students’ valuation for college experiences.
\newblock {\em Journal of Public Economics}, 224:104926.

\bibitem[Bansak et~al., 2016]{bansak2016}
Bansak, K., Hainmueller, J., and Hangartner, D. (2016).
\newblock How economic, humanitarian, and religious concerns shape european
  attitudes toward asylum seekers.
\newblock {\em Science}, 354(6309):217--222.

\bibitem[Batista et~al., 2025]{batista2025}
Batista, C., Costa, D.~M., Freitas, P., Lima, G., and Reis, A.~B. (2025).
\newblock What matters for the decision to study abroad? a lab-in-the-field
  experiment in cape verde.
\newblock {\em Journal of Development Economics}, 173:103401.

\bibitem[Ben-Akiva et~al., 2019]{benakiva2019}
Ben-Akiva, M., McFadden, D., Train, K., et~al. (2019).
\newblock Foundations of stated preference elicitation: Consumer behavior and
  choice-based conjoint analysis.
\newblock {\em Foundations and Trends{\textregistered} in Econometrics},
  10(1-2):1--144.

\bibitem[B{\'e}nabou and Tirole, 2016]{benabou2016}
B{\'e}nabou, R. and Tirole, J. (2016).
\newblock Mindful economics: The production, consumption, and value of beliefs.
\newblock {\em Journal of Economic Perspectives}, 30(3):141--164.

\bibitem[Bernheim et~al., 2021]{bernheim2022}
Bernheim, B.~D., Bj{\"o}rkegren, D., Naecker, J., and Pollmann, M. (2021).
\newblock Causal inference from hypothetical evaluations.
\newblock Technical report, National Bureau of Economic Research.

\bibitem[Bhuller et~al., 2022]{bhuller2022}
Bhuller, M., Eisenhauer, P., and Mendel, M. (2022).
\newblock Sequential choices, option values, and the returns to education.
\newblock {\em arXiv preprint arXiv:2205.05444}.

\bibitem[Blass et~al., 2010]{blass2010}
Blass, A.~A., Lach, S., and Manski, C.~F. (2010).
\newblock Using elicited choice probabilities to estimate random utility
  models: Preferences for electricity reliability.
\newblock {\em International Economic Review}, 51(2):421--440.

\bibitem[Boyer et~al., 2020]{boyer2020}
Boyer, M.~M., De~Donder, P., Fluet, C., Leroux, M.-L., and Michaud, P.-C.
  (2020).
\newblock Long-term care insurance: Information frictions and selection.
\newblock {\em American Economic Journal: Economic Policy}, 12(3):134--69.

\bibitem[Burdett, 2012]{burdett2012}
Burdett, K. (2012).
\newblock Towards a theory of the labor market with a public sector.
\newblock {\em Labour economics}, 19(1):68--75.

\bibitem[Carneiro et~al., 2003]{carneiro2003}
Carneiro, P., Hansen, K.~T., and Heckman, J.~J. (2003).
\newblock 2001 lawrence r. klein lecture estimating distributions of treatment
  effects with an application to the returns to schooling and measurement of
  the effects of uncertainty on college choice.
\newblock {\em International Economic Review}, 44(2):361--422.

\bibitem[Cavalcanti and Santos, 2020]{cavalcanti2020}
Cavalcanti, T. and Santos, M. (2020).
\newblock {(MIS)Allocation Effects of an Overpaid Public Sector}.
\newblock {\em Journal of the European Economic Association}, 19(2):953--999.

\bibitem[Chernozhukov et~al., 2010]{chernozhukov2010}
Chernozhukov, V., Fern{\'a}ndez-Val, I., and Galichon, A. (2010).
\newblock Quantile and probability curves without crossing.
\newblock {\em Econometrica}, 78(3):1093--1125.

\bibitem[Chernozhukov et~al., 2013]{chernozhukov2013}
Chernozhukov, V., Fern{\'a}ndez-Val, I., and Melly, B. (2013).
\newblock Inference on counterfactual distributions.
\newblock {\em Econometrica}, 81(6):2205--2268.

\bibitem[Chernozhukov et~al., 2020]{chernozhukov2020}
Chernozhukov, V., Fern{\'a}ndez-Val, I., Newey, W., Stouli, S., and Vella, F.
  (2020).
\newblock Semiparametric estimation of structural functions in nonseparable
  triangular models.
\newblock {\em Quantitative Economics}, 11(2):503--533.

\bibitem[Chernozhukov and Hansen, 2005]{chernozhukov2005}
Chernozhukov, V. and Hansen, C. (2005).
\newblock An iv model of quantile treatment effects.
\newblock {\em Econometrica}, 73(1):245--261.

\bibitem[Christiaensen and Premand, 2017]{christiaensen2017}
Christiaensen, L. and Premand, P. (2017).
\newblock C{\^o}te d'ivoire jobs diagnostic: employment, productivity, and
  inclusion for poverty reduction.

\bibitem[Cummings, 1986]{cummings1986}
Cummings, R.~G. (1986).
\newblock Valuing environmental goods.
\newblock {\em An assessment of the contingent valuation method}, pages
  104--107.

\bibitem[Cunha et~al., 2004]{cunha2004}
Cunha, F., Heckman, J., and Navarro, S. (2004).
\newblock Separating heterogeneity from uncertainty an aiyagari-laitner
  economy.
\newblock In {\em Goldwater Conference on Labor Markets in Arizona}, volume
  608.

\bibitem[Cunha et~al., 2005]{cunha2005}
Cunha, F., Heckman, J., and Navarro, S. (2005).
\newblock Separating uncertainty from heterogeneity in life cycle earnings.
\newblock {\em oxford Economic papers}, 57(2):191--261.

\bibitem[Cunha and Heckman, 2007]{cunha2007}
Cunha, F. and Heckman, J.~J. (2007).
\newblock Identifying and estimating the distributions of ex post and ex ante
  returns to schooling.
\newblock {\em Labour Economics}, 14(6):870--893.

\bibitem[Delavande and Zafar, 2019]{delavande2019}
Delavande, A. and Zafar, B. (2019).
\newblock University choice: The role of expected earnings, nonpecuniary
  outcomes, and financial constraints.
\newblock {\em Journal of Political Economy}, 127(5):2343--2393.

\bibitem[Duflo et~al., 2021]{DufloEtal2021WP}
Duflo, E., Dupas, P., and Kremer, M. (2021).
\newblock The impact of free secondary education: Experimental evidence from
  ghana.
\newblock National Bureau of Economic Research No. 28937.

\bibitem[Eisenhauer et~al., 2015]{eisenhauer2015}
Eisenhauer, P., Heckman, J.~J., and Vytlacil, E. (2015).
\newblock The generalized roy model and the cost-benefit analysis of social
  programs.
\newblock {\em Journal of Political Economy}, 123(2):413--443.

\bibitem[Evdokimov, 2010]{evdokimov2010}
Evdokimov, K. (2010).
\newblock Identification and estimation of a nonparametric panel data model
  with unobserved heterogeneity.
\newblock Working Paper, Princeton University.

\bibitem[Folke and Rickne, 2022]{folke2022}
Folke, O. and Rickne, J. (2022).
\newblock Sexual harassment and gender inequality in the labor market.
\newblock {\em The Quarterly Journal of Economics}, 137(4):2163--2212.

\bibitem[Gindling et~al., 2020]{gindling2020}
Gindling, T., Hasnain, Z., Newhouse, D., and Shi, R. (2020).
\newblock Are public sector workers in developing countries overpaid? evidence
  from a new global dataset.
\newblock {\em World Development}, 126:104737.

\bibitem[Girsberger and M\'eango, 2022]{girsberger2022}
Girsberger, E.~M. and M\'eango, R. (2022).
\newblock The puzzle of educated unemployment in west africa.
\newblock IZA Discussion Paper No. 15721.

\bibitem[Giustinelli, 2022]{giustinelli2022}
Giustinelli, P. (2022).
\newblock Expectations in education: Framework, elicitation, and evidence.
\newblock {\em Available at SSRN 4318127}.

\bibitem[Giustinelli and Shapiro, 2023]{giustinelli2023}
Giustinelli, P. and Shapiro, M.~D. (2023).
\newblock Seate: Subjective ex ante treatment effect of health on retirement.
\newblock Technical report.
\newblock Unpublished manuscript.

\bibitem[Giustinelli and Shapiro, 2024]{giustinelli2024}
Giustinelli, P. and Shapiro, M.~D. (2024).
\newblock Seate: Subjective ex ante treatment effect of health on retirement.
\newblock {\em American Economic Journal: Applied Economics}, 16(2):278--317.

\bibitem[Gong et~al., 2020]{gong2020}
Gong, Y., Stinebrickner, T., and Stinebrickner, R. (2020).
\newblock Perceived and actual option values of college enrollment.
\newblock {\em Journal of Applied Econometrics}, 35(7):940--959.

\bibitem[Gong et~al., 2022]{gong2022}
Gong, Y., Stinebrickner, T., Stinebrickner, R., and Yao, Y. (2022).
\newblock The role of non-pecuniary considerations: Locations decisions of
  college graduates from low income backgrounds.
\newblock Technical report, University of Western Ontario, Centre for Human
  Capital and Productivity (CHCP).

\bibitem[Hainmueller et~al., 2014]{hainmueller2014}
Hainmueller, J., Hopkins, D.~J., and Yamamoto, T. (2014).
\newblock Causal inference in conjoint analysis: Understanding multidimensional
  choices via stated preference experiments.
\newblock {\em Political analysis}, 22(1):1--30.

\bibitem[Heckman et~al., 2006]{heckman2006}
Heckman, J.~J., Lochner, L.~J., and Todd, P.~E. (2006).
\newblock Earnings functions, rates of return and treatment effects: The mincer
  equation and beyond.
\newblock {\em Handbook of the Economics of Education}, 1:307--458.

\bibitem[Heckman and Navarro, 2007]{heckman2007}
Heckman, J.~J. and Navarro, S. (2007).
\newblock Dynamic discrete choice and dynamic treatment effects.
\newblock {\em Journal of Econometrics}, 136(2):341--396.

\bibitem[Hudomiet et~al., 2021]{hudomiet2021}
Hudomiet, P., Hurd, M.~D., Parker, A.~M., and Rohwedder, S. (2021).
\newblock The effects of job characteristics on retirement.
\newblock {\em Journal of pension economics \& finance}, 20(3):357--373.

\bibitem[Hudomiet et~al., 2018]{hudomiet2018}
Hudomiet, P., Hurd, M.~D., and Rohwedder, S. (2018).
\newblock The causal effects of economic incentives, health and job
  characteristics on retirement: estimates based on subjective conditional
  probabilities.
\newblock In {\em 2018 Working Longer and Retirement Conference}.

\bibitem[IOM, 2023]{iom2023}
IOM (2023).
\newblock National study of the labour market in c\^ote d’ivoire.
\newblock International Organization for Migration (IOM), Geneva.

\bibitem[Juster, 1966]{juster1966}
Juster, F.~T. (1966).
\newblock Consumer buying intentions and purchase probability: An experiment in
  survey design.
\newblock {\em Journal of the American Statistical Association},
  61(315):658--696.

\bibitem[Juster and Shay, 1964]{juster1964}
Juster, F.~T. and Shay, R.~P. (1964).
\newblock Consumer sensitivity to finance rates.
\newblock In {\em Consumer sensitivity to finance rates: An empirical and
  analytical investigation}, pages 6--46. NBER.

\bibitem[Karr, 1993]{karr1993}
Karr, A.~F. (1993).
\newblock {\em Probability}.
\newblock Springer New York, New York, NY.

\bibitem[Kesternich et~al., 2013]{kesternich2013}
Kesternich, I., Heiss, F., McFadden, D., and Winter, J. (2013).
\newblock Suit the action to the word, the word to the action: Hypothetical
  choices and real decisions in medicare part d.
\newblock {\em Journal of Health Economics}, 32(6):1313--1324.

\bibitem[Kettlewell et~al., 2024]{kettlewell2024}
Kettlewell, N., Walker, M.~J., and Yoo, H.~I. (2024).
\newblock Alternative models of preference heterogeneity for elicited choice
  probabilities.
\newblock {\em Available at SSRN}.

\bibitem[Ko{\c{s}}ar and O'Dea, 2023]{kocsar2023}
Ko{\c{s}}ar, G. and O'Dea, C. (2023).
\newblock Expectations data in structural microeconomic models.
\newblock In {\em Handbook of Economic Expectations}, pages 647--675. Elsevier.

\bibitem[Ko{\c{s}}ar et~al., 2022]{kocsar2022}
Ko{\c{s}}ar, G., Ransom, T., and Van~der Klaauw, W. (2022).
\newblock Understanding migration aversion using elicited counterfactual choice
  probabilities.
\newblock {\em Journal of Econometrics}, 231(1):123--147.

\bibitem[Krosnick, 1991]{krosnick1991}
Krosnick, J.~A. (1991).
\newblock Response strategies for coping with the cognitive demands of attitude
  measures in surveys.
\newblock {\em Applied cognitive psychology}, 5(3):213--236.

\bibitem[Lee et~al., 2015]{lee2015}
Lee, S. Y.~T., Shin, Y., and Lee, D. (2015).
\newblock The option value of human capital: Higher education and wage
  inequality.
\newblock Technical report, National Bureau of Economic Research.

\bibitem[Lenzner et~al., 2010]{lenzner2010}
Lenzner, T., Kaczmirek, L., and Lenzner, A. (2010).
\newblock Cognitive burden of survey questions and response times: A
  psycholinguistic experiment.
\newblock {\em Applied cognitive psychology}, 24(7):1003--1020.

\bibitem[Lewbel et~al., 2011]{lewbel2011}
Lewbel, A., McFadden, D., and Linton, O. (2011).
\newblock Estimating features of a distribution from binomial data.
\newblock {\em Journal of Econometrics}, 162(2):170--188.

\bibitem[Low, 2024]{low2024}
Low, C. (2024).
\newblock Pricing the biological clock: The marriage market costs of aging to
  women.
\newblock {\em Journal of Labor Economics}, 42(2):395--426.

\bibitem[Maestas et~al., 2023]{maestas2023}
Maestas, N., Mullen, K.~J., Powell, D., Von~Wachter, T., and Wenger, J.~B.
  (2023).
\newblock The value of working conditions in the united states and implications
  for the structure of wages.
\newblock {\em American Economic Review}, 113(7):2007--2047.

\bibitem[Mangal, 2024]{mangal2024}
Mangal, K. (2024).
\newblock {How Much Are Government Jobs in Developing Countries Worth?}
\newblock {\em The World Bank Economic Review}, page lhad047.

\bibitem[Manski, 1999]{manski1999}
Manski, C.~F. (1999).
\newblock Analysis of choice expectations in incomplete scenarios.
\newblock {\em Journal of Risk and Uncertainty}, 19:49--66.

\bibitem[Manski, 2004]{manski2004}
Manski, C.~F. (2004).
\newblock Measuring expectations.
\newblock {\em Econometrica}, 72(5):1329--1376.

\bibitem[McFadden, 2017]{mcfadden2017}
McFadden, D. (2017).
\newblock Stated preference methods and their applicability to environmental
  use and non-use valuations.
\newblock In {\em Contingent valuation of environmental goods}, pages 153--187.
  Edward Elgar Publishing.

\bibitem[McFadden and Train, 2000]{mcfadden2000}
McFadden, D. and Train, K. (2000).
\newblock Mixed mnl models for discrete response.
\newblock {\em Journal of applied Econometrics}, 15(5):447--470.

\bibitem[Meango, 2023]{meango2023b}
Meango, R. (2023).
\newblock Using probabilistic stated preference analyses to understand actual
  choices.
\newblock {\em arXiv preprint arXiv:2307.13966}.

\bibitem[Meango, 2025]{meango2025}
Meango, R. (2025).
\newblock Is the distribution of resolvable uncertainty type i extreme value? a
  test for random coefficient models using choice probabilities.
\newblock {\em arXiv preprint arXiv:2503.13901}.

\bibitem[M\'{e}ango and Poinas, 2023]{meango2022}
M\'{e}ango, R. and Poinas, F. (2023).
\newblock The option value of overstaying.
\newblock CESifo Working Paper 10536.

\bibitem[Melly and Santangelo, 2015]{melly2015}
Melly, B. and Santangelo, G. (2015).
\newblock The changes-in-changes model with covariates.
\newblock {\em Universit{\"a}t Bern, Bern}.

\bibitem[Stange, 2012]{stange2012}
Stange, K.~M. (2012).
\newblock An empirical investigation of the option value of college enrollment.
\newblock {\em American Economic Journal: Applied Economics}, 4(1):49--84.

\bibitem[Trachter, 2015]{trachter2015}
Trachter, N. (2015).
\newblock Stepping stone and option value in a model of postsecondary
  education.
\newblock {\em Quantitative Economics}, 6(1):223--256.

\bibitem[Ugalde, 2022]{ugalde2022}
Ugalde, M.~P. (2022).
\newblock Gender, grade sensitivity, and major choice.
\newblock unpublished manuscript.

\bibitem[Wiswall and Zafar, 2015]{wiswall2015}
Wiswall, M. and Zafar, B. (2015).
\newblock Determinants of college major choice: Identification using an
  information experiment.
\newblock {\em The Review of Economic Studies}, 82(2):791--824.

\bibitem[Wiswall and Zafar, 2018]{wiswall2018}
Wiswall, M. and Zafar, B. (2018).
\newblock Preference for the workplace, investment in human capital, and
  gender.
\newblock {\em The Quarterly Journal of Economics}, 133(1):457--507.

\end{thebibliography}


\newpage
\appendix
\section*{Appendix}
\section{A comparison with the revealed preference approach to the identification of ex ante returns}\label{app:comparison}
In a series of influential papers, \cite{carneiro2003, cunha2004, cunha2005} make a distinction between \textit{ex ante} and \textit{ex post} returns, where their focus is mainly on \textit{earnings returns}, the increase in the agent's lifetime earnings from a human capital investment (college education) \citep[see also a review in ][]{cunha2007}. They emphasise the difference between components of earnings variability that are forecastable and acted upon at the time students decide to go to college (heterogeneity) and components that are unforecastable. More specifically, the analyst observes the investment decision at time $t=0$, say $D$, and the associated ex post stream of income of each individual, say $Y_t$ at time $t>0$, but observes neither the potential outcomes $Y_t(0), Y_t(1)$, nor the direct costs to college education $C$,  or the full information set $\mathcal{I}_0$ at the time of decision.  The agent is assumed to take the investment if the expected gains from schooling are greater than or equal to the expected costs $\mathbb{E}(Y(1) - Y(0) - C \vert \mathcal{I}_0)$, where $Y(d)$ is the lifetime earnings with education $d$. This literature decomposes the variability of the ex post returns $Y(1) - Y(0) - C$ into its component that relate to $\mathbb{E}(Y(1) - Y(0) - C \vert \mathcal{I}_0)$, the (forecastable) heterogeneity, and $Y(1) - Y(0) - C  - \mathbb{E}(Y(1) - Y(0) - C \vert \mathcal{I}_0)$, the unforecastable component (`luck').

In the revealed preference approach, identifying predictable components that enter the information set $\mathcal{I}_0$ from unpredictable components relies critically on assumptions about the market structure facing agents and their preferences. For example, \cite{carneiro2003} considers an environment of complete autarky. \cite{cunha2005} assumes complete markets. The main identification condition is that the choice of education $D$ is not associated with the forecast error $Y(1) - Y(0) - C - \mathbb{E}(Y(1) - Y(0) - C \vert \mathcal{I}_0)$. Operationalising this condition requires an estimate of the joint distribution of potential outcome $F_{Y(1) - Y(0)}$, which the analyst can obtain by embedding a factor model structure in the generalised Roy model \citep{carneiro2003}.

Similar to the aforementioned literature, the interest in this paper is in the distribution of ex ante returns, pecuniary and nonpecuniary, $\mathbb{E}(Y(1) - Y(0) - C \vert \mathcal{I}_0)$ in their notations. Stated preference data are different in nature as they pertain to ex ante perceptions of agents. To be clear, for the objects we identify, it is not necessary for the analyst to observe the respondents perception of potential outcome \textit{at the time of decision}, $F_{Y(1), Y(0) \vert \mathcal{I}_0}$. In the empirical application, we do observe the perceived distribution of earnings \textit{at the time of elicitation}, which is close to the time of decision. What is instrumental within the choice experiment is that the analyst observes stated counterfactual choices for different, exogenous, combinations of these outcomes, say $D(y_1,y_0)$. Hence, this bypasses the problem of estimating $F_{Y(1) - Y(0)}$, which is controlled for by the analyst. When agents are allowed to express uncertainty about their decision, the analyst observes the probability that agents make the investment, say $\Pr (D(y_1,y_0) = 1)$. The results in the present paper is that the latter contains information rich enough to identify the distribution of ex ante returns for counterfactual realisations of earnings $(y_1,y_0)$, that is, $\mathbb{E}(y_1 - y_0 - C \vert \mathcal{I}_0,Y(1)=y_1,Y(0)=y_0)$ in \cite{cunha2007}'s notations. Crucially, identification can be achieved without relying on assumptions about the market structure facing agents and their preferences, or on a particular model structure. 

If in addition to stated preference data, information about realised outcomes $Y_t$ and actual choices $D$ are available, the analyst can recover the ex post returns $Y(1) - Y(0) - C$ with usual instrumental variables methodologies \citep[e.g.][]{eisenhauer2015}. This would allow decomposing the variability of returns in the spirit of \cite{cunha2005}. This case is left for future research.

Stated preference data permit an additional layer of complexity. Symmetrically to \cite{cunha2007}'s distinction between the time of decision (where \textit{outcomes are anticipated}) and the time of outcome realisation, the analyst can distinguish the time of elicitation (where the \textit{decision is anticipated}) from the time of decision. At the time of elicitation, say $t=-1$, the agent is not certain of their returns, and holds beliefs about the distribution of $\mathbb{E}(Y(1) - Y(0) - C \vert \mathcal{I}_0)$ given their information set $\mathcal{I}_{-1}$. This is the resolvable uncertainty, the uncertainty that agents expect to be resolved at the time of decision.  The result in this paper is that choice experiments eliciting the probability to make an investment are rich enough to measure the magnitude of the resolvable uncertainty.  Quantifying the resolvable uncertainty is essential to understand the option value for waiting to collect more information rather than committing to one choice ex ante \citep{heckman2007, stange2012, gong2020, meango2022}.

\section{Threshold-crossing model without a definition of a random utility model}\label{app:threshold-crossing}
Consider the threshold-crossing model defined by eqautions (\ref{eq:choice_equation_threshold}) and (\ref{eq:char_S}). Under the assumption that the mapping $y_1 \mapsto S(.,y_1,.)$ is continuous and strictly increasing, the model uniquely characterise $S$. The key idea is that threshold-crossing models are usually identified up to a scale. However, $S$ is linear in $y_1$. Thus, it fixes the scale of the threshold-crossing model. To show this formally, we can show that if two strictly monotone functions share the same sign and satisfy (\ref{eq:char_S}), but are different at some point, a contradiction arises.

First, we show the existence of at least one function $S$. Suppose that $y_1 \in [0,y_{\max}]$. For any individual with unobserved heterogeneity $\eta^*$, denote by $D(x,\eta^*) := D_i(x)$ the choice between option 0 and 1 when the choice attributes are characterised by $x$. Define: $R(y_0,z,\eta^*) = \inf_{y_1} D(y_0,y_1,z,\eta^*) = 1$. If $D(x,\eta^*) $ is always $0$ for all $y_1$, $R(y_0,z,\eta^*) = y_{\max}$. If $D(x,\eta^*) $ is always $1$ for all $y_1$, $R(y_0,z,\eta^*) = 0$. Then, define $S(y_0,y_1,z,\eta^*) = y_1 - R(y_0,z,\eta^*)$. $S(.)$ satisfies all the conditions.

To show unicity, suppose that $\tilde{S}$ are strictly monotone functions of $y_0$ and satisfy equations (\ref{eq:choice_equation_threshold}),  - $S$ and $\tilde{S}$ always have the same sign - and (\ref{eq:char_S}). However, there exists a given $(\bar{x},\eta^*)$ such that $S(\bar{x},\eta^*) \neq \tilde{S}(\bar{x},\eta^*)$. Under equation (\ref{eq:char_S}) and by monotonicity with respect to $y_1$:
$$\tilde{S}(x,\eta^*) = y_1 - \tilde{S}^{-1}(0,\bar{y}_0,\bar{z},\eta^*),$$ where $S^{-1}$ is the inverse of $\tilde{S}$ with respect to its second argument. The above expression implies that $\tilde{S}$ must be a linear function of $y_1$, just like $S$. Moreover, because $S(\bar{x},\eta^*) \neq \tilde{S}(\bar{x},\eta^*)$, it must be that $R(\bar{y}_0,\bar{z},\eta^*) \neq \tilde{S}^{-1}(0,\bar{y}_0,\bar{z},\eta^*)$. Thus, there exists some $\epsilon$, such that $|R(\bar{y}_0,\bar{z},\eta^*) - \tilde{S}^{-1}(0,\bar{y}_0,\bar{z},\eta^*)|>\epsilon.$ Yet, because $S$ and $\tilde{S}$ share the same sign, by equation (\ref{eq:choice_equation_threshold}), it must be that $R(\bar{y}_0,\bar{z},\eta^*)$ and $\tilde{S}^{-1}(0,\bar{y}_0,\bar{z},\eta^*)$ share the same sign.

Suppose without loss of generality that $\tilde{S}^{-1}(0,\bar{y}_0,\bar{z},\eta^*) > R(\bar{y}_0,\bar{z},\eta^*) \ge 0$. We can define $\bar{y}_1' = R(\bar{y}_0,\bar{z},\eta^*) + \epsilon$, leading to $S(\bar{y}_0,\bar{y}_1',\bar{z},\eta^*) = -\epsilon < 0$, which implies that $D(\bar{y}_0,\bar{y}_1',\bar{z},\eta^*) = 0$. However, $\tilde{S}(\bar{y}_0,\bar{y}_1',\bar{z},\eta^*) = \tilde{S}^{-1}(0,\bar{y}_1,\bar{z},\eta^*) - R(\bar{y}_0,\bar{z},\eta^*) + \epsilon \ge 0$, which implies that $D(\bar{y}_0,\bar{y}_1',\bar{z},\eta^*)=1$. This is a contradiction.

If $ R(\bar{y}_0,\bar{z},\eta^*) > \tilde{S}^{-1}(0,\bar{y}_0,\bar{z},\eta^*) \ge 0$, take $\bar{\bar{y}}_0' = \tilde{S}^{-1}(0,\bar{y}_0,\bar{z},\eta^*) + \epsilon.$

\section{Proofs}\label{app:proofs}
This section collects the proofs from the main text. Few notations: for any variables $X$, $Y$, we denote by $F_X(x)$ and $F_Y(y)$ the marginal cumulative distribution functions, $F_{Y\vert X}(y \vert x)$, the conditional distribution function of $Y$ given $X$, and $Q_{Y\vert X}(y \vert x)$ the conditional quantile function of $Y$ given $X$, which is the generalised inverse of $F_{Y|X}$. 

\subsection{Proof of Lemma \ref{lem:characterisation}}
Start with the reservation wage. From equations(\ref{eq:choice_equation}) - (\ref{eq:definition_R}), we have for any pair $(x,\eta^*) \in \mathcal{X} \times \mathcal{H}$ and $r \in [0,y_{\max}]$:$$\{\eta^*: R(y_0,z,\eta^*) \le r\} = \{\eta^*: r-R(y_0,z,\eta^*) \ge 0\} = \{\eta^*: S(y_0,r,z,\eta^*) \ge 0\}.$$
Hence: $F_{R,i}(x,\eta^*) = \Pr (S(y_0,r,z,\eta^*) \ge 0 \vert \eta_i) = m(y_0,r,z,\eta_i)$. The last equality follows by the definition of $m$ in Assumption \ref{ass:ceteris_paribus}.

We show the result for $F_{S,i}$ using the characterisation of equation (\ref{eq:char_S}).
By equation (\ref{eq:char_S}) and the monotonicity of $S$, for any pair $(x,\eta^*) \in \mathcal{X} \times \mathcal{H}$ and $s$ such that $(y_1 - s) \in [0,y_{\max}]$: $S(y_0,y_1- S(x,\eta^*),z,\eta^*) = 0$ if and only if $S(x,\eta^*) = y_1 - S^{-1}(y_0,0,z,\eta^*),$ where $S^{-1}$ is the inverse of $S$ with respect to its second argument. It follows that: $\bigl\{{\eta^*: S(x,\eta^*) \le s}\bigr\} = \bigl\{{\eta^*: y_1 - S^{-1}(y_0,0,z,\eta^*) \le s}\bigr\} = \bigl\{{\eta^*: (y_1- s) - S^{-1}(y_0,0,z,\eta^*) \le 0}\bigr\} =  \bigl\{{\eta^*: S(y_0,y_1-s,z,\eta^*)  \ge 0}\bigr\}$. Taking the conditional expectation, it follows that:
    \begin{eqnarray*}
        F_{S,i}(s;x) &=& \Pr (S(x,\eta_i^*) \le s \vert \eta_i), \text{ by definition.}\\
        &=& \Pr (S(y_0,y_1-s,z,\eta_i^*)  \le 0 \vert \eta_i)\\
        &=& 1 - m(y_0,y_1-s,z,\eta_i), \text{ by definition of } m(.).
    \end{eqnarray*}

\subsection{Proof of Theorem \ref{th:characterisation}}
We show the result for the ex ante returns. The proof for the reservation wage is essentially similar. From Lemma \ref{lem:characterisation}, we can write:
\begin{eqnarray*}
F_{Q,R}(s;\tau, F_{\tilde{X}}) &=& \Pr\left({Q_{S,i}(\tau;\tilde{X}) \le s }\right) \\
&=& \Pr\left[{F_{S,i}(s;\tilde{X}) \ge \tau}\right] \\
&=& \mathbb{E} \left[{I\left\{{1-m(y_0,y_1-s,z,n)\ge \tau}\right\}}\right]\\
&=& \int_{\mathcal{X}}\int_{\mathcal{H}} I\left\{{m(y_0,y_1-s,z,n)\le 1 - \tau}\right\} dF_{\eta \vert X }(n \vert y_0,y_1,z)dF_{\tilde{X}}(y_0,y_1,z)\\
&=& \int_{\mathcal{X}}\int_{\mathcal{H}} I\left\{{m(y_0,y_1-s,z,n)\le 1 - \tau}\right\} dF_{\eta \vert }(n)dF_{\tilde{X}}(y_0,y_1,z).
\end{eqnarray*}
The second equality is by definition of the quantile function and the distribution function, the third equality is by definition, the fourth equality makes the expectation explicit, and the last one holds because of the monotonicity of $S$ with respect to $y_1$. Notice now that the RHS is monotonically decreasing with the quantity $m(y_0,y_1-s,z,n)$, and depends on $\eta$ only through $m$. Thus, one can rewrite the integral as:
\begin{eqnarray*}
F_{Q,S}(s;\tau, F_{\tilde{X}}) 
&=& \int_{\mathcal{X}}\int_0^1 I\left\{{Q_{m(X,\eta) \vert X}(a \vert y_0,y_1-s,z)\le 1 - \tau}\right\} da\; dF_{\tilde{X}}(y_0,y_1,z)\\
 &=& \int_{\mathcal{X}} \int_{0}^1 I\left\{{Q_{P \vert X}(a \vert y_0,y_1-s,z)\le 1 - \tau}\right\} da\;dF_{\tilde{X}}(y_0,y_1,z)
\end{eqnarray*}
The RHS is identified from the observed joint distribution of $(P,X)$ and any given distribution $F_{\tilde{X}}$. Hence the distribution of quantiles (LHS) is identified.

\subsection{Proof of Theorem \ref{th:characterisation_2}}
We show that: $ q(x,a) = Q_{m(x,\eta)}(a)$, the quantile treatment response (QTR) function. Note that for any $a \in (0,1)$ and any $x \in \mathcal{X}$
    \begin{eqnarray*}
Q_{m(x,\eta) }(a) &=& Q_{m(X,\eta) \vert X}(a \vert x) \text{ by independence as from Assumption \ref{ass:independence},}\\
        &=& Q_{P \vert X}(a \vert x) = q(x,a)
    \end{eqnarray*}
The QTR is very convenient because it translates a possibly infinite dimensional problem (the dimension of $\eta$) to a uni-dimensional problem (the dimension of $\alpha$). Indeed, for any respondent $i$, there exists $\alpha_i$, a realisation of a uniformly distributed random variable such that $P_{i} = m(X_{i},\eta_i) =q(X_{i},\alpha_i)$, where $\alpha_i = F_{P_{i} \vert X_{i}}(P_{i} \vert X_{i})$. \vskip4pt
Recall that: \[
\begin{array}{l}
A^{\mu}(x,a) =  \int_{\mathcal{S}} \left[{ q(t(s,x),a) - 1\{s\le 0\}}\right]ds \\
A^\tau(x,a) = \int_{\mathcal{S}} 1 \left\{{\left[{ 1- q(t(s,x),a)}\right] \le \tau}\right\} - 1\{s\le 0\} ds,\\
\end{array}
\]
$\mu_S(x,\eta)$ and $Q_{S}(\tau;x,\eta)$ can be seen as random objects, which are strictly increasing functionals of $m(x,\eta)$. $A^\mu(x,a)$ and $A^\tau(x,a)$ that replace $q(x,a)$ for $m(x,n)$ represent their quantile functions. 

Thus equations (\ref{eq:def_mu}) can be rewritten:
\begin{eqnarray*}
&&\Pr\left({\mu_S(X,\eta) \le y }\right)
=  \int_{\mathcal{X}} \int_{0}^1  1 \left\{{A^{\mu}(x,a) \le y }\right\}  da\; dF_{\tilde{X}}(x), \\ 
\end{eqnarray*}
Things are more challenging with Mean Treatment Effects ( $mWTP$) and Quantile Effects ($qWTP$). This is because the difference of two quantile functions does not in general equal the quantile of the difference. One exception is when there is rank invariance. This would imply that $\eta$ is uni-dimensional, a hypothesis we do not wish to impose in the general case.
\vskip4pt
We adopt here the potential outcome notation of \cite{chernozhukov2005}, that define a collection of random variable $\{\alpha_x\}_\mathcal{X}$, such that $P_x = q(x,\alpha_x)$, where $\alpha_x = F_{m(x,\eta)}(m(x,\eta)) \sim U(0,1)$ and $P = q(X,\alpha_X)$. To simplify notation, denote $x' = x+h$. Then, replacing $\mu_S$ and $Q_S$ by their quantile functions in equations gives: 
\begin{eqnarray*}
\Pr\left({mWTP(h;X,\eta) \le y }\right) &=&\int_{\mathcal{X}} \int_{(0,1)^2}  1 \left\{{A^{\mu}(x',a_{x'}) - A^{\mu}(x,a_x)  \le y }\right\}  dF_{\alpha_{x'},\alpha_x}(a_{x'},a_x) dF_{\tilde{X}}(x), \label{eq:quantile_ME} \\
\Pr\left({qWTP(h,\tau;X,\eta) \le y }\right)
&=& \int_{\mathcal{X}} \int_{(0,1)^2}  1 \left\{{A^\tau(x',a_{x'}) - A^\tau(x,a_x) \le y }\right\}   dF_{\alpha_{x'},\alpha_x}(a_{x'},a_x) dF_{\tilde{X}}(x), \label{eq:quantile_QTE}
\end{eqnarray*}

Compared to the previous quantities, the unknown here is $F_{\alpha_{x'},\alpha_x}$. This joint distribution has marginal distributions that are uniformly distributed but not necessarily independent. The idea is to use scenarios $(P_1,X_1,P_2,X_2)$, to learn about the joint distribution of $F_{\alpha_{x'},\alpha_x}$. More specifically, recall that $V_k = F_{P_k \vert X_k}(P_k \vert X_k), k = 1,2$.

Note that:
\begin{eqnarray*}
    F_{P_k \vert X_k} (p \vert x) &=& \Pr \left({P_k \le p \vert X_k = x}\right)\\
    &=& \Pr \left({m(X_k, \eta)\le p \vert X_k = x}\right)\\
    &=& \Pr \left({m(x, \eta)\le p  }\right), \text{ since } X_k \independent \eta,\\
    &=& \int I \left\{{m(x,n) \le p}\right\} dF_{\eta}(n).
\end{eqnarray*}
Applying the conditional distribution to the pair $(P,X)$ gives the following expression for $V_k, k = 1,2$:
\begin{equation}
    V_k = \int I \left\{{m(X_k,n)  \le P}\right\} dF_{\eta}(n) 
    =\int I \left\{{m(X_k,n) \le  m (X_k, \eta)}\right\} dF_{\eta}(n)
    = \varphi(X_k,\eta).
\end{equation}
Note that for any $x \in \mathcal{X}$: $$\varphi(x,\eta) = \int I \left\{{m(x,n) \le  m (x, \eta)}\right\} dF_{\eta}(n) = F_{m(x,\eta)}(m(x,\eta)) = \alpha_x.$$
The conditional joint distribution of $(V_1,V_2)$ is associated to the following characteristic function:
\begin{eqnarray} \label{eq:joint_characteristic_function}
 \nonumber  && \phi_{V_{1}, V_{2} \vert X_{1}, X_{2}}(s_1,s_2 \vert X_{1}=x, 
    X_{2}=x')\\ &&=  \nonumber \mathbb{E}\left({\exp \left({i s_1 V_{1} + i s_2 V_{2}}\right) \vert X_{1}=x, X_{2}=x'}\right),\\
   \nonumber  &&=  \mathbb{E}\left({\exp \left({i s_1 \varphi(X_1,\eta) + i s_2 \varphi(X_2,\eta)}\right) \vert X_{1}=x, X_{2}=x'}\right),\\
      \nonumber  &&=  \mathbb{E}\left({\exp \left({i s_1 \varphi(X_1,\eta)}\right)\vert X_{1}=x}\right)\mathbb{E}\left({\exp \left({i s_2 \varphi(X_2,\eta)}\right)\vert X_{2}=x'}\right),\text{ since } X_k \independent (X_{3-k},\eta), k=1,2\\
      \nonumber  &&=  \mathbb{E}\left({\exp \left({i s_1 \varphi(x,\eta)}\right)}\right)\mathbb{E}\left({\exp \left({i s_2 \varphi(x',\eta)}\right)}\right),\text{ since } \eta \independent X_1,X_2\\
      \nonumber  &&=  \mathbb{E}\left({\exp \left({i s_1 \alpha_{x}}\right)}\right)\mathbb{E}\left({\exp \left({i s_2 \alpha_{x'}}\right)}\right),\\  
  \nonumber  &&=  \mathbb{E}\left({\exp \left({i s_1 \alpha_{x} + i s_2 \alpha_{x'}}\right) }\right)
\end{eqnarray}
Hence, the conditional joint distribution of $(V_1,V_2)$, $F_{V_1,V_2 \vert X_1,X_2}(. \vert x,x')$, identifies the joint distribution $F_{\alpha_{x'},\alpha_x}$. The former distribution is identified from the stated choices in the pair of scenarios.

\subsection{Proof for the lower bound on the cost elasticity}
A first-order Taylor expansion of  gives:
$$\text{Elasticity } \approx \varepsilon\left({1 + \dfrac{1+\varepsilon}{\varepsilon y_0}}\Bigl[{F_S^{-1}\bigl[{F_S(0)}\bigr] + \varepsilon F_S(0).(F_S^{-1})'(F_S(0))}\Bigr]\right) = \varepsilon\left({1+\dfrac{1+\varepsilon}{y_0}\dfrac{F_S(0)}{F_S'(0)}}\right).$$
The last equality follows by noting that $F_S^{-1}\bigl[{F_S(0)}\bigr] = 0$ and $(F_S^{-1})'(F_S(0)) = 1/F_S'(0)$.
To recover a lower bound, it is necessary to find a lower bound for $\dfrac{F_S(0)}{F_S'(0)} = \left({\dfrac{d \log(F_S(s))}{ds}}\right)^{-1}$. Recalling equation (\ref{eq:best_predictor_equal}), we have:
\begin{equation}\label{eq:best_predictor_equal_2}
\dfrac{F_S'(s)}{F_S(s)}= \dfrac{ \int_0^1 F'_{Q,S}(s;\tau)\omega_\tau d\tau}{\int_0^1 F_{Q,S}(s;\tau)\omega_\tau d\tau} = \int_0^1 \tilde{w}(s;\tau)\dfrac{ F'_{Q,S}(s;\tau)}{F_{Q,S}(s;\tau)}d\tau.
\end{equation}
where $\tilde{w}(s;\tau) = \dfrac{F_{Q,S}(s;\tau)\omega_\tau}{\int_0^1 F_{Q,S}(s;\tau)\omega_\tau d\tau}$ are positive weights that sum up to one. 
This quantity is maximised by:
$\max_{\tau \in (0,1)} \dfrac{ F'_{Q,S}(s;\tau)}{F_{Q,S}(s;\tau)}$. The inverse provides the required minimum.

\section{Perception of the labour market}\label{app:perception}
Perceptions in this section are elicited by asking the students to think about 20 students with similar characteristics as theirs (same age, gender, education, and family background).\footnote{For example, for the likelihood of receiving a job offer from the public sector: \textit{Imagine 20 young Ivorians your age, i.e. [age variable]. Imagine they have an education similar to yours, i.e. [degree and field of study], all [male / female] like you, with similar family backgrounds. Think of these young people as they enter the job market. Out of these 20 young people, how many do you think will receive at least one job offer in the public sector? }} To elicit the perceived sector-specific wage distribution, the students are instructed to think instead of one wage offer for each in a given sector.\footnote{More precisely, the question states: \textit{Imagine that these twenty young people each receive a job offer in the public sector. These job offers may vary according to the administration or public company offering them. We'll show you categories representing salary ranges, and ask you to allocate the twenty job offers to each range.}} Perceptions of students in either track are broadly similar, aside from a level difference in wage. Therefore, they are pooled in the following exposition.

\begin{figure}
    \centering
    \includegraphics[width=0.6\linewidth]{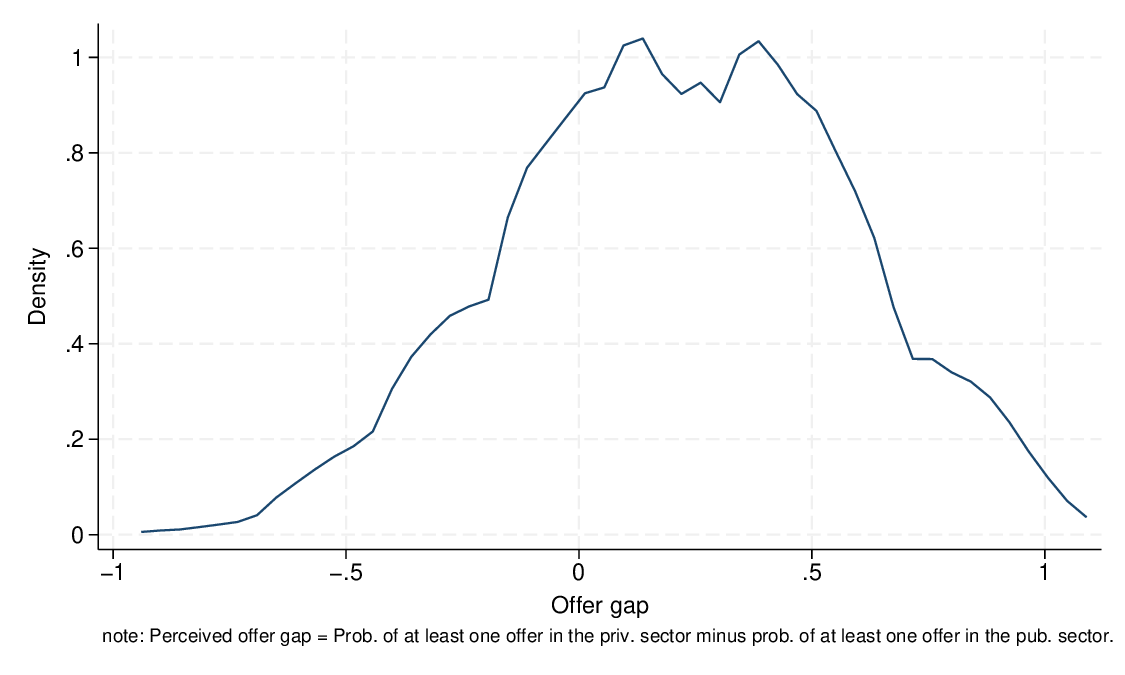}
    \caption{Perceived offer gap distribution}
    \label{fig:offer_gap}
    \floatfoot{Note: The figure represents the probability distribution of the average perceived offer gap between two sectors. The offer gap is defined as the perceived probability of receiving a private sector offer minus the same perceived probability for a public sector offer. }
\end{figure}

First, three quarters of students perceive that jobs are equally or more abundant in the private sector than in the public sector. The average likelihood of receiving a job offer stands at 0.50 in the public sector, and 0.71 in the private sector. Figure \ref{fig:offer_gap} shows the distribution of perceived offer rate gap between the public and the private sector (offer rate in private minus public), which is tilted toward positive values.  The median individual perceived a 25-percentage-point higher chance to receive a job offer from the private sector. Yet, a quarter of the population perceives better opportunities in the public sector.
\begin{figure}
    \centering
        \subfigure[Perceived wage distribution by sector]{ 
        \includegraphics[width=0.7\linewidth]{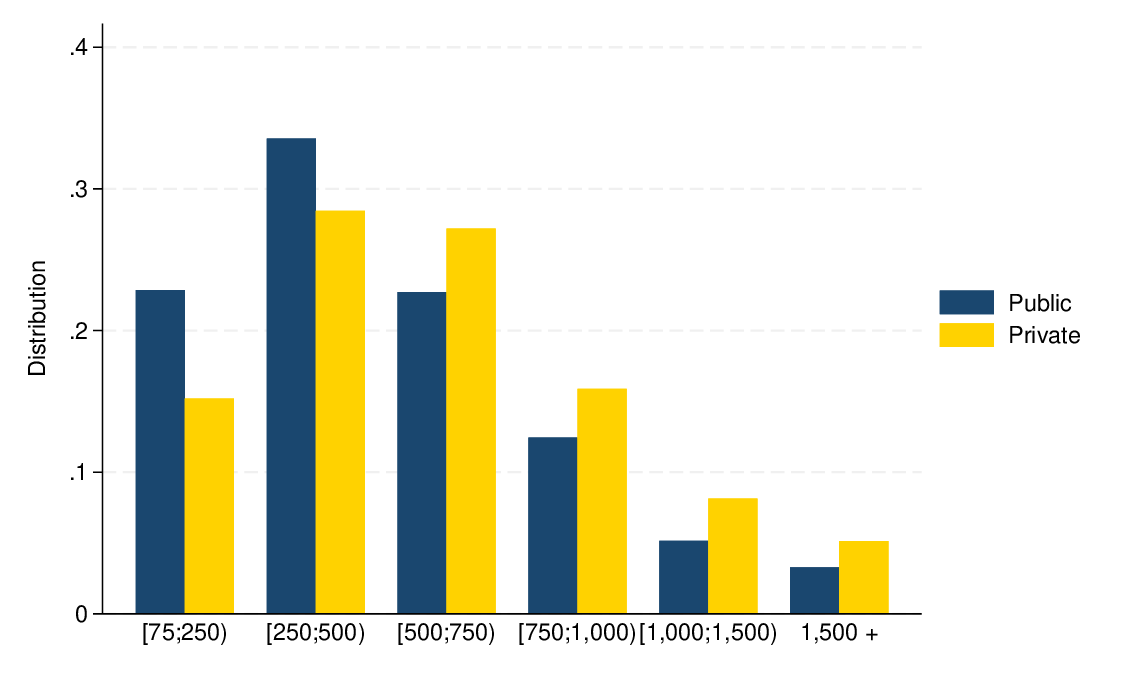}
        }
                \subfigure[Average perceived wage gap]{ 
    \includegraphics[width=0.7\linewidth]{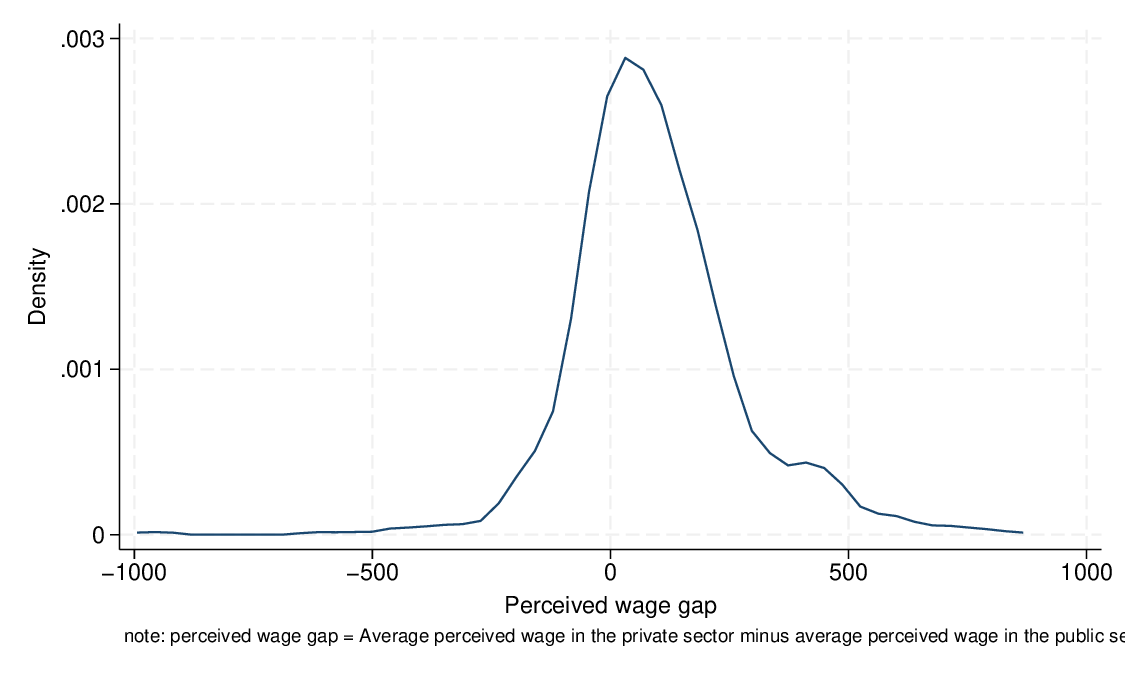}
    }
    \caption{Perceived wage by sector and average perceived wage gap}
    \label{fig:wage_perception}
            \floatfoot{Note: In Panel (a), each bar represents the reported a probability of a wage offer following within a given bin by sector averaged on the sample. The intervals are in 1,000 CFA Francs. Panel (b) represents the probability distribution of the average perceived wage gap between two sectors. For each individual, for each sector, an average perceived wage is calculated by taking the median of the interval and the mass associated with the interval. The wage gap is defined as the average perceived wage in the private sector minus the average perceived wage in the public sector. }
\end{figure}

Students also believe that the private sector provides jobs with higher pay.  Figure \ref{fig:wage_perception} shows in Panel (a) the distribution of wages by sector averaged across all respondents. The average distribution of wages in the private sector is clearly shifted to right, with less offers in the bottom categories and more offers in the middle and top categories. Panel (b) constructs an average perceived wage for each individual and each sector, and computes the distribution of the average wage gap in the population. Four out of five respondents perceive that jobs are, on average, equally or better paid in the private sector than in the public sector. The perceived average gain amounts to about 96,600 CFA Franc, that is 16.7 percent of the perceived average wage in the public sector. Note that about two third of the population perceives the wage distribution in the public sector to be more concentrated than the one in the private sector.

\begin{figure}
    \centering
    \includegraphics[width=0.6\linewidth]{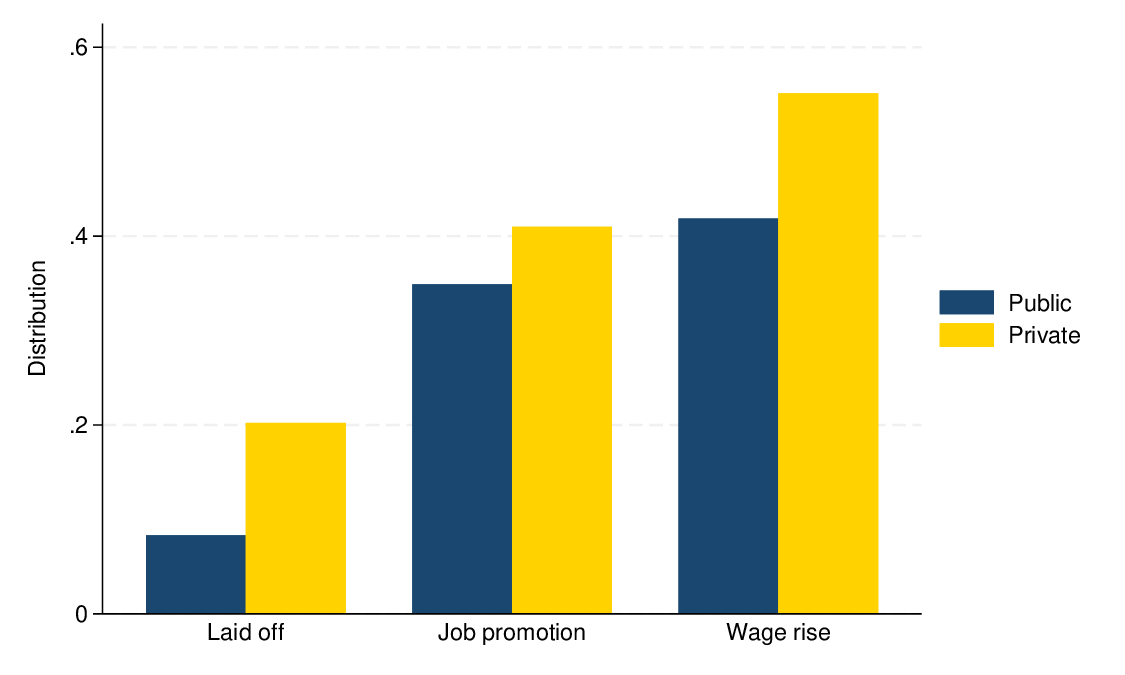}
    \caption{Other perceived characteristics by sector}
    \label{fig:non_wage}
        \floatfoot{Note: Each bar represents the reported a probability by sector averaged on the sample. `Laid off' represents the probability of being laid off within two years, `job promotion', the probability of being promoted to a more senior position, `wage rise', the probability of receiving a 20 percent wage rise within two years.}
\end{figure}
Furthermore, students perceive private-jobs as less secure, but offering more possibility for career and wage progression (Figure \ref{fig:non_wage}). The perceived job destruction rate after two years in the private sector is, on average, 20 percent, more than double the job destruction rate in the public sector (8 percent). Concomitantly, opportunities for moving up the job ladder are seen as better in the private sector, with 6 percentage points (pp) gap, on average in the chance of obtaining a career advancement and a 13 pp gap in the chance of obtaining a 20 percent wage rise within two years.\footnote{All differences are significant at standard levels.} 

\begin{figure}
    \centering
        \subfigure[Verbal description of the public sector]{ 
        \includegraphics[width=0.7\linewidth]{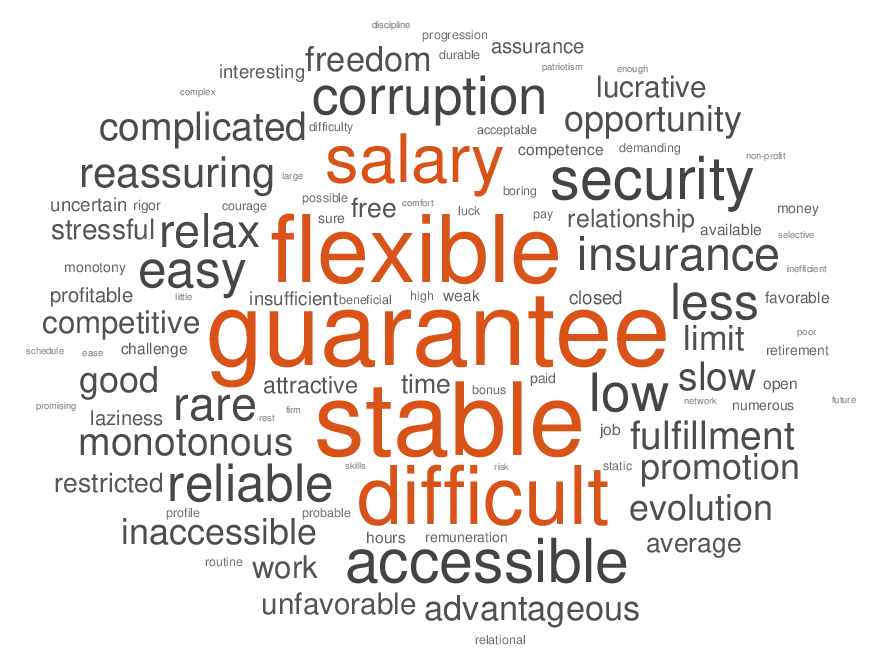}
        }
                \subfigure[Verbal description of the private sector]{ 
    \includegraphics[width=0.7\linewidth]{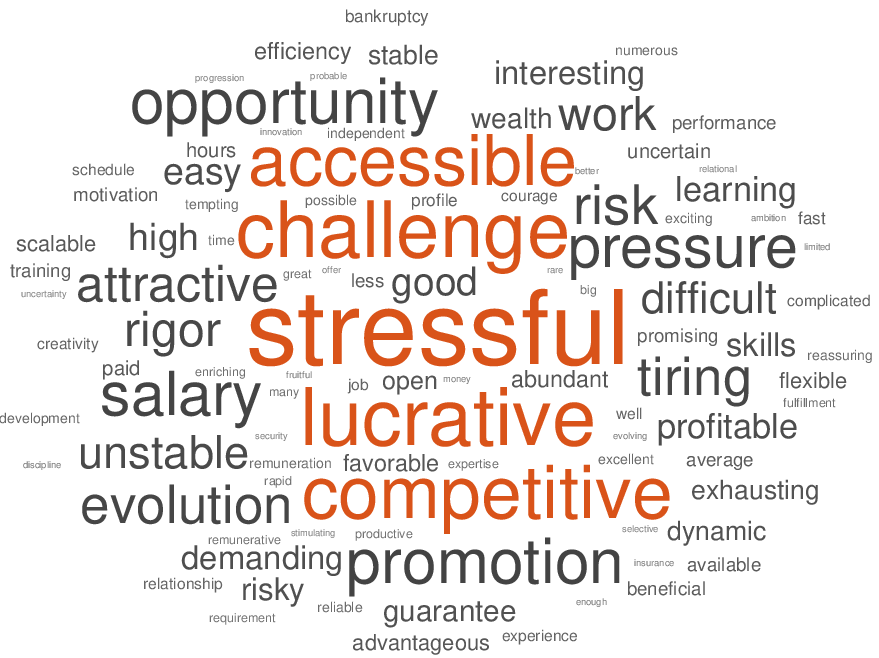}
    }
    \caption{What students think about each sector in three words.}
            \floatfoot{Note: The word cloud represent the words used by the students to describe each sector. Each respondent could use three words. The bigger the font size, more often this word is used in the verbal descriptions. The words in orange represent the top 5 words in each sector.}
    \label{fig:trois_mots}
\end{figure}

Finally, respondents are asked to describe in three words their opportunities in each sector (figure \ref{fig:trois_mots}). The top-three words describing public sector opportunities are `guarantee', `stable', and `flexible'.\footnote{Within the Top-10 words, one finds: `security', `reliable', `insurance'. The word `corruption' also appears in the Top-10 list, while it is never mentioned in relationship with the private sector.} By contrast, the top three words describing the private sector opportunities are `stressful', `lucrative', and `competitive.'

The general insight from these results is that public and private sector jobs are perceived very differently in the population. One the one hand, the private sector appears as a dynamic sector with more abundant, better paid jobs, and more opportunity for progression. However, these benefits are tied with stressful work conditions. On the other hand, the public sector appears as a secure, stable alternative, even if less well paid. It is important to note that beyond these general trends, there is heterogeneity in individual perceptions.

It is difficult to compare the stated beliefs to actual labour market statistics. Our efforts to construct a comparable sample of workers from existing (survey) data was unsuccessful. For example, the latest publicly available labour force survey ERI/ESI 2017 contains only 69 workers aged between 25 and 34 with university education and available wage information, and may not represent well students from these elite universities.
Comparing with other data sources is still instructive. The average wage perception of those high-ability student securely places them in the top quartile of the wage distribution in C\^ote d'Ivoire \citep[compare with][p.124]{christiaensen2017}, which reflects well their ability. Looking at the distribution of their wage perception, students seem aware that even for them, offers below the mean wage in STEM occupations can be frequent (23 percent for public sector jobs, 15 percent for private sector jobs).\footnote{According to the ILO database, the 2019 average wage in STEM occupations (for all education levels) was 227,876 CFA Franc.} The perceived wage penalty for public sector jobs may seem counter-intuitive given the received knowledge of a wage premium in the public sector. \cite{girsberger2022} using a regional survey data conducted in 2003 for francophone countries in West Africa, including C\^ote d'Ivoire, find that a public sector wage premium exists only for low education groups, not for tertiary educated. However, destruction rates in the public sector are significantly lower than in the private sector. \cite{gindling2020}, comparing 68 countries including several low-income countries, also conclude that high skilled public sector employees in several of those countries pay a wage penalty for working in the public sector. Hence, the evidence gathered suggest that students' perception about the labour market are not too far off. Following their career development and comparing those to their initial beliefs is a fruitful research avenue that we hope to pursue in future. 

\subsection{Preferences over sectors}\label{app:subsec:preferences}
Respondents are asked first to rank five type of employers according to their attractiveness:  public administration, public sector firms, small and medium-sized (private) enterprises (SMEs), large-sized enterprises, and international institutions.\footnote{This includes regional institutions, as the WAEMU or ECOWAS and their agencies, panafrican institutions, for example, the African Union and its agencies, or intercontinental institution, for example, the United Nations and its agencies.}
\begin{figure}
    \centering
    \includegraphics[width=0.6\linewidth]{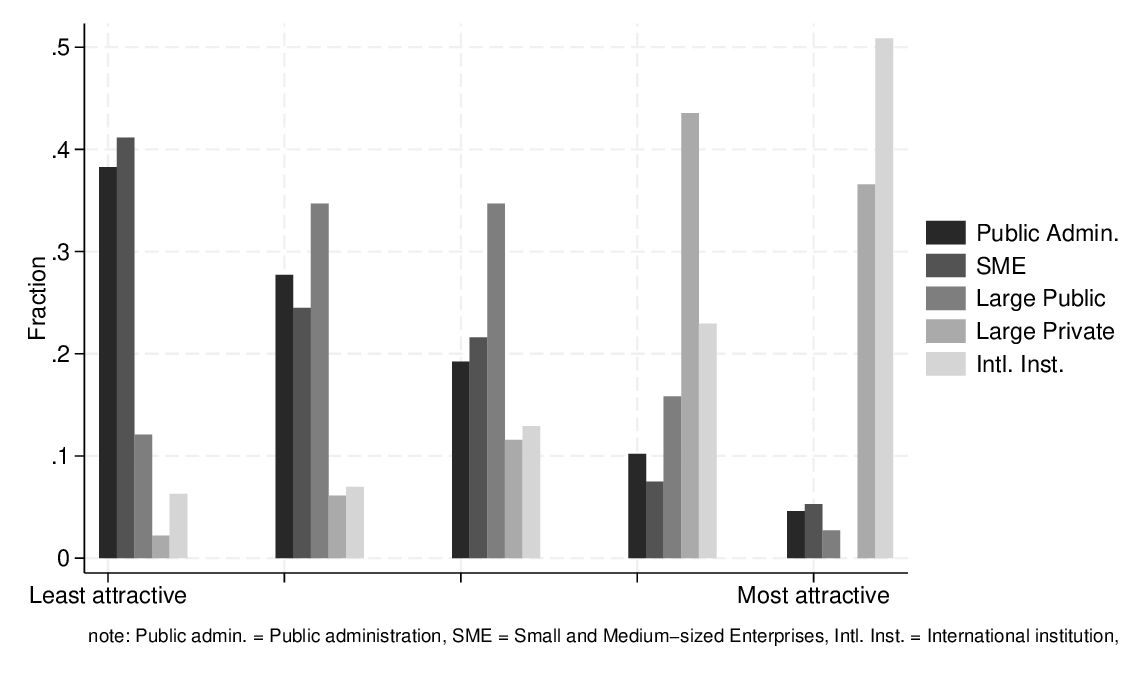}
    \caption{Attractiveness of sectors}
    \label{fig:attractiveness}
    \floatfoot{Note: The figure represents the average ranking of employers. Ranks go from 1 (Most attractive) to 5 (least attractive). International institutions and large private firms are, in general, the most attractive, SMEs and public administration are the least attractive.}
\end{figure}
Figure \ref{fig:attractiveness} presents the result of this exercise. International institutions are at the top of the ranking, being ranked as the most attractive employers by more than half of the sample. They are closely followed by large private firms, which are ranked first by close to four out of ten respondents. Large public firms are generally in the middle position. Public administration and SMEs share the bottom place with seventy percent of the population ranking them as one of the two least attractive employers. Thus, it appears that the divide between public and private is not enough to explain preferences and the ranking is influenced also by the size of the firm.

\section{Further details about the survey}\label{app:survey}
This section presents the instructions for the choice experiment. They were presented using a short video.

\begin{quote}
\textbf{Choice experiment instructions.} Now we're going to show you a series of hypothetical scenarios. In these scenarios, you receive two job offers for a position that corresponds to your degree. One offer comes from the public sector, the other from the private sector. In these scenarios, these two offers differ in the starting salary they offer, the number of hours of work expected, the risk of losing the job, and the progression in position. \vskip12pt
For example, let's compare these two offers:\vskip12pt
\textbf{Offer 1:}\\
Your employer: Public administration\\
Starting salary 500,000 \\
Number of hours per week 40 \\
Risk of losing job within two years 2/100 \\
Number of people who become managers with salary +20\%: 20/100
\vskip12pt
\textbf{Offer 2:}\\
Your employer: SME \\
Starting salary 800,000 \\
Number of hours per week 60 \\
Risk of losing job within two years 10/100 \\
Number of people who become managers with salary +20\%: 1/10 \\
\vskip12pt
The starting salary in the public sector offer is 500,000 FCFA per month for a job in public administration. It is normal to work 40 hours a week. 2 out of 100 people lose their job within two years. Unemployment lasts 6 months. 20 people out of 100 become managers in their administration or company after two years.

The starting salary in the private sector is 800,000 FCFA per month for a job in an SME. It's normal to work 60 hours a week. 20 people out of 100 lose their job within two years. Unemployment lasts 6 months. 10 people out of 100 become managers in their administration or company after two years.

We are going to ask you to consider several such scenarios. Considering all the factors that might affect your decision to enter the job market (e.g. family background, your parents' wishes, macroeconomic conditions), we'd like to know which offer you'd be most likely to choose. We'll ask you how likely you would be to choose the [randomize= private/public] sector.
Between different scenarios, the only variations are indicated by the scenario. For example, the level of responsibility or the number of days off in the public sector job remain the same. Only the stated characteristics change. The same goes for the private sector offer. You can consider that these offers are your only options, and that you would not receive another offer within two years if you turned down both offers.

Q1 - Q5. Consider the following two offers, what is the chance that you will choose the offer from the [randomize= private/public] sector.
    
\end{quote} 

Table  \ref{tab:support} presents the support of choice attributes in the choice experiment. The values are chosen randomly for the second to the fifth scenario. In the first scenario, the employer are public administration and SME, the weekly hours worked 40, the chance of losing job is 5 percent, the chance of job job promotion, 10 percent, and the starting salary 750,000 CFAF (25 percent above the average perceived wage). 
\begin{table}[htbp]
  \centering
  \caption{Support of choice attributes in the choice experiment}
    \begin{tabular}{ll} 
    \hline \hline
    Public [Randomised between the values] &  \\ \hline
    Employer & Public administration, Public sector company \\
    Number of hours per week & 35, 40 \\
    Chance of losing job within two years  & 2/100, 5/100, 10/100 \\
    Chance of job promotion within two years & 5/100, 10/100, 20/100 \\
    Starting salary & from 300K to 1,000K with steps of 50K \\
          &  \\ \hline
    Private [Randomised between the values] &  \\ \hline
    Employer & SME, Large-sized enterprise \\
    Number of hours per week & 40, 50, 60 \\
    Chance of losing job within two years  & 10/100, 20/100, 30/100 \\
    Chance of job promotion within two years & 5/100, 10/100, 20/100 \\
    Starting salary & from 300K to 1,000K with steps of 50K \\ \hline \hline
    \end{tabular}%
  \label{tab:support}%
      \floatfoot{Note: Each row reports the possible values that a choice attribute can assume. Values are randomly assigned among individuals. }
\end{table}%
\newpage
\section{Additional figures}\label{app:additional_figures}

\begin{figure}[htbp]
    \includegraphics[width=0.6\linewidth]{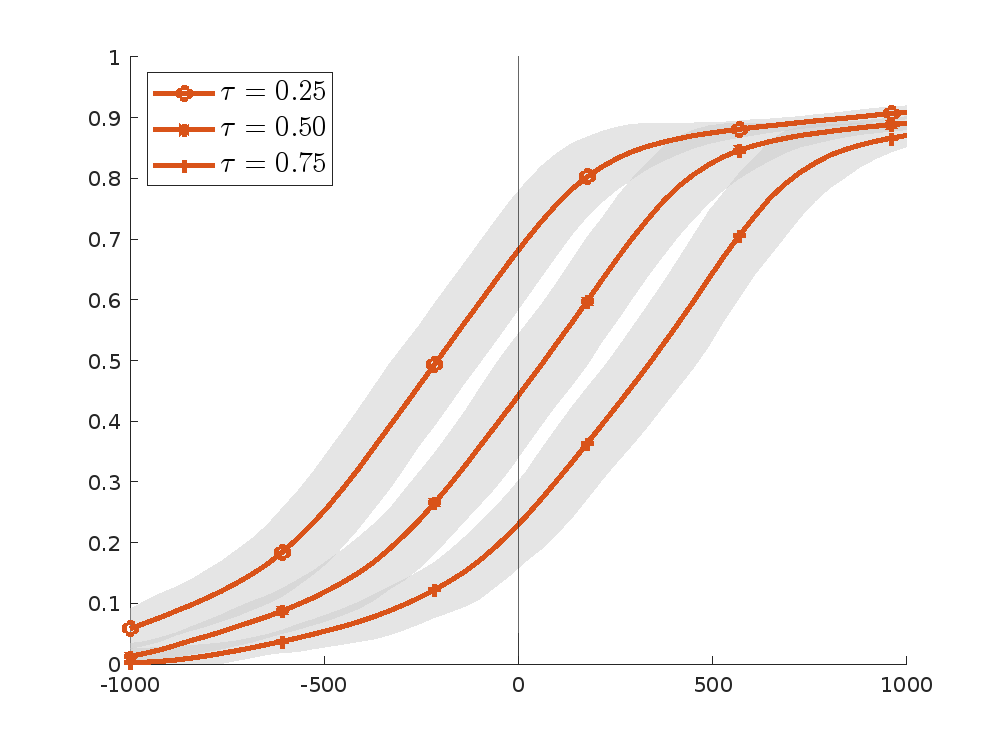}
    \caption{Distribution of quantiles $\hat{F}_{Q,S}\left({.;\tau, F_{\tilde{X}}}\right), \tau =0.25, 0.50, 0.75$ for individual-specific perceptions and equating the perceived wage to the wage in the public sector.}
    \floatfoot{Note: The dark-grey area shows the 90 percent pointwise confidence interval. $F_{\tilde{X}}$ is such that the wage offers are both equal to the individual-specific perceived average wage in the public sector. Since the wage is the same in both sectors, the figure shows the distribution of returns that comes solely from the amenities.} 
    \label{fig:distr_quantile_FQ_indiv}
\end{figure}

\begin{figure}[htbp]
    \centering
        \includegraphics[width=0.65\linewidth]{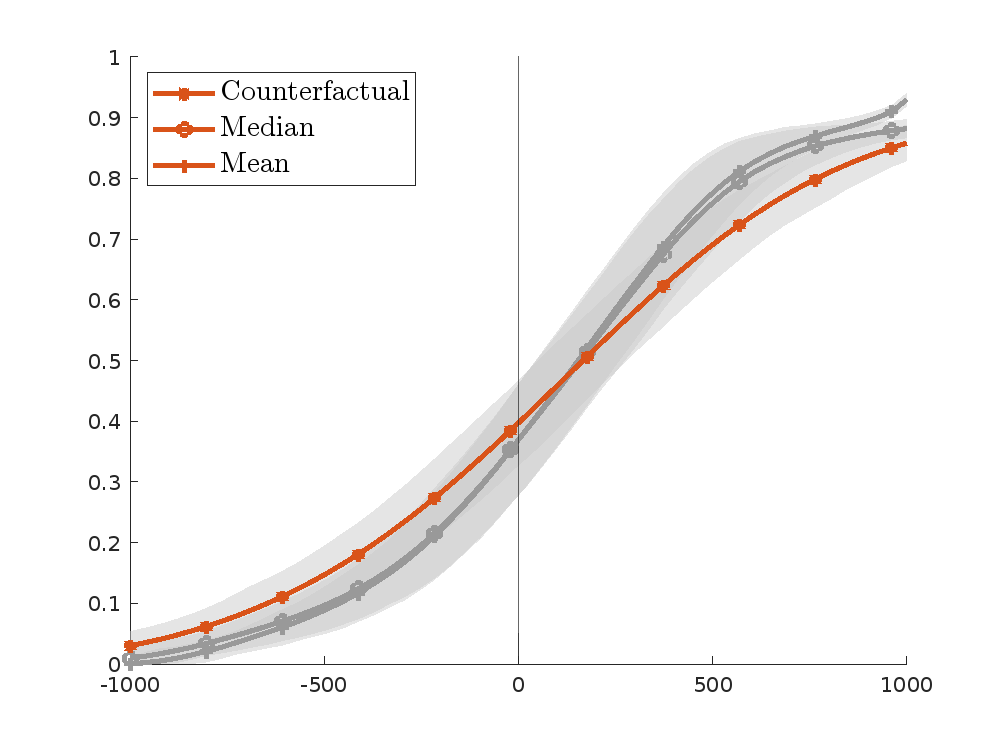}
    \caption{Comparing the predicted distribution of realised returns to the distribution of average returns and median returns.}
    \floatfoot{Note: The predicted distribution of realised returns (`counterfactual') is calculated by equation (\ref{eq:best_predictor_equal}), using quantiles $\tau = 0.05, 0,10, \ldots, 0.95$. The distributions of mean and median returns are calculated using Theorem \ref{th:characterisation} and \ref{th:characterisation_2}. $F_{\tilde{X}}$ is set for individual-specific average offers.}
    \label{fig:compare_distributions}
\end{figure}

\begin{figure}
    \centering
    \subfigure[$F_{Q,R}(s;\tau, F_{\tilde{X}})$]{
    \includegraphics[width=0.45\linewidth]{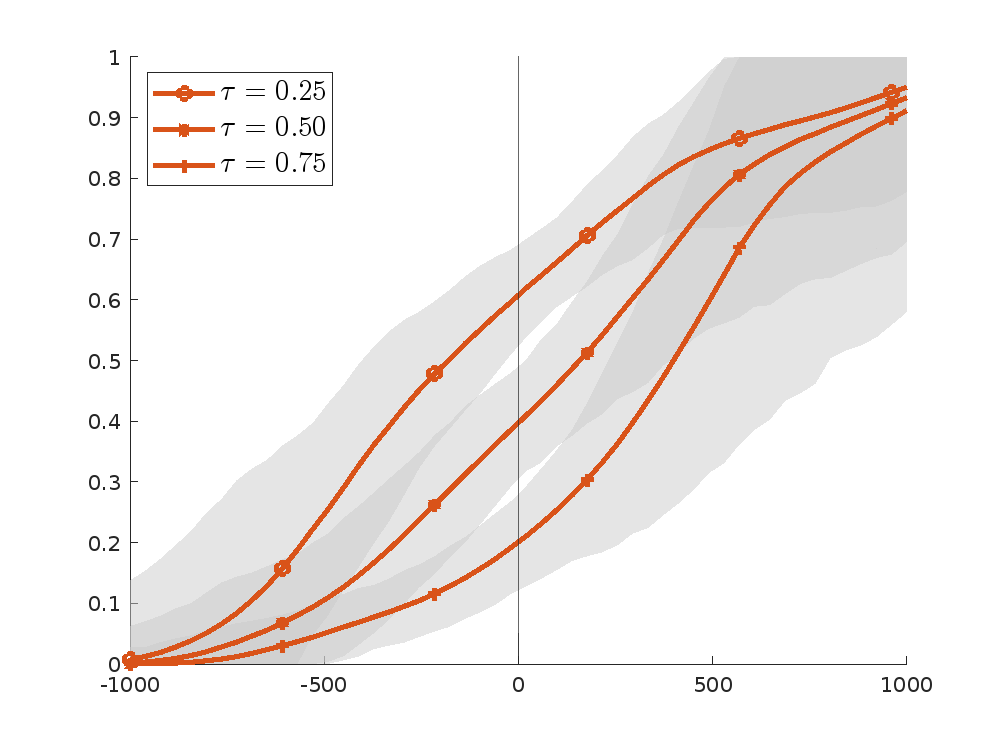}
    }
    \subfigure[Distribution of IQR]{
    \includegraphics[width=0.45\linewidth]{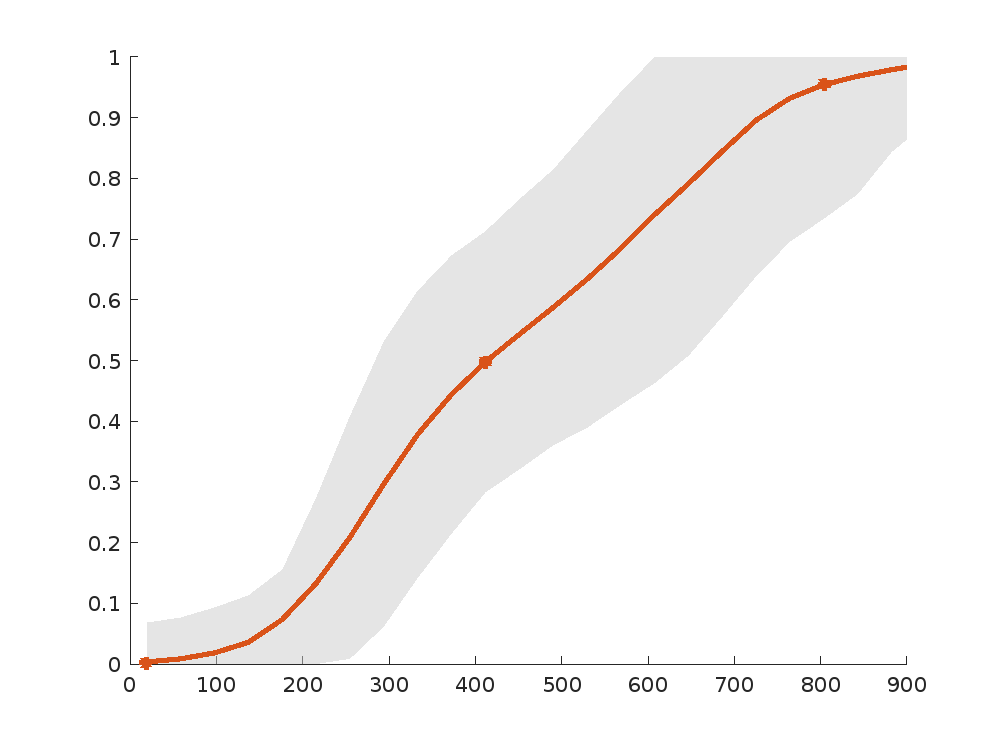}
    }
    \subfigure[Weighted counterfactual distributions $\bar{F}_S$]{
    \includegraphics[width=0.45\linewidth]{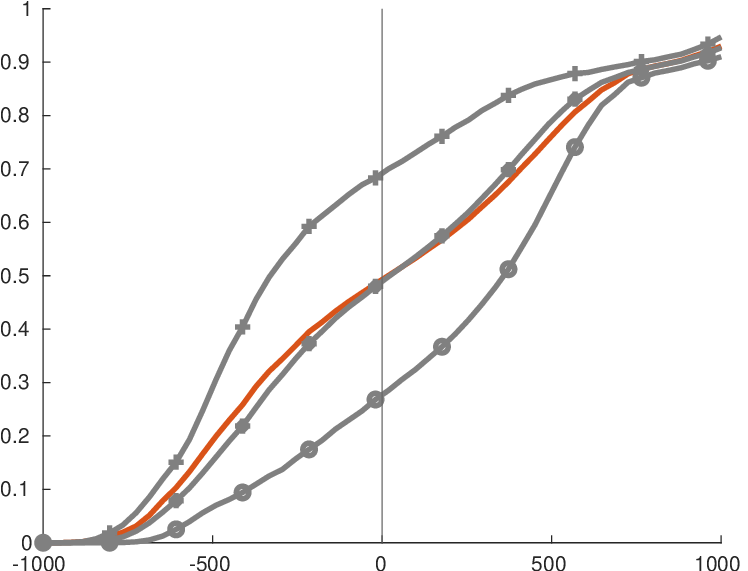}
    }
        \subfigure[Estimated transfers]{
    \includegraphics[width=0.45\linewidth]{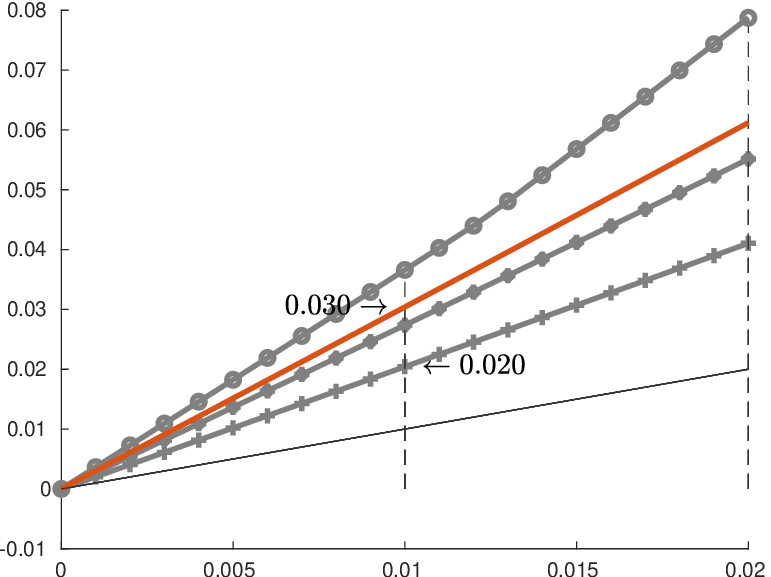}
    }
    \caption{Results from Model (2)}
        \floatfoot{Note: This figure reproduces the analysis for the more flexible specification of $r(x)$ that includes up to cubic functions of the wages in each sector, and interaction terms of the public sector wage with all other choice attributes. The results are less precise but qualitatively similar to those obtained with Model (1).}
    \label{fig:model_2}
\end{figure}

\begin{figure}
    \centering
    \subfigure[Being laid off]{
    \includegraphics[width=0.45\linewidth]{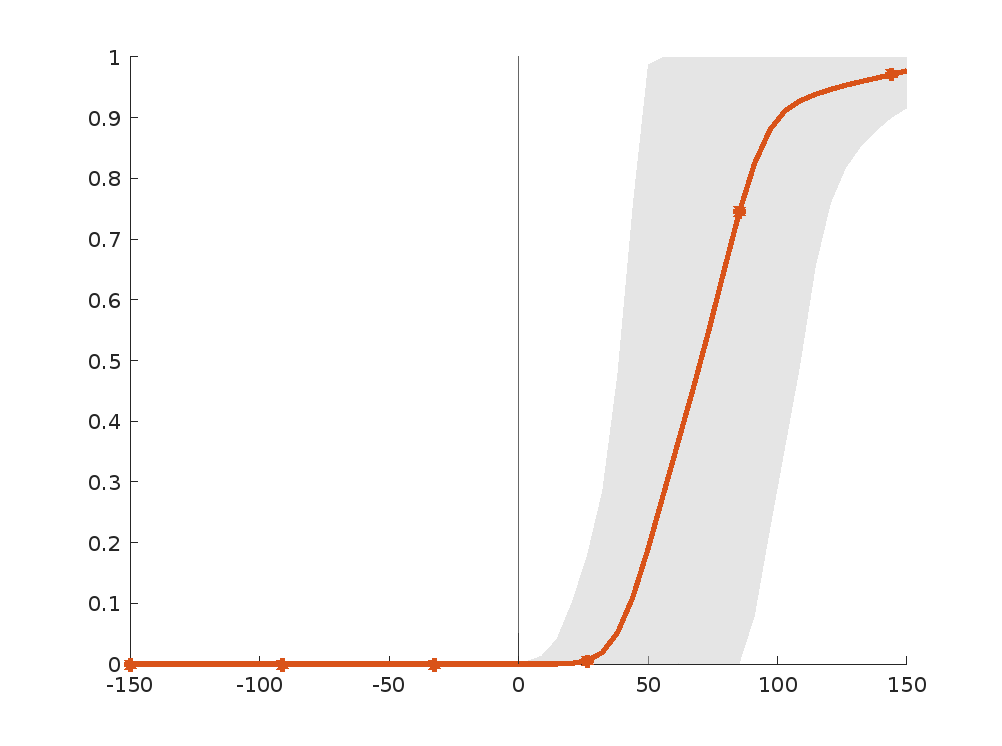}
    }
    \subfigure[Job promotion]{
    \includegraphics[width=0.45\linewidth]{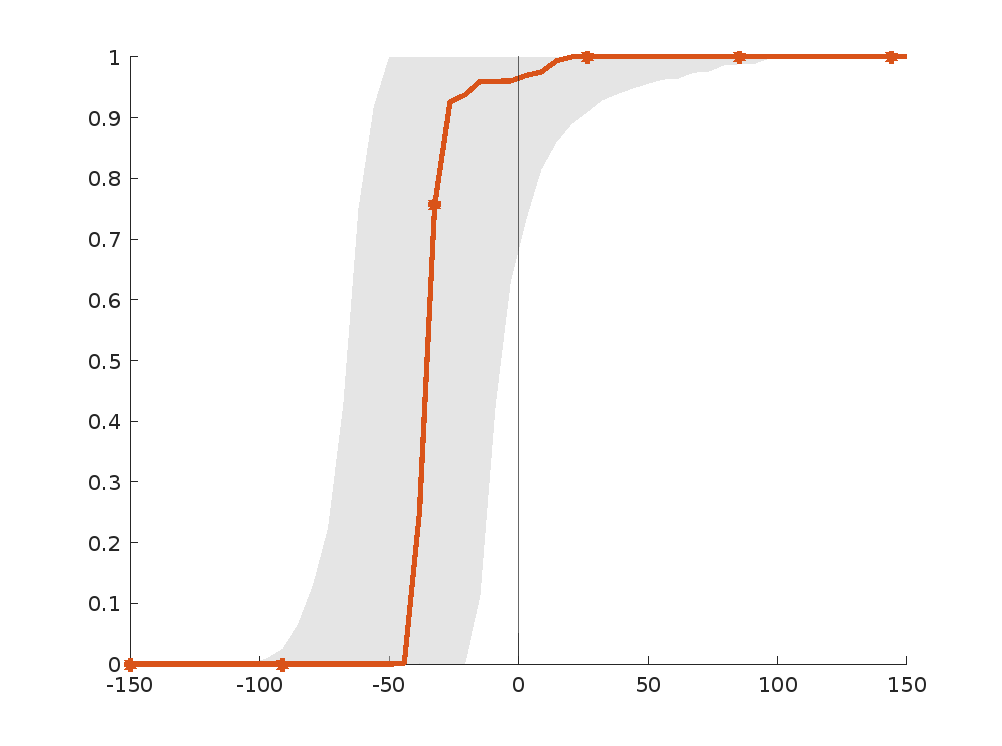}
    }
        \subfigure[Working hours]{
    \includegraphics[width=0.45\linewidth]{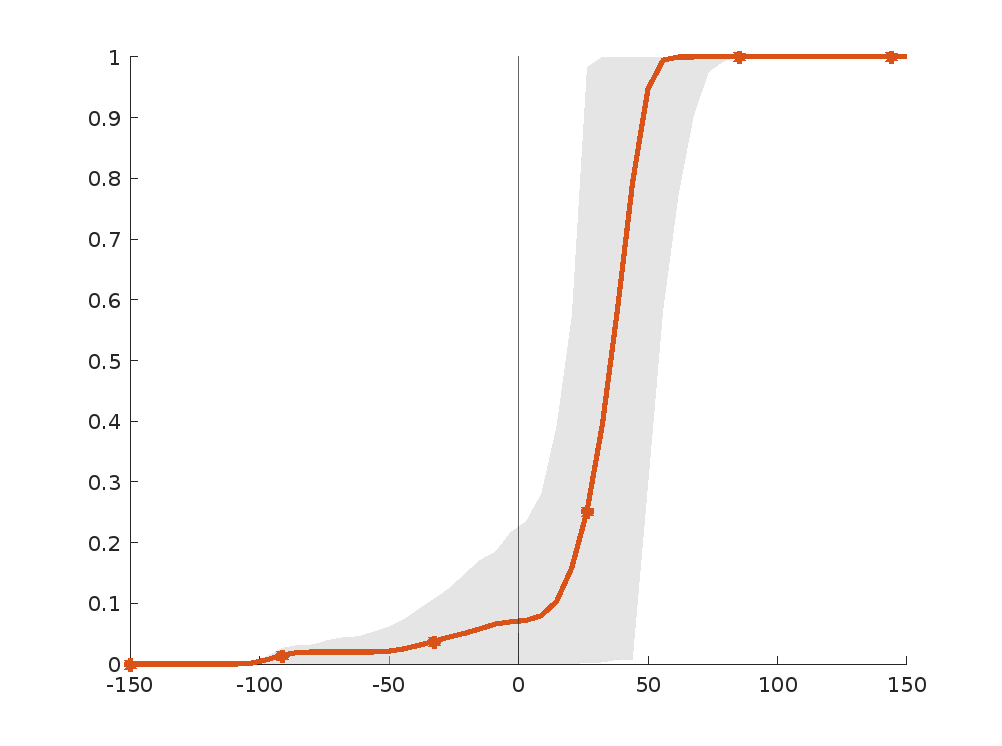}
    }
    \caption{Results from Model (2) - $mWTP$}
        \floatfoot{Note: This figure reproduces the analysis for the more flexible specification of $r(x)$ that includes up to cubic functions of the wages in each sector, and interaction terms of the public sector wage with all other choice attributes. The results are qualitatively similar to those obtained with Model (1).}
    \label{fig:model_2_wtp}
\end{figure}


\end{document}